\documentclass[]{article}
\usepackage{multicol}
\usepackage[utf8]{inputenc}
\usepackage{graphicx}
\usepackage{longtable}
\usepackage{csvsimple}
\usepackage{color}
\usepackage{algorithm}
\usepackage{algpseudocode}
\usepackage{amsmath}
\usepackage{xcolor}
\usepackage{hyperref}
\usepackage{authblk}
\usepackage{subcaption}
\usepackage{geometry}
\usepackage[export]{adjustbox}
\usepackage{caption}
\usepackage[backend=biber,style=numeric,sorting=none, maxbibnames=99]{biblatex}
\usepackage{wrapfig}
\usepackage{comment}
\usepackage{array}
\usepackage{xcolor}
\usepackage{pdfpages}
\usepackage{makecell}

\usepackage{tabularx}  
\usepackage{booktabs}  

\DeclareBibliographyDriver{article}{%
  \printnames{author}%
  \newunit\newblock
  \printfield{title}%
  \newunit\newblock
  \printfield{journaltitle}%
  \setunit{\addcomma\space}%
  \printfield{volume}%
  \setunit{\addcomma\space}%
  \iffieldundef{pages}{\printfield{eid}}{\printfield{pages}}%
  \setunit{\space}%
  \printtext{(\printfield{year})}%
  \iffieldundef{doi}{}{%
    \setunit{\addcomma\space}%
    \printfield{doi}}%
  \finentry
}

\definecolor{airforceblue}{rgb}{0.36, 0.54, 0.66}

\newcommand\rew[1]{{\color{black}#1}}
\newcommand\rev[1]{{\color{black}#1}} 

\title{Proximity-based cities emit less mobility-driven CO$_2$}

\author[1,2,3,*]{Francesco Marzolla}
\author[2,3,4,5]{Hygor P. M. Melo}
\author[2,3]{Matteo Bruno}
\author[1,2,3,6]{Vittorio Loreto}
\affil[1]{Sapienza Univ. of Rome, Physics Dept, Piazzale A. Moro, 2, 00185, Rome, Italy}
\affil[2]{Sony Computer Science Laboratories - Rome, Joint Initiative CREF-SONY, Centro Ricerche Enrico Fermi, Via Panisperna 89/A, 00184, Rome, Italy}
\affil[3]{Centro Ricerche Enrico Fermi (CREF), Via Panisperna 89/A, 00184, Rome, Italy}
\affil[4]{Postgraduate Program in Applied Informatics, University of Fortaleza, 60811-905, Fortaleza, CE, Brazil}
\affil[5]{Núcleo de Ciência de Dados e Inteligência Artificial (NCDIA), University of Fortaleza, 60811-905, Fortaleza, CE, Brazil}
\affil[6]{Complexity Science Hub, Josefst\"{a}dter Strasse 39, A 1080, Vienna, Austria}
\affil[*]{francesco.marzolla@uniroma1.it}

\date{}

\begin{document}

\maketitle

\begin{abstract}
In the quest for more environmentally sustainable urban areas, the concept of the 15-minute city has been proposed to encourage active mobility, primarily through walking and cycling. An urban area is considered a ``15-minute city" if every resident can access essential services within a 15-minute walk or bike ride from their home. However, there is an ongoing debate about the effectiveness of this model in reducing car usage and carbon emissions. In this study, we conduct a large-scale data-driven analysis to evaluate the impact of service proximity to homes on CO$_2$ emissions. By examining nearly 400 cities worldwide, we discover that, within the same city, areas with services located closer to residents produce less CO$_2$ emissions per capita from transportation. We establish a clear relationship between the proximity of services and CO$_2$ emissions for each city. Additionally, we quantify the potential reduction in emissions for 30 cities if they optimise the location of their services. This optimisation maintains each city's total number of services while redistributing them to ensure equal accessibility throughout the entire urban area. Our findings indicate that improving the proximity of services can significantly reduce expected urban emissions related to transportation.
\end{abstract}
\begin{multicols}{2}

\section*{Introduction}
Many countries worldwide have pledged to achieve carbon neutrality by 2050~\cite{IEA2023}. In a world where most people live in cities and the urban population is rising~\cite{WorldUrbanizationProspects}, building more sustainable urban 
environments is crucial to achieving this goal~\cite{IEA2023}. 

Urban mobility is a key aspect to address in this effort, as how people move within cities significantly contributes to environmental challenges. The transport sector is responsible for 21\% of the CO$_2$ emissions of the World~\cite{IEA2023}. Road transport, in particular, is the source of 16\% of the CO$_2$ emitted worldwide~\cite{IEA2023}, of 28\% of the CO$_2$ emitted in the European Union~\cite{etde}, and of 31\% of that emitted in the U.S.~\cite{underwoodDoesSharingBackfire2018}. Mobility in cities accounts for around 40\% of these emissions and is responsible for up to 70\% of other transport-related pollutants~\cite{Nanaki2017Environmental}. \rew{Cars, in particular, are estimated to emit around 3 billion tons of carbon dioxide per year globally, which corresponds to 8\% of total CO$_2$ emissions and to over one-third of emissions for transport~\cite{IEA2023}.

One promising solution to promote sustainable mobility over car-based systems in urban environments is the 15-minute city model.} This approach aims to design cities where residents can access their daily needs, such as work, shopping, healthcare, and leisure, within a 15-minute walk or bike ride from their homes~\cite{moreno2021introducing, hill2024beyond}.

Cities in which services are more easily accessible on foot are indeed found to be correlated with lower greenhouse gas emissions~\cite{marzolla2024compact15minutecitiesgreener}. This evidence should not surprise, since by promoting dense, mixed-use development and prioritising non-motorised modes of transportation, 15-minute cities are designed to reduce car dependency~\cite{allam202215,  khavarian202315, allam2022theoretical, sepehriXminuteCitiesGrowing2025}. 

However, only sometimes distributing services more uniformly over the areas of cities resulted in lowering greenhouse gas emissions. A case study on Beijing between 2000 and 2009~\cite{wangChangingUrbanForm2014} found that switching to a more decentralised urban form led to increased commuting distance and car usage, resulting in higher CO$_2$ emissions. Even building infrastructures for active mobility can be ineffective: a case study in three UK municipalities~\cite{BRAND2014284} found that newly built walking and cycling infrastructures increased physical activity but did not significantly reduce CO$_2$ emissions from motorised travel. Neither living near one of the infrastructures nor using it predicted changes in CO$_2$ emissions from motorised travel. Thus, the impact of different urban planning strategies on CO$_2$ emissions from transport still needs to be understood entirely. 


\rew{Some of the key characteristics of the 15-minute city have a positive impact on reducing emissions: in particular, high density, both in terms of population and Points Of Interest (POIs), and land-use mixing contribute to developing sustainable urban environments~\cite{choi2020exploring, Guneralp_2020}. There is, in general, robust evidence that denser urban areas are associated with lower transport emissions~\cite{gudipudi2016city, baur2014urban, pop_density_CO2}. Residential density, together with transit accessibility and intersection density, is positively related to active transportation and negatively related to motorised transportation~\cite{FRANK2010S99}. Similarly, shorter journeys are observed in cities with a higher density of POIs~\cite{noulasTaleManyCities2012}, and recent studies have found that residents of 15-minute areas tend to reach closer destinations~\cite{abbiasov202415}.
Trip lengths are generally shorter in locations with higher densities, or feature mixed land use~\cite{ewing2001travel, FRANK2010S99}. This
holds for both the home end (i.e., residential neighbourhoods) and the
non-home end (i.e., activity centres) of trips~\cite{ewing2001travel}. 
Local density and land-use patterns also affect mode choice. Public transport use depends primarily on local densities and
secondarily on the degree of land-use mixing~\cite{ewing2001travel}. Walking
depends as much on the degree of land-use mixing as on local densities~\cite{ewing2001travel}.
In general, any drop in
automobile trips with greater accessibility, density, or mix is roughly
matched by a rise in transit or walking-biking trips~\cite{ewing2001travel}.
A case study on Quebec City (Canada)~\cite{des2017greenhouse} showed that residential density and land-use mixing significantly lower greenhouse gas emissions for transport and motorisation rates, promoting active mobility.
Also a case study on the Seoul metropolitan area highlights the same positive impact of land-use mixing on active mobility~\cite{choi2020exploring}. 
They explain this finding by noting that in highly mixed-use areas, people can carry out various activities without having to travel far, as a wide range of facilities are typically located nearby.
Thus, mixed-use development is a fairly effective way to encourage sustainable transportation~\cite{choi2020exploring}.
Switching to active mobility contributes significantly to reducing greenhouse gas emissions: analysing data from seven European cities, Brand et al.~\cite{brand2021climate} found that an average person who ``shifted travel modes'' from car to bike decreased their life cycle CO$_2$ emissions by 3.2 kg/day.}

\rew{Focusing directly on the impact of the 15-minute city, a case study in the Lisbon
Metropolitan Area~\cite{colacco2025does} showed that this urban planning strategy increases non-motorised travel among its residents, promoting sustainable mobility, especially if coupled with high density. 
Nonetheless, a systematic, large-scale, empirical evaluation of the impact of implementing the 15-minute concept on mobility, and therefore on road transport emissions, is still lacking.}

In this paper, we aim to address the gap between proximity of services and transport emissions and investigate, through a data-driven study on a large number of cities, whether zones of cities that are more adherent to the 15-minute ideal emit less CO$_2$ from transport than zones with less proximity of services inside the same city. We also quantify the expected variation in transport emissions of cities \rew{if services were distributed in an efficient way to boost local accessibility~\cite{bruno2024universal}.}
We find that, inside several cities worldwide, zones with services more in proximity, i.e., more adherent to the 15-minute paradigm, emit less CO$_2$ for transport. We also find that most cities, if they underwent the relocation of services to adhere more to the 15-minute paradigm, would lower their transport emissions. The predicted change in CO$_2$ emissions for cities undergoing the \rew{idealised relocation} of services sheds light on the effectiveness of an actual implementation of the 15-minute paradigm in reducing transport emissions.

\section*{Methods}

\subsection*{Data collection and preprocessing}
\rew{To assess how pedestrian-friendly urban designs contribute to sustainable mobility, we examine 396 cities in various countries. For all the cities, we have the population distribution, the boundaries of urban areas, and per capita CO$_2$ emissions from transport.}

Population distribution data in cities were obtained from WorldPop~\cite{soton444004}. The boundaries of urban areas were sourced from shapefiles provided by the Organisation for Economic Co-operation and Development (OECD)~\cite{dijkstra2019eu}, specifically focusing on the defined core city. In instances where OECD data were unavailable, we used the core city boundaries from the Global Human Settlement files~\cite{fua2}.

The per capita CO$_2$ emissions from the road transport sector in 2021 were derived from the \rew{Emission Database for Global Atmospheric Research (EDGAR)~\cite{EDGAR, monica2022co2}, from the EU Joint Research Centre.} This dataset provides a gridded estimation of air pollutant emissions worldwide, categorised by sector and year, covering the period from 1970 to 2021. Emissions are measured in terms of the mass of pollutants emitted per unit of time and area. The estimates in the EDGAR dataset are based on fuel combustion data~\cite{monica2022co2} and include corrections for land use, land-use change, and forestry, as well as adjustments for reduction factors due to installed abatement systems. Importantly, this analysis does not account for CO$_2$ emissions resulting from biomass or biofuel combustion (short-cycle carbon). The EDGAR dataset relies on the International Energy Agency (IEA) data for CO$_2$ emissions from fossil fuel combustion~\cite{IEA}, which provides estimates from 1970 to 2019, broken down by country and sector. These emissions estimates are subsequently extended using a Fast Track approach, informed by British Petroleum statistics for 2020 and 2021~\cite{monica2022co2}. \rew{The spatial resolution of the dataset is of 0.1° of latitude times 0.1° of longitude. Therefore, on latitude the resolution is constant in length, being the spacing between datapoints in that direction 11 km; conversely, in longitude it ranges, among the cities studied, from 4.6 km at 65.55° north in Oulu, Finland, to 10 km at 21.25° north in Honolulu, in the U.S..

For every element of a hexagonal grid with elements of lateral size $l=200$ m, superimposed on urban areas under study, we also estimated the proximity of services.} The proximity of services to residential areas in cities has been measured in various studies~\cite{bruno2024universal, logan2022x, nicolettiDisadvantagedCommunitiesHave2023, olivariAreItalianCities2023, vale2023accessibility, guzman2024proximity}. In this work, we use the proximity time defined by Bruno et al. in~\cite{bruno2024universal}.
\rew{Denoted as $s$, it is a metric of pedestrian accessibility, which measures the average time a person needs to walk from a certain starting point to meet daily needs in the city~\cite{bruno2024universal}. A low proximity time, in a certain city's area, indicates that residents there can access services quickly and, therefore, good accessibility. 
In more detail, for each hexagon in each grid, we compute the average walking time (in minutes) from its centroid to reach one of the 20 closest Points of Interest (POIs) that satisfy a given daily need. These needs are assumed to be: education, healthcare, dining, supplies, public transportation, cultural activities, physical exercise, other services. For each category of services, we then have a time to access that kind of service on foot. The proximity time $s$ is the average of these category-specific times, measuring therefore the average time a resident would need to walk to access everyday services on foot.
The locations and categories of the POIs are derived from OpenStreetMap (OSM)~\cite{OSM}. OSM is a collaborative platform where private contributors can map the geography of places where they reside, have travelled or even have studied remotely with satellite images.
The quality of these crowd-sourced data can be assessed along multiple dimensions~\cite{senaratne2017review}, one of which is completeness, quantifying how many of the POIs in a region have been mapped onto the platform. 
Although the positional and thematic accuracies of OSM datasets are generally comparable to those of official reference data~\cite{zhang2022assessing},  completeness varies across different study areas~\cite{barrington2017world}.
For this reason, our analysis included only cities located in countries classified as high-income by the World Bank~\cite{WB_classification}, where the completeness of OSM data is sufficiently high~\cite{herfort2023spatio} to allow accurate measurement of proximity time~\cite{bruno2024universal}. There is indeed an observed decline in the coverage of the OSM mapping of POIs in regions with lower economic status~\cite{Polonia}, with completeness of OSM data being particularly low in lower-income countries in the case of informal settlements~\cite{Lagos_slum}. Additionally, lower-income countries tend to have lower motorisation rates~\cite{carsmoney}. This suggests that the ability to afford a car plays a more significant role in transportation choices in those countries than it does in high-income countries. As a result, the correlation between emissions and service proximity is weaker in lower-income countries~\cite{marzolla2024compact15minutecitiesgreener}.}

Our datasets consist, therefore, of two types of grids: a coarser rectangular grid derived from the EDGAR dataset for emissions and a finer hexagonal grid for measuring proximity time. For each element of the emission grid, we computed the population-weighted average of the proximity time values $s$ from the hexagons whose centroids are located within the corresponding rectangular element.
We included in the study only emission grid elements that contained at least five hexagons from the accessibility grid. We further filtered the data, retaining only those cities for which we had at least nine mapped emission grid elements. \rew{Our initial dataset with proximity time values encompassed 699 cities;} after these two rounds of filtering, we ended up with a sample of 396 cities. \rew{These cities are listed in the Supplementary Information.
From this point onward, all analyses were conducted at the emission grid level.}

\subsection*{Statistical analysis of proximity time vs emissions}

\rew{For each city, we computed the Pearson correlation coefficient between the logarithms of proximity time $s$ and road traffic emissions per capita $C_\text{pc}$, using one data point per cell.}

We then computed the p-value for the correlation between emissions and proximity time at the intra-city level. To calculate it, we generated a null scenario by shuffling 1000 times the data grid inside the same city, controlling for population. \rew{ At each shuffling iteration, the emissions the element $i$ of the grid are assigned to an element $j$ with probability 
\begin{equation}
    p_{ij} \propto \frac{1}{|P_i-P_j|} \, ,
\end{equation}}
where $P_i$ represents the population residing in the element $i$ of the grid and $P_j$ the population residing in the element $j$.

\rew{We collected in a histogram the Pearson correlation coefficients between the logarithms of proximity time $s$ and CO$_2$ emissions $C_\text{pc}$ at the emission grid level for all the 1000 realisations of the shuffling. This distribution provides the null-case scenario of no correlation. We then collected in a histogram with the same binning the actual correlation coefficients, coming from real data points. The bins were taken of variable length to have at least 5 cities in each of them in the null-case distribution. To assess if the correlation between $s$ and $C_\text{pc}$ inside cities is statistically significant, we finally computed the $\chi^{(2)}$ of the frequencies registered for each histogram bin with respect to the probabilities of falling in each bin in the random scenario, estimated by an appropriate normalisation of the former histogram described. We obtained a value of $\chi^{(2)} = 10 \, 112$, with 30 degrees of freedom.}

Subsequently, we performed a linear least-squares regression between the logarithms of proximity time $s$ and emissions $C_\text{pc}$, separately for each city. We then computed the following normalised residual $\lambda$ of each point log$\,(C_{pc, i}) := y_i$ respect to the linear regression prediction $\hat{y}_i$
\[
\lambda_i = \frac{|y_i - \hat{y}_i|}{\sigma} \, ,
\]
with 
\[
\sigma = \sqrt{ \frac{\sum_j (y_j - \hat{y}_j)^2}{n-2}
} \, .
\]
Data points having $\lambda_i > 4$ ($4.9\%$ of the points) have been considered outliers, and therefore excluded. Logarithms of CO$_2$ emissions $C_\text{pc}$ have then been linearly fitted again versus logarithms of proximity time $s$ for each city. 
\rew{Linearly fitting the logarithms of $C_\text{pc}$ and $s$ is equivalent to fitting to data a power law of the form~\cite{marzolla2024compact15minutecitiesgreener}
\begin{equation}
    C_\text{pc} = A s^\gamma \, ,
    \label{powerlaw_methods}
\end{equation}
where $A$ and $\gamma$ are fitting parameters.
In the Supplementary Information, the plots of these fittings and maps showing the grid elements' spatial location corresponding to data points are collected.} 
The distribution of the exponents $\gamma$  of Eq. (\ref{powerlaw_methods}) obtained from such fittings, one for each city, is not Gaussian at 95$\%$ C.L. (KS test). For this reason, the confidence interval of the estimation of the general exponent $\gamma$ across cities has been computed by integrating the histogram collecting all city-specific exponents $\gamma$ around the mean value until obtaining a 68$\%$ confidence interval.

\subsection*{Relocation of services}

\rew{To model optimised scenarios for 15-minute cities, Bruno et al.~\cite{bruno2024universal} have introduced a framework for the relocation of services to equalise accessibility, simulating this relocation in various cities worldwide. They relocate services to obtain an equal number of services per capita in each 15-minute radius in the city, implementing the philosophy of proximity by moving services where more residents need them. }

\rew{The algorithm first calculates, for each cell in a city grid, the number of residents who can reach that area within 15 minutes of walking. It also determines the capacity of a service, which is defined as the total resident population of the city divided by the number of services of a specific type available (for example, if there are 100 restaurants in a city with a population of 10,000, then each restaurant has a capacity of serving 100 people). The algorithm then iteratively allocates a point of interest (POI) in the area that is accessible by the largest number of residents. It reduces the demand for service in that 15-minute area by the capacity of the service. By repeating this process multiple times until all services in the city have been addressed, the algorithm produces an optimal distribution where the number of services per capita remains roughly constant across each 15-minute area.}

\rew{
The steps of the relocation procedure are as follows:
\begin{itemize}
    \item Create a grid of the city where each cell has a resident population ;
    \item Compute for each cell the 15-minute neighbourhood as the set of cells that can be reached in 15 minutes walking;
    \item Compute the average capacity of a POI as the resident population divided by the number of POIs in the city;
\end{itemize}
Then, iteratively, for all POIs of each category of services to be relocated:
\begin{itemize}
    \item Identify the cell with the highest population in its 15-minute neighbourhood;
    \item Allocate one POI within the selected 15-minute neighbourhood, with a probability proportional to the resident population of each cell in the neighbourhood;
    \item Subtract an amount of "satisfied demand" population from the cells in the 15-minute neighbourhood, summing up to the average capacity of the POI and proportionally to the population resident in the cells.
\end{itemize}}

\rew{A more detailed version can be found in the original paper~\cite{bruno2024universal}.}

\subsection*{Estimation of optimised emissions}

For each city $i$ considered, the emissions \rew{$C_{\text{pc, } ij}^\text{opt} = C_{\text{pc}} (s_{ij}^\text{opt})$} of a grid element $j$ when optimised for proximity to have proximity time $s_{ij} ^\text{opt}$, is estimated as follows.
Assuming that Eq.~(\ref{powerlaw_methods}) holds, a naive estimator would be
\rew{
\begin{equation}
    C_{\text{pc}}^\text{exp} (s^\text{opt}) = A_i \cdot \, ({s^\text{opt}})^{\gamma_i} \, .
    \label{naive}
\end{equation}
}
We build from this and define the following quantity, which quantifies how much the real emissions of a zone $j$ differ from the ones expected by Eq.~(\ref{powerlaw_methods}), or equivalently Eq.~(\ref{naive}), based on its proximity time:
\rew{
\begin{equation}
    \Delta_{ij} = \frac{C_{\text{pc, }ij} - C_\text{pc}^\text{exp}(s_{ij})}{C_\text{pc}^\text{exp}(s_{ij})} \, ,
    \label{Delta}
\end{equation}
where $s_{ij}$ is the real value, non-optimised, of proximity time of the grid element $j$ of city $i$.
We finally estimate the emissions of the element $j$ of city $i$ after the relocation as 
\begin{equation}
    C_{\text{pc,}ij}^\text{opt} = (1 + \Delta_{ij}) \, C_\text{pc}^\text{exp}(s_{ij}^\text{opt}) \, . 
\end{equation}
}

\section*{Results}

\subsection*{Emissions inside a city}
\begin{figure*}[htbp]
\centering
\includegraphics[width=\textwidth]{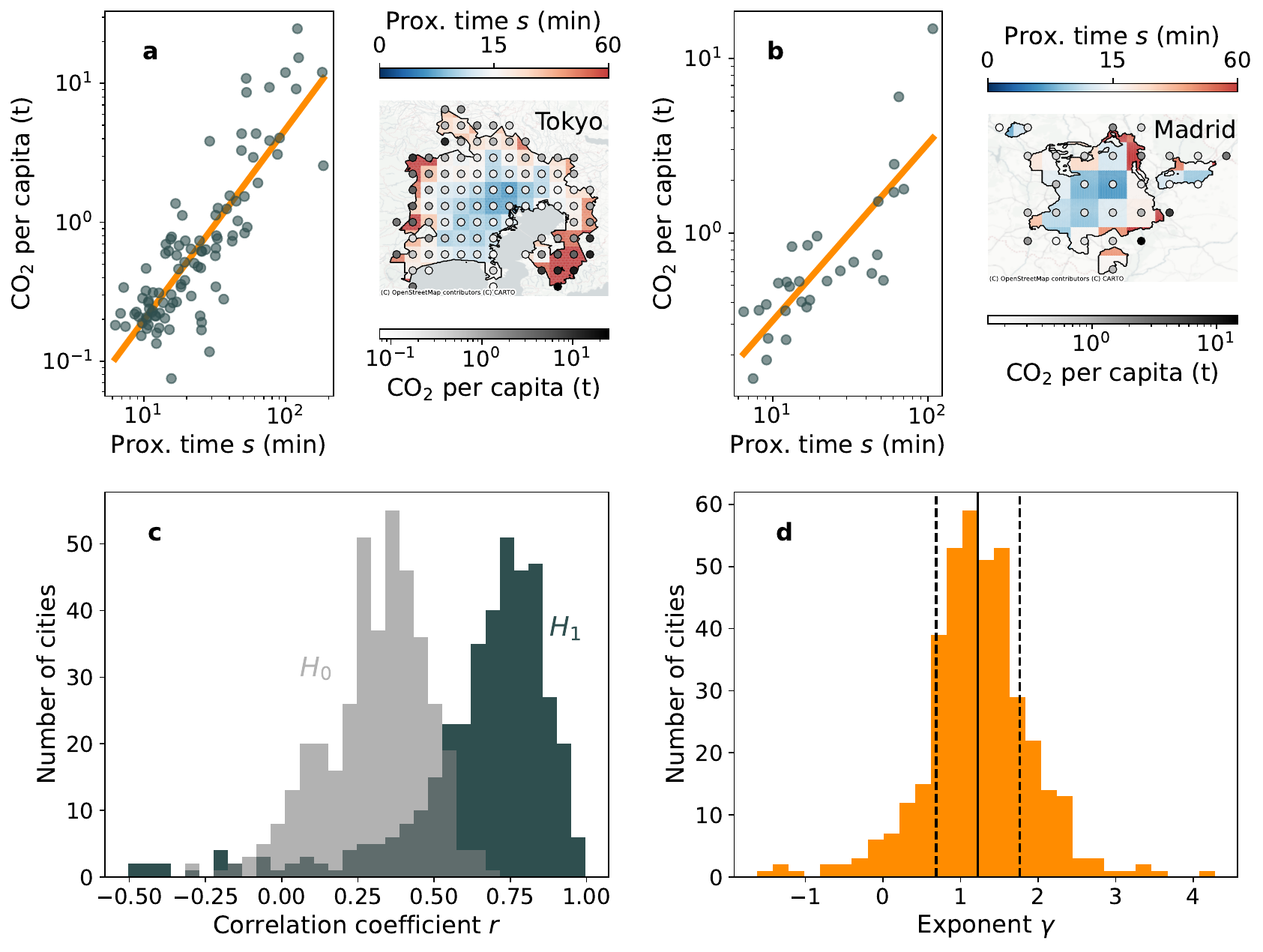}
\caption{\textbf{Correlation between proximity and CO$_2$ emissions within cities.} 
\rev{Panels (a) and (b) show, for Tokyo and Madrid, the relationship between population-weighted proximity time $s$ and per capita CO$_2$ emissions from road transport ($C_\text{pc}$) in 2021. Each dot represents a 0.1° $\times$ 0.1° grid cell; maps on the right display the same grid coloured by proximity time and emissions. Orange lines are power-law fits (Eq.~\ref{powerlaw_methods}). Panel (c) shows the distribution ($H_1$) of correlation coefficients between $\log C_\text{pc}$ and $\log s$ across all cities, compared to a population-controlled randomisation ($H_0$, grey). Panel (d) shows the distribution of power-law exponents $\gamma$, with the mean (solid) and 68\% confidence interval (dashed).
}
}
	\label{fig:quadratiNord}
\end{figure*}

\rew{Figures~\ref{fig:quadratiNord}a and~\ref{fig:quadratiNord}b show proximity time $s$ and per capita CO$_2$ emissions from road transport $C_\text{pc}$ for Tokyo and Madrid at the emission grid scale.} The two quantities are positively correlated, and peripheral areas are the ones emitting the most per capita, with the same areas also having worse accessibility (\rew{marked by} higher proximity time).
\rew{In Supplementary Information, analogous plots are collected for all the 396 cities considered. Together with data points, in the planes $C_\text{pc}$ vs $s$, are also shown the fitting power laws, of the form of Eq. (\ref{powerlaw_methods}).}

The histogram \rew{labelled $H_1$} in Fig.~\ref{fig:quadratiNord}c shows the distribution of the correlation coefficients \rew{between the logarithms of proximity time $s$ and road emissions $C_\text{pc}$ across} cities. \rew{The histogram labelled $H_0$ shows the expected distribution of correlation coefficients across cities in the case of no correlation between proximity of services and road emissions.
The comparison of these two distributions} allows us to conclude \rew{at a $p$-value $p=0.00$} that proximity time $s$ and road emissions $C_\text{pc}$ inside a city correlate.

The histogram in Fig.~\ref{fig:quadratiNord}d shows the distribution of the \rew{fitted values of the} exponent $\gamma$ of the power law in Eq.~(\ref{powerlaw_methods}) \rew{across cities}. From such distribution, we can estimate \rev{the} average exponent of the power law linking proximity time $s$ and CO$_2$ emissions per capita for road transport $C_\text{pc}$ inside a city as
\begin{equation}
    \gamma =1.2\pm0.5 \, .
    \label{gamma_intracity}
\end{equation} 
This exponent differs from zero, and the difference is statistically significant. 
This evidence reveals a trend \rew{linking} areas with shorter proximity times to lower carbon emissions from transport. \rew{A shorter proximity time reflects a higher degree of service proximity, which represents the core principle of the 15-minute city paradigm.}
\subsection*{Expected emissions' variation after accessibility optimisation}
\begin{figure*}[htbp]
	\centering
    \includegraphics[width=\textwidth]{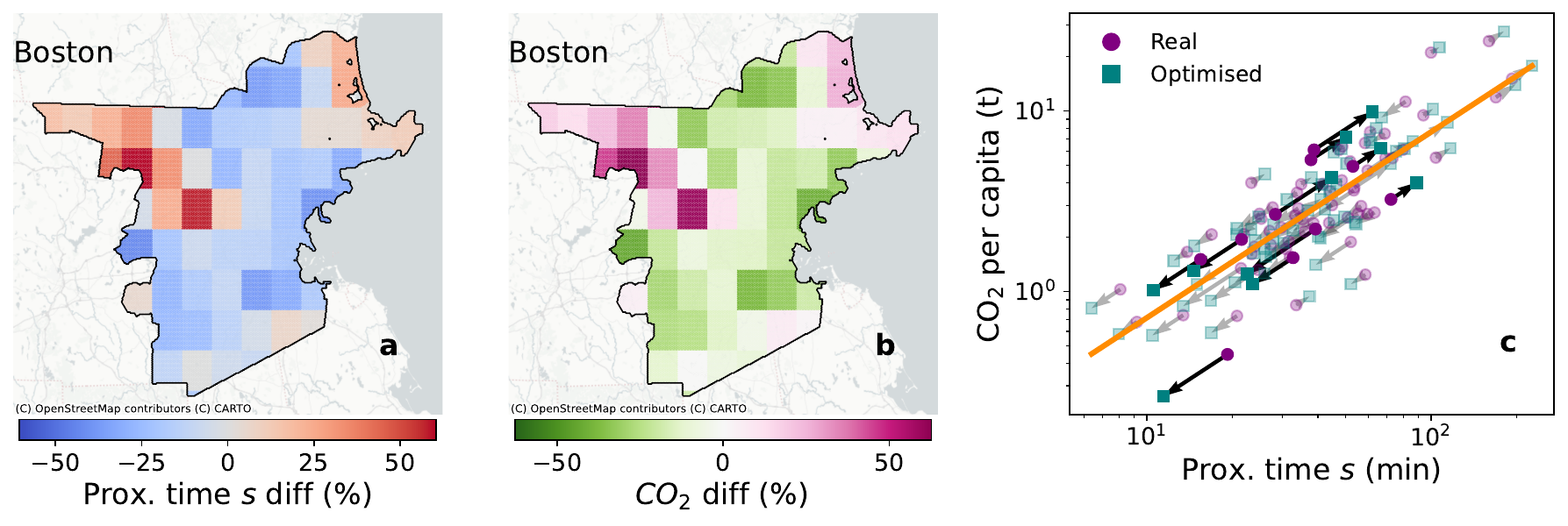}
    \caption{\rev{\textbf{Optimising walking accessibility and its expected impact on transport emissions in Boston.} Panels (a) and (b) show Boston maps with colour gradients indicating, respectively, the percentage change in proximity time and in expected transport-related CO$_2$ emissions after optimisation. Panel (c) compares current (circles) and optimised (squares) grid elements, linked by arrows; the orange line is a power-law fit of emissions versus proximity time.}}

    \label{fig:variation_city}
\end{figure*}
In Fig.~\ref{fig:variation_city}a, we depict the effect \rew{of the POIs relocation for optimising services proximity} 
for the city of Boston, US. The colour scheme encodes the \rev{percentage} variation in proximity time after the relocation.
In Fig.~\ref{fig:variation_city}b, we show the expected outcome of the \rew{service relocation} in terms of emissions: \rev{the colour scheme encodes the percentage variation in CO$_2$ emissions for transport per capita.} 

This expected emission variation is also shown in Fig.~\ref{fig:variation_city}c, \rew{where dots encode proximity times and emissions of Boston grid elements before the relocation step and squares after it.} 
In the Supplementary Information, we show \rew{analogous figures for} all 30 cities for which we computed the optimised scenario.
\newcolumntype{Y}{>{\centering\arraybackslash}X}
\begin{table*}[ht]
\centering
\caption{City values for the proximity time and the transport emissions per capita before and after the POIs relocation procedure, with the power law exponent of the fit between the two quantities.}
\label{tab:cities}
\begin{tabularx}{\textwidth}{l|YYYYYY}
\textbf{\small City} & \thead{Power-law\\ exponent} & \thead{Prox. time \\(min:sec)} & \thead{Prox. time, \\optimised \\(min:sec)} & \thead{CO$_{2}$ p.c. \\(t)} & \thead{CO$_{2}$ p.c., \\optimised\\ (t)} & \thead{CO$_{2}$ \\ variation}\\
\toprule
\begin{filecontents*}{tabella_risultati_main.csv}
Oslo,1.1, 9:52, 7:10,0.14,0.11,-20\%
Lisbon,2.0, 9:44, 6:26,0.27,0.16,-42\%
Edinburgh,2.1, 8:17, 5:55,0.71,0.35,-50\%
Helsinki,1.6,12:58, 9:26,0.33,0.21,-36\%
Milan,1.3, 6:42, 4:55,0.32,0.15,-53\%
Prague,-0.3,11:11, 8:07,0.41,0.45,+10\%
Budapest,0.1,11:12, 7:20,0.33,0.33,-2\%
Auckland,-0.5,14:01, 9:44,0.52,0.61,+18\%
Munich,0.8, 7:46, 5:38,0.39,0.31,-21\%
Vienna,0.5, 8:31, 5:34,0.35,0.28,-18\%
Sapporo,1.5,14:59,11:07,0.38,0.24,-36\%
Barcelona,1.8, 9:02, 5:49,0.25,0.13,-48\%
Warsaw,-0.2,10:04, 6:25,0.48,0.52,+8\%
Montreal,0.8,13:10, 8:10,0.48,0.40,-16\%
Rotterdam,-0.0,37:20,21:35,0.84,0.84,+0.4\%
Athens,0.7,11:06, 9:02,0.34,0.29,-14\%
Fukuoka,1.8,18:38,13:41,0.52,0.31,-40\%
Milwaukee,0.8,17:10,10:35,1.41,0.97,-31\%
Amsterdam,0.6,18:37,12:40,0.89,0.73,-18\%
Berlin,1.4, 8:25, 5:44,0.41,0.24,-40\%
Rome,1.4,13:34, 8:52,0.63,0.37,-41\%
Madrid,1.0,11:16, 7:40,0.45,0.32,-29\%
Paris,0.4, 7:56, 6:03,0.32,0.29,-10\%
Atlanta,0.3,50:25,42:13,1.36,1.31,-4\%
Minneapolis,1.0,26:45,19:51,2.43,1.91,-21\%
Osaka,1.5,12:20, 8:58,0.39,0.26,-33\%
Boston,1.0,23:11,19:42,1.80,1.61,-11\%
Tokyo,1.4,12:17, 9:13,0.29,0.21,-25\%
Seoul,0.7,14:59,11:13,0.47,0.39,-16\%
Dallas,0.4,36:26,25:50,1.70,1.52,-10\%
\end{filecontents*}
\csvreader[
    no head,
    late after line=\\\hline,
    late after last line=\\
]{tabella_risultati_main.csv}{}%
{%
\csvcoli & \csvcolii & \csvcoliii & \csvcoliv & \csvcolv & \csvcolvi & \csvcolvii }
\end{tabularx}
\end{table*}
%
%
%
\begin{figure*}[htbp]
  \centering
\includegraphics[width=\textwidth]{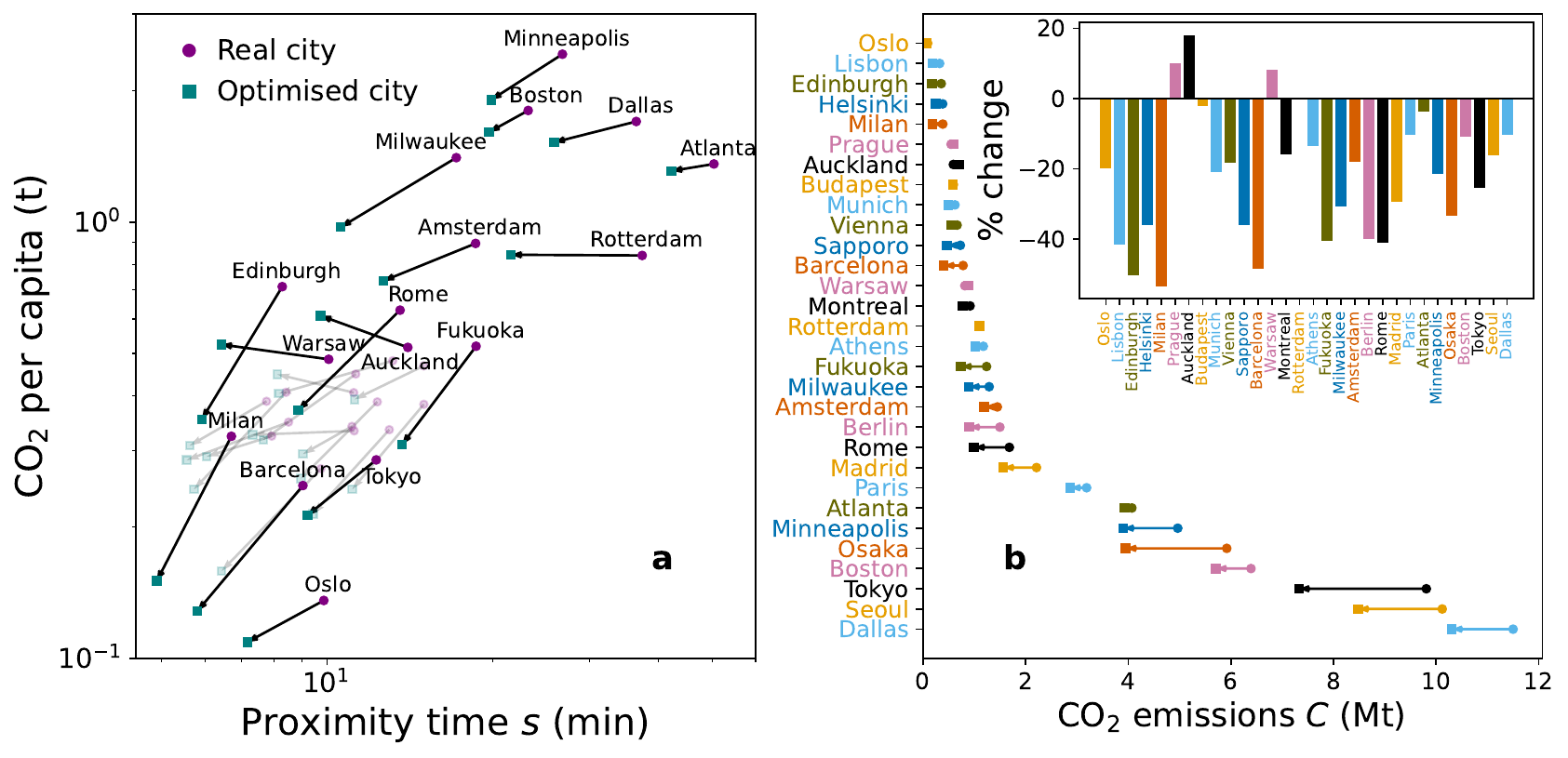}
\caption{\rev{\textbf{Emissions of accessibility-optimised cities.} Panel (a) shows the trajectories of cities in the log-log plane of per capita transport CO$_2$ emissions versus proximity time, before and after service relocation. Panel (b) shows the expected change in cumulative transport CO$_2$ emissions for the cities included in the optimisation process.}}

\label{fig:frecce}
\end{figure*}
Figure~\ref{fig:frecce}a shows the expected trajectories of cities traced in the $C_\text{pc}$ vs $s$ plane while they undergo the proximity optimisation process. We can see a tendency of the arrows to be directed towards the bottom left of the plane, decreasing both their proximity time and, consequently, their emissions. While the translation from right to left of the coordinates describing cities has to be expected as an intended consequence of the relocation algorithm, the tendency of going towards lower values of CO$_2$ per capita is due to the link between higher proximity of services and lower emissions for transport, which is present in most of the cities under study. The shift towards lower emissions under proximity optimisation of cities gives insight into the possible effectiveness of implementing the 15-minute paradigm in those cities in fostering more sustainable urban mobility.

Finally, Fig.~\ref{fig:frecce}b shows the variation in total CO$_2$ \rew{yearly emitted by each city for road transport}, denoted $C$. Twenty-seven out of thirty cities experience a reduction in emissions under proximity optimisation. \rev{Table \ref{tab:cities} reports proximity time and transport emissions before and after accessibility optimisation, the percentage variation of CO$_2$ emissions, and the coefficient $\gamma$ of the power law linking CO$_2$ emissions and proximity time within each city (Eq.~\ref{powerlaw_methods}). Cities are listed in the same order as in Fig.~\ref{fig:frecce}b to facilitate comparison.
In cities where per-capita CO$_2$ does not decrease, accessibility and emissions are anti-correlated. This may occur when external factors outweigh the city’s internal structure and functionality. For instance, in Rotterdam the presence of the port could significantly influence transport-related emissions. Additional cases are discussed in the SI.

}

\section*{Discussion and limitations}
The 15-minute city concept offers a new approach to optimising resource allocation within urban areas. It emphasises bringing activities closer to neighbourhoods instead of requiring people to travel to centralised locations for those activities. This shift in perspective liberates residents from the necessity of quickly reaching a downtown area, which is typically seen as the sole hub of city life.

When services are located close to home, there is less need to use the fastest means of transport to reach them, at least for everyday needs. As a result, cars can be replaced with active transportation methods for nearby activities. This study aims to determine whether the necessity for car usage is indeed reduced in 15-minute neighbourhoods. Given that only about $3\%$ of cars worldwide are electric~\cite{IEA_veicoli, IEA_veicoli_elettrici}, we interpret the observed variations in CO$_2$ road emissions as indicators of changes in car usage. Our findings show a reduction in road transport emissions in areas with nearby services, suggesting that the 15-minute city model leads to either reduced car use or shorter trips, ultimately decreasing emissions. This evidence is consistent with previous research linking accessibility to changes in mobility behaviour~\cite{abbiasov202415}.

In~\cite{marzolla2024compact15minutecitiesgreener}, it was found that CO$_2$ emissions per capita for transport at the level of the whole city scale linearly with proximity time, i.e. Eq.~\ref{powerlaw_methods} is valid among cities, with $\gamma = 1.01 \pm 0.06$. Here, we add that Eq.~\ref{powerlaw_methods} is also valid inside cities, among different zones of the same urban area, with $\gamma = 1.2\pm0.5$, which is compatible with linearity at $1 \sigma$. Unlike~\cite{marzolla2024compact15minutecitiesgreener}, our analysis excludes emissions from rail transport to prevent biases at the small scale considered. Our findings indicate that the lack of proximity to services contributes to the higher transport emissions generated in the suburbs compared to city centres, aligning with the results found in~\cite{jones2014spatial} for US cities. Most of the cities examined exhibit a core-periphery structure in their proximity time distribution, with downtown areas showing lower proximity times than peripheral regions.

In the second part of the study, we predicted the impact on emissions of optimal implementation of the 15-minute paradigm in cities, following the framework proposed in~\cite{bruno2024universal}. As the authors note, applying this strategy in practice is not always straightforward. In sprawling, car-dependent cities, such as many located in the United States or Australia, low population density and rigid land-use planning present significant challenges. Zoning laws often enforce functional separation, leading to structural lock-in~\cite{khavarian2023garden}. 
In contrast, European cities typically have more adaptable infrastructure to the 15-minute model ~\cite{bruno2024universal,guzman2024proximity}. Interest in applying this concept is also growing in the Global South, which is experiencing increasing urbanization~\cite{allam202215_net_zero,swilling2018weight}. However, the implementation of the 15-minute city paradigm in these regions faces obstacles such as informal settlements, economic disparities, and insufficient infrastructure~\cite{guzman2024proximity}. 

It is important to note that the implementation of the 15-minute city must consider potential issues such as socio-economic segregation~\cite{abbiasov202415}. 
\rew{Additionally, the idea of complete self-sufficiency within every neighbourhood implies a level of decentralisation that may be impractical, particularly for higher-order services like hospitals or universities~\cite{mouratidis2024time}. However, a more moderate form of decentralisation, focused on essential daily needs, is feasible. Accessibility to higher-order services should be addressed through public transport, fundamental to address also specific mobility needs, such as those of elderly people or people with disabilities~\cite{calafiore202220}. The implementation of the 15-minute city should also ensure the quality and value of nearby services.~\cite{hill2024beyond}. Social mixing and inclusion are key goals of the 15-minute city, which aims to achieve these objectives by providing opportunities to underserved neighbourhoods and incorporating a variety of housing types within the same area. This approach seeks to attract residents with different income levels to live together~\cite{allam2022theoretical}, but it can potentially generate social threats such as gentrification and social tensions, particularly when it is underpinned by a pathologisation of poverty~\cite{casarin2023rethinking}.
Moreover, the novelty of the 15-minute paradigm is often overstated, as it builds on long-standing urban planning principles~\cite{d2024liveable, howard1898, CNU,  perry1929neighbourhood, krier1977, haaland2015challenges, d2013simulating}.  

We identify two major limitations of our work. Firstly, our analysis focuses exclusively on cities located in high-income countries. As previously mentioned, this choice is motivated by the lower quality of available data in lower-income countries and by the more pronounced influence of private vehicle affordability on mobility behaviours in those countries, which tends to weaken the relationship between urban form and sustainable travel choices. The second limitation we acknowledge concerns the resolution of the emissions dataset, which is based on a relatively coarse-grained grid. Each grid cell likely includes multiple neighbourhoods with varying levels of proximity. Nevertheless, cells that include neighbourhoods with higher average proximity times tend to exhibit higher emissions. We therefore believe that this limitation does not compromise the ability of our methodology to reveal the relationship between service proximity and transport-related emissions.
}

\section*{Conclusions}
This study explores the impact of \rew{proximity of services, central in the} the 15-minute city model\rew{,} on urban emissions. By analysing nearly 400 cities in high-income countries, we find that disparities in the proximity of services within cities are significantly linked to mobility-related carbon emissions, with areas designed around the 15-minute city concept\rew{, therefore featuring a higher degree of services proximity,} demonstrating lower transportation emissions. Thus, mobility tends to be more sustainable in zones that adhere closely to this framework.

Additionally, our analysis of 30 cities in the second part of the study indicates that most would see a reduction in transport emissions if they adopted the proposed implementation of the 15-minute city, as outlined in ~\cite{bruno2024universal}. This approach aims to relocate services closer to residents throughout the city, thereby reducing inequalities in access to essential services \rew{providing an idealised scenario of increased adherence to the 15-minute paradigm for each city}. While the practical application of the 15-minute city model must be tailored to the unique context of each city, \rev{and further work will be needed to identify the densification and land-use objectives needed to achieve such a model,} our findings suggest that it can effectively decrease transportation emissions and promote more sustainable mobility.

\rew{ 
\section*{Acknowledgments}
Hygor P. M. Melo acknowledges the support of Fundação Edson Queiroz, Universidade de Fortaleza, and Fundação Cearense de Apoio ao Desenvolvimento Científico e Tecnológico.

\section*{Data availability}
The population distribution data used can be downloaded from WorldPop~\cite{WorldPop}. The boundaries of urban areas can be sourced from OECD~\cite{OECD_link}, and, in instances where OECD data are unavailable, from the Global Human Settlement files~\cite{GHS_link}. The EDGAR dataset is also publicly available \cite{EDGAR}. To compute proximity times, POIs data are publicly available on OpenStreetMap~\cite{OSM}, while travel times were calculated using the Open Source Routing Machine
(OSRM)~\cite{luxen2011real}.

\section*{Code availability}
The code used for the analysis and visualisations is available from the corresponding author upon reasonable request.

\section*{Author Contributions}
Research design and study concept: F.M., M.B., H.P.M.M.. Data analysis: F.M.. Result interpretation: all authors. Manuscript drafting: F.M.. Manuscript review and editing: all authors.
All authors have read and approved the manuscript.

 \section*{Competing Interests}
 The authors declare no competing interests.
 }

\printbibliography

\end{multicols}


\newpage 
\appendix



\section*{Supplementary material}

# Supplementary Information for
# Proximity-based cities emit less mobility-driven CO$_2$

## Results tables

Supplementary Table 1: Cities' metrics: population, area, average proximity time, average annual CO$_2$ emissions for road transport per capita, correlation coefficient between proximity time and emissions inside the city, parameters of power-law fitting of the form of Eq. (2) in the main text.

| City | Country | Population | A (km$^2$) | $s$ (min) | C$_{\text{pc}}$ (t) | $r$ | $A$ | $\gamma$ |
|---|---|---|---|---|---|---|---|---|
| Abbotsford | CAN | 154993 | 382 | 19.7 | 1.95 | 0.922 | 0.083 | 1.066 |
| Aberdeen | GBR | 232211 | 185 | 11.2 | 0.81 | 0.915 | 0.004 | 1.936 |
| Ada | USA | 487270 | 2552 | 28.8 | 1.49 | 0.781 | 0.005 | 1.603 |
| Akita | JPN | 285456 | 867 | 19.8 | 1.3 | 0.851 | 0.006 | 1.746 |
| Alachua | USA | 240090 | 2343 | 51.2 | 3.54 | 0.713 | 0.097 | 0.952 |
| Albacete | ESP | 164625 | 1121 | 20.4 | 1.47 | 0.554 | 0.125 | 0.912 |
| Albany | USA | 449673 | 1917 | 34.7 | 3.61 | 0.932 | 0.04 | 1.214 |
| Albuquerque | USA | 749581 | 2510 | 29.2 | 1.88 | 0.734 | 0.039 | 1.129 |
| Allen | USA | 360427 | 1704 | 45.5 | 2.87 | 0.647 | 0.042 | 1.08 |
| Amsterdam | NLD | 1519130 | 1019 | 19.3 | 1.17 | 0.328 | 0.111 | 0.596 |
| Aomori | JPN | 265822 | 790 | 29.8 | 1.26 | 0.835 | 0.0 | 2.377 |
| Asahikawa | JPN | 323796 | 684 | 22.6 | 0.82 | 0.866 | 0.008 | 1.398 |
| Ashford | GBR | 134237 | 578 | 25.9 | 1.97 | 0.787 | 0.184 | 0.729 |
| Athens | GRC | 3514199 | 1917 | 15.5 | 0.34 | 0.588 | 0.056 | 0.743 |
| Atlanta | USA | 2995863 | 3582 | 50.4 | 1.41 | 0.208 | 0.353 | 0.323 |
| Atlantic City | USA | 272180 | 1347 | 54.0 | 2.94 | 0.763 | 0.008 | 1.445 |
| Auckland | NZL | 1475194 | 682 | 14.6 | 0.52 | -0.202 | 1.301 | -0.46 |
| Austin | USA | 1902418 | 5356 | 41.3 | 1.68 | 0.701 | 0.027 | 1.102 |
| Badajoz | ESP | 137210 | 1256 | 28.7 | 1.41 | 0.732 | 0.007 | 1.482 |
| Barcelona | ESP | 3496615 | 536 | 9.3 | 0.25 | 0.801 | 0.004 | 1.765 |
| Bari | ITA | 343308 | 346 | 32.3 | 0.36 | 0.812 | 0.001 | 1.686 |
| Basingstoke And Deane | GBR | 182496 | 632 | 22.7 | 1.7 | 0.748 | 0.037 | 1.204 |
| Bath And North East Somerset | GBR | 184138 | 352 | 15.2 | 1.08 | 0.747 | 0.014 | 1.576 |
| Bedford | GBR | 168255 | 476 | 25.0 | 1.19 | 0.591 | 0.028 | 1.155 |

Supplementary Table 1: Cities' metrics: population, area, average proximity time, average annual $CO_2$ emissions for road transport per capita, correlation coefficient between proximity time and emissions inside the city, parameters of power-law fitting of the form of Eq. (2) in the main text.

| City | Country | Population | A (km$^2$) | s (min) | C$_{pc}$ (t) | r | A | γ |
|---|---|---|---|---|---|---|---|---|
| Bell | USA | 371600 | 2634 | 65.9 | 2.64 | 0.845 | 0.003 | 1.558 |
| Benton (Ar) | USA | 272614 | 2261 | 68.4 | 2.93 | 0.827 | 0.016 | 1.205 |
| Benton (Wa) | USA | 203639 | 3317 | 37.5 | 4.74 | 0.819 | 0.0 | 2.42 |
| Berks | USA | 421660 | 2235 | 58.0 | 2.95 | 0.743 | 0.001 | 2.024 |
| Berlin | DEU | 3683056 | 1078 | 8.4 | 0.42 | 0.837 | 0.016 | 1.444 |
| Besançon | FRA | 175992 | 409 | 25.1 | 1.11 | 0.566 | 0.086 | 0.782 |
| Bielefeld | DEU | 351932 | 258 | 11.8 | 0.77 | 0.895 | 0.01 | 1.731 |
| Boras | SWE | 110333 | 964 | 42.0 | 1.16 | 0.432 | 0.019 | 1.101 |
| Boston | USA | 3554142 | 4814 | 23.2 | 1.87 | 0.799 | 0.067 | 1.03 |
| Boulder | USA | 295402 | 1796 | 24.2 | 2.2 | 0.852 | 0.009 | 1.698 |
| Bratislava | SVK | 399346 | 366 | 13.8 | 0.76 | 0.682 | 0.049 | 0.943 |
| Braunschweig-Salzgitter Wolfsburg | DEU | 479962 | 621 | 15.3 | 1.09 | 0.686 | 0.02 | 1.474 |
| Brazos | USA | 230776 | 1400 | 40.9 | 2.04 | 0.845 | 0.006 | 1.496 |
| Bremen | DEU | 580387 | 320 | 18.0 | 0.59 | -0.039 | 0.594 | -0.032 |
| Brevard | USA | 591679 | 2419 | 50.7 | 1.97 | 0.846 | 0.002 | 1.705 |
| Brisbane | AU | 2288460 | 3045 | 25.3 | 1.15 | 0.283 | 0.332 | 0.291 |
| Brivelagaillarde | FRA | 81277 | 343 | 42.8 | 1.67 | 0.707 | 0.025 | 1.125 |
| Broome | USA | 168562 | 1835 | 53.0 | 7.15 | 0.893 | 0.03 | 1.318 |
| Brown | USA | 258295 | 1376 | 49.4 | 3.48 | 0.794 | 0.018 | 1.307 |
| Budapest | HUN | 1789480 | 522 | 11.2 | 0.34 | 0.077 | 0.293 | 0.054 |
| Butte | USA | 216494 | 3724 | 103.1 | 3.73 | 0.498 | 0.015 | 1.268 |
| Caceres | ESP | 85713 | 1732 | 21.8 | 2.11 | 0.537 | 0.119 | 0.998 |
| Caddo | USA | 230706 | 2192 | 79.5 | 4.1 | 0.56 | 0.031 | 1.1 |
| Calgary | CAN | 1376109 | 842 | 20.4 | 1.03 | 0.852 | 0.001 | 2.416 |
| Cameron | USA | 407059 | 2333 | 67.3 | 2.6 | 0.518 | 0.044 | 0.953 |
| Carlisle | GBR | 114580 | 979 | 28.8 | 2.09 | 0.756 | 0.031 | 1.212 |
| Cartagena | ESP | 217826 | 542 | 32.7 | 1.43 | 0.036 | 0.468 | 0.107 |
| Cass | USA | 169209 | 4540 | 45.5 | 6.89 | 0.793 | 0.008 | 1.614 |
| Centre | USA | 152681 | 2630 | 54.7 | 6.95 | 0.636 | 0.204 | 0.895 |
| Chalonsursaône | FRA | 104446 | 365 | 33.8 | 1.03 | 0.648 | 0.006 | 1.436 |
| Champaign | USA | 209653 | 2572 | 38.3 | 4.4 | 0.696 | 0.072 | 1.079 |
| Charleroi | BEL | 509839 | 1433 | 26.2 | 2.22 | 0.657 | 0.174 | 0.794 |
| Charleston | USA | 357038 | 2198 | 76.2 | 1.62 | 0.63 | 0.009 | 1.045 |

Supplementary Table 1: Cities' metrics: population, area, average proximity time, average annual $CO_2$ emissions for road transport per capita, correlation coefficient between proximity time and emissions inside the city, parameters of power-law fitting of the form of Eq. (2) in the main text.

| City | Country | Population | A (km$^2$) | s (min) | C$_{pc}$ (t) | r | A | $\gamma$ |
|---|---|---|---|---|---|---|---|---|
| Charlotte | USA | 1133706 | 1408 | 37.7 | 1.57 | 0.284 | 0.346 | 0.4 |
| Chatham | USA | 276369 | 919 | 51.6 | 2.15 | 0.145 | 0.148 | 0.482 |
| Cheshire West And Chester | GBR | 338895 | 908 | 20.4 | 1.73 | 0.546 | 0.082 | 1.053 |
| Chicago | USA | 8627748 | 10100 | 24.7 | 1.37 | 0.742 | 0.049 | 1.004 |
| Christchurch | NZL | 387695 | 347 | 16.4 | 0.74 | 0.255 | 0.049 | 0.741 |
| Cincinnati | USA | 886535 | 1489 | 29.8 | 2.38 | 0.632 | 0.133 | 0.815 |
| Colchester | GBR | 190515 | 334 | 20.3 | 0.83 | 0.801 | 0.053 | 0.946 |
| Collier | USA | 392080 | 2762 | 66.9 | 2.16 | 0.795 | 0.0 | 2.492 |
| Cologne | DEU | 1417693 | 566 | 11.9 | 0.67 | 0.883 | 0.0 | 3.381 |
| Columbus | USA | 1216783 | 1404 | 28.6 | 1.78 | 0.676 | 0.068 | 0.914 |
| Comanche | USA | 123905 | 2533 | 90.2 | 4.28 | 0.675 | 0.001 | 1.894 |
| Cordoba | ESP | 301731 | 1027 | 43.8 | 0.76 | 0.831 | 0.001 | 1.723 |
| Coventry | GBR | 729392 | 813 | 15.3 | 0.98 | 0.866 | 0.003 | 2.049 |
| Cracow | POL | 757815 | 326 | 15.0 | 0.63 | -0.079 | 0.688 | -0.066 |
| Cumberland (Me) | USA | 272230 | 2394 | 55.9 | 4.2 | 0.344 | 0.615 | 0.452 |
| Cumberland (Nc) | USA | 312068 | 1599 | 71.3 | 2.34 | 0.626 | 0.006 | 1.305 |
| Cuyahoga | USA | 1347032 | 1784 | 35.4 | 1.83 | 0.314 | 0.321 | 0.475 |
| Dacorum | GBR | 151379 | 212 | 16.9 | 1.22 | 0.947 | 0.003 | 2.035 |
| Dallas | USA | 6760664 | 9237 | 36.4 | 1.76 | 0.392 | 0.425 | 0.374 |
| Dane | USA | 527053 | 3166 | 35.3 | 3.59 | 0.775 | 0.05 | 1.175 |
| Dauphin | USA | 273019 | 1397 | 68.7 | 3.86 | 0.463 | 1.053 | 0.384 |
| Davidson | USA | 658462 | 1351 | 47.9 | 2.47 | 0.593 | 0.049 | 0.973 |
| Debrecen | HUN | 256651 | 461 | 20.6 | 0.63 | 0.737 | 0.02 | 1.125 |
| Delaware | USA | 106422 | 1012 | 43.6 | 4.03 | 0.708 | 0.042 | 1.199 |
| Denver | USA | 2807196 | 8616 | 25.8 | 2.0 | 0.776 | 0.041 | 1.191 |
| Derry | GBR | 156129 | 1212 | 40.1 | 1.1 | 0.64 | 0.025 | 1.02 |
| Dessau | DEU | 98145 | 244 | 16.3 | 1.16 | 0.732 | 0.028 | 1.336 |
| Detroit (Greater) | USA | 3562468 | 5228 | 28.9 | 1.8 | 0.029 | 1.418 | 0.029 |
| Doncaster | GBR | 317746 | 566 | 39.2 | 1.56 | 0.608 | 0.001 | 2.029 |
| Douglas (Ks) | USA | 113634 | 1204 | 36.5 | 4.61 | 0.921 | 0.028 | 1.402 |
| Douglas (Ne) | USA | 550147 | 863 | 25.3 | 1.97 | 0.812 | 0.001 | 2.242 |
| Dresden | DEU | 1342488 | 5777 | 17.1 | 1.68 | 0.849 | 0.051 | 1.201 |
| Durham | USA | 289781 | 767 | 33.6 | 2.39 | 0.436 | 0.136 | 0.758 |

Supplementary Table 1: Cities' metrics: population, area, average proximity time, average annual $CO_2$ emissions for road transport per capita, correlation coefficient between proximity time and emissions inside the city, parameters of power-law fitting of the form of Eq. (2) in the main text.

| City | Country | Population | A (km$^2$) | $s$ (min) | $C_{pc}$ (t) | $r$ | $A$ | $\gamma$ |
|---|---|---|---|---|---|---|---|---|
| Dusseldorf | DEU | 807055 | 316 | 10.3 | 0.57 | 0.901 | 0.033 | 1.259 |
| East Baton Rouge | USA | 443877 | 1151 | 53.9 | 1.46 | 0.725 | 0.078 | 0.733 |
| East Staffordshire | GBR | 121594 | 389 | 31.1 | 0.86 | 0.742 | 0.001 | 1.892 |
| Ector | USA | 146792 | 1557 | 46.6 | 2.99 | 0.773 | 0.0 | 2.816 |
| Edinburgh | GBR | 503142 | 262 | 8.3 | 0.71 | 0.956 | 0.006 | 2.09 |
| Edmonton | CAN | 1183322 | 1980 | 20.0 | 1.51 | 0.808 | 0.05 | 1.115 |
| El Paso (Co) | USA | 692916 | 4801 | 40.2 | 1.78 | 0.717 | 0.039 | 0.993 |
| Erfurt | DEU | 217929 | 269 | 15.0 | 0.91 | 0.868 | 0.081 | 0.874 |
| Erie (Ny) | USA | 842385 | 2709 | 39.6 | 2.75 | 0.651 | 0.111 | 0.893 |
| Erie (Pa) | USA | 248682 | 2079 | 43.5 | 4.07 | 0.729 | 0.011 | 1.49 |
| Escambia | USA | 280260 | 1812 | 56.6 | 2.61 | 0.82 | 0.001 | 1.912 |
| Falkirk | GBR | 166718 | 293 | 19.0 | 1.93 | 0.847 | 0.001 | 2.485 |
| Fayette | USA | 321264 | 729 | 32.9 | 2.09 | 0.79 | 0.001 | 2.216 |
| Ferrara | ITA | 129419 | 403 | 39.8 | 1.74 | 0.819 | 0.116 | 0.821 |
| Flagler-Daytona Beach | USA | 173334 | 1141 | 73.0 | 2.56 | 0.681 | 0.0 | 2.907 |
| Foggia | ITA | 139634 | 492 | 16.8 | 1.38 | 0.647 | 0.009 | 1.661 |
| Forsyth | USA | 363154 | 1068 | 54.4 | 3.01 | 0.558 | 0.105 | 0.803 |
| Frankfurt Am Main | DEU | 966758 | 369 | 11.8 | 0.62 | 0.897 | 0.005 | 2.064 |
| Fresno (Greater) | USA | 1027641 | 12664 | 60.9 | 2.59 | 0.687 | 0.0 | 2.205 |
| Fuji | JPN | 363913 | 610 | 26.8 | 1.06 | 0.548 | 0.057 | 0.948 |
| Fukui | JPN | 248304 | 485 | 35.5 | 1.46 | 0.759 | 0.009 | 1.366 |
| Fukuoka | JPN | 2389948 | 1066 | 18.7 | 0.52 | 0.61 | 0.002 | 1.755 |
| Fukushima | JPN | 268582 | 582 | 32.5 | 0.98 | 0.714 | 0.002 | 1.676 |
| Gdansk | POL | 690944 | 396 | 17.7 | 0.5 | 0.409 | 0.088 | 0.526 |
| Genesee | USA | 394597 | 1679 | 64.2 | 2.34 | 0.219 | 0.587 | 0.311 |
| Glasgow | GBR | 1212675 | 1056 | 17.9 | 1.35 | 0.756 | 0.001 | 2.276 |
| Greene | USA | 281138 | 1749 | 44.6 | 3.27 | 0.706 | 0.063 | 1.046 |
| Greenville | USA | 486236 | 2006 | 37.6 | 2.21 | 0.79 | 0.048 | 1.014 |
| Guadalajara | ESP | 104583 | 220 | 21.7 | 1.02 | 0.666 | 0.07 | 0.814 |
| Guildford | GBR | 145627 | 270 | 19.1 | 1.64 | 0.574 | 0.029 | 1.423 |
| Guilford | USA | 521155 | 1698 | 51.3 | 2.85 | 0.694 | 0.116 | 0.761 |
| Hachinohe | JPN | 208393 | 304 | 21.5 | 1.08 | 0.933 | 0.004 | 1.755 |
| Hakodate | JPN | 253938 | 647 | 29.2 | 0.37 | 0.453 | 0.116 | 0.394 |

4

Supplementary Table 1: Cities' metrics: population, area, average proximity time, average annual $CO_2$ emissions for road transport per capita, correlation coefficient between proximity time and emissions inside the city, parameters of power-law fitting of the form of Eq. (2) in the main text.

| City | Country | Population | A (km$^2$) | s (min) | C$_{pc}$ (t) | r | A | $\gamma$ |
|---|---|---|---|---|---|---|---|---|
| Halifax | CAN | 412150 | 4908 | 44.4 | 2.13 | 0.31 | 0.131 | 0.691 |
| Hamamatsu | JPN | 622424 | 319 | 16.7 | 0.57 | 0.603 | 0.017 | 1.103 |
| Hamburg | DEU | 1826359 | 737 | 11.8 | 0.52 | 0.834 | 0.04 | 1.009 |
| Hamilton (Tn) | USA | 322324 | 1470 | 71.6 | 2.93 | 0.726 | 0.036 | 1.012 |
| Hampden | USA | 447569 | 1622 | 39.7 | 3.39 | 0.853 | 0.008 | 1.589 |
| Harrison | USA | 172443 | 1503 | 104.4 | 2.46 | 0.771 | 0.001 | 1.741 |
| Hartford | USA | 893744 | 1944 | 40.3 | 2.66 | 0.799 | 0.031 | 1.151 |
| Helsingborg | SWE | 140577 | 346 | 21.6 | 0.71 | 0.563 | 0.032 | 0.838 |
| Helsinki | FIN | 946261 | 768 | 17.4 | 0.41 | 0.865 | 0.004 | 1.646 |
| Hidalgo | USA | 968156 | 3631 | 95.1 | 1.38 | 0.713 | 0.0 | 2.325 |
| Himeji | JPN | 549842 | 560 | 28.4 | 0.8 | 0.172 | 0.194 | 0.386 |
| Hiroshima | JPN | 1317949 | 1406 | 26.9 | 0.85 | 0.944 | 0.003 | 1.671 |
| Honolulu | USA | 946094 | 1046 | 42.0 | 0.79 | -0.204 | 1.357 | -0.337 |
| Houston | USA | 5642449 | 6737 | 43.1 | 1.49 | 0.669 | 0.033 | 0.987 |
| Indian River | USA | 159120 | 971 | 46.4 | 2.01 | 0.858 | 0.0 | 2.405 |
| Indianapolis | USA | 1299957 | 2082 | 41.7 | 1.76 | 0.668 | 0.054 | 0.898 |
| Ingham | USA | 258437 | 1433 | 33.2 | 2.86 | 0.531 | 0.459 | 0.509 |
| Isesaki | JPN | 363407 | 604 | 25.5 | 1.05 | 0.987 | 0.01 | 1.422 |
| Jackson (Mo) | USA | 1441523 | 3216 | 34.8 | 2.55 | 0.661 | 0.079 | 0.974 |
| Jackson (Or) | USA | 205302 | 6705 | 48.6 | 5.64 | 0.69 | 0.015 | 1.464 |
| Jacksonville | USA | 910885 | 1948 | 47.9 | 2.11 | 0.554 | 0.039 | 0.949 |
| Jaen | ESP | 105038 | 347 | 26.3 | 0.95 | 0.873 | 0.015 | 1.244 |
| Jefferson (Al) | USA | 579666 | 2808 | 62.0 | 3.13 | 0.341 | 0.351 | 0.475 |
| Jefferson (Ky) | USA | 758175 | 1027 | 35.1 | 1.83 | 0.362 | 0.534 | 0.313 |
| Jefferson (Tx) | USA | 238450 | 1965 | 77.5 | 2.58 | 0.715 | 0.001 | 1.897 |
| Jerez De La Frontera | ESP | 205312 | 967 | 21.7 | 1.28 | 0.574 | 0.117 | 0.959 |
| Johnson | USA | 143558 | 1571 | 38.8 | 4.75 | 0.574 | 0.08 | 1.133 |
| Kagoshima | JPN | 530884 | 524 | 38.7 | 0.54 | -0.069 | 1.892 | -0.095 |
| Kalamazoo | USA | 239460 | 1479 | 48.4 | 2.48 | 0.677 | 0.005 | 1.472 |
| Kanazawa | JPN | 563361 | 563 | 20.9 | 0.68 | 0.87 | 0.001 | 2.251 |
| Kankakee | USA | 115089 | 1742 | 69.1 | 5.42 | 0.818 | 0.067 | 1.048 |
| Kent | USA | 592801 | 2252 | 42.2 | 2.54 | 0.464 | 0.52 | 0.42 |
| Kern | USA | 999672 | 16724 | 92.0 | 3.67 | 0.573 | 0.002 | 1.826 |

5

Supplementary Table 1: Cities' metrics: population, area, average proximity time, average annual $CO_2$ emissions for road transport per capita, correlation coefficient between proximity time and emissions inside the city, parameters of power-law fitting of the form of Eq. (2) in the main text.

| City | Country | Population | A (km$^2$) | $s$ (min) | $C_{pc}$ (t) | $r$ | $A$ | $\gamma$ |
|---|---|---|---|---|---|---|---|---|
| Kitakyushu | JPN | 980261 | 508 | 28.7 | 0.56 | -0.455 | 164.661 | -1.623 |
| Kitchener | CAN | 498106 | 319 | 17.7 | 0.82 | -0.505 | 26.638 | -1.289 |
| Knox | USA | 454107 | 1361 | 54.2 | 2.24 | 0.55 | 0.08 | 0.79 |
| Koriyama | JPN | 313218 | 734 | 31.9 | 1.54 | 0.613 | 0.03 | 1.251 |
| Kumamoto | JPN | 815454 | 536 | 27.2 | 0.65 | 0.546 | 0.085 | 0.668 |
| Kushiro | JPN | 171969 | 1205 | 44.5 | 0.56 | 0.591 | 0.014 | 0.979 |
| Kyoto | JPN | 1467248 | 828 | 10.5 | 0.48 | 0.864 | 0.023 | 1.37 |
| København | DNK | 1119969 | 392 | 14.9 | 0.47 | 0.491 | 0.026 | 0.94 |
| Lackawanna | USA | 190470 | 1184 | 43.9 | 4.92 | 0.533 | 0.343 | 0.683 |
| Lafayette | USA | 246881 | 691 | 61.8 | 2.06 | 0.676 | 0.011 | 1.236 |
| Lafayette (In) | USA | 187487 | 1282 | 39.4 | 2.36 | 0.81 | 0.008 | 1.469 |
| Lancaster (Ne) | USA | 303520 | 2166 | 30.6 | 2.51 | 0.716 | 0.009 | 1.548 |
| Lancaster (Pa) | USA | 544923 | 2535 | 55.7 | 2.91 | 0.725 | 0.071 | 0.922 |
| Lane | USA | 352527 | 11131 | 61.4 | 4.86 | 0.644 | 0.113 | 0.986 |
| Larimer | USA | 323162 | 5717 | 37.0 | 3.04 | 0.87 | 0.01 | 1.443 |
| Las Cruces | USA | 229947 | 6689 | 111.7 | 5.17 | 0.471 | 0.004 | 1.605 |
| Las Vegas | USA | 2649719 | 10470 | 37.3 | 1.1 | 0.497 | 0.051 | 0.944 |
| Lee | USA | 813542 | 2061 | 74.1 | 1.56 | -0.427 | 21.597 | -0.624 |
| Leeds | GBR | 2162520 | 1667 | 14.6 | 1.01 | 0.503 | 0.06 | 1.014 |
| Lehigh | USA | 684246 | 1878 | 43.0 | 2.51 | 0.748 | 0.113 | 0.831 |
| Leipzig | DEU | 538550 | 297 | 10.3 | 0.67 | 0.967 | 0.097 | 0.816 |
| Liege | BEL | 803074 | 1966 | 19.4 | 2.29 | 0.86 | 0.081 | 1.115 |
| Lille | FRA | 1063111 | 258 | 20.5 | 0.41 | 0.837 | 0.016 | 1.47 |
| Linn | USA | 216945 | 1841 | 39.4 | 3.29 | 0.727 | 0.047 | 1.149 |
| Lisbon | PRT | 1196211 | 1090 | 9.8 | 0.5 | 0.912 | 0.002 | 2.0 |
| Liverpool | GBR | 1259404 | 555 | 15.2 | 0.74 | 0.884 | 0.0 | 3.498 |
| Ljubljana | SVN | 302352 | 275 | 14.3 | 0.97 | 0.78 | 0.139 | 0.697 |
| Lleida | ESP | 142737 | 211 | 17.3 | 0.79 | 0.528 | 0.176 | 0.705 |
| Lorca | ESP | 96801 | 1662 | 57.3 | 2.24 | 0.324 | 0.145 | 0.709 |
| Los Angeles (Greater) | USA | 18026424 | 48918 | 27.6 | 1.31 | 0.803 | 0.005 | 1.585 |
| Lubbock | USA | 302840 | 2292 | 31.2 | 3.31 | 0.79 | 0.003 | 1.881 |
| Lubeck | DEU | 222577 | 209 | 14.2 | 0.8 | 0.294 | 0.105 | 0.721 |
| Lucas | USA | 408236 | 904 | 33.4 | 2.16 | 0.889 | 0.01 | 1.438 |

Supplementary Table 1: Cities' metrics: population, area, average proximity time, average annual $CO_2$ emissions for road transport per capita, correlation coefficient between proximity time and emissions inside the city, parameters of power-law fitting of the form of Eq. (2) in the main text.

| City | Country | Population | A (km$^2$) | $s$ (min) | $C_{pc}$ (t) | $r$ | $A$ | $\gamma$ |
|---|---|---|---|---|---|---|---|---|
| Lugo | ESP | 84129 | 328 | 20.1 | 1.13 | 0.369 | 0.478 | 0.385 |
| Luzerne | USA | 281499 | 2268 | 63.8 | 4.46 | 0.62 | 0.014 | 1.341 |
| Madison | USA | 361224 | 1982 | 61.0 | 2.49 | 0.637 | 0.197 | 0.626 |
| Madrid | ESP | 5396933 | 1292 | 11.4 | 0.44 | 0.823 | 0.03 | 1.014 |
| Mahoning | USA | 206764 | 1090 | 47.0 | 4.21 | 0.614 | 0.333 | 0.699 |
| Maidstone | GBR | 173604 | 395 | 26.0 | 1.79 | 0.575 | 0.033 | 1.206 |
| Malaga | ESP | 722306 | 440 | 12.0 | 0.51 | 0.904 | 0.001 | 2.516 |
| Manchester | GBR | 2905881 | 1274 | 14.0 | 0.91 | 0.667 | 0.015 | 1.532 |
| Mannheim-Ludwigshafen | DEU | 561554 | 308 | 12.2 | 0.65 | 0.202 | 0.505 | 0.102 |
| Marion (Fl) | USA | 381418 | 4065 | 126.6 | 3.4 | 0.638 | 0.007 | 1.287 |
| Marion (Or) | USA | 332844 | 2935 | 44.4 | 2.96 | 0.797 | 0.015 | 1.338 |
| Matsumoto | JPN | 230324 | 810 | 28.9 | 1.38 | 0.794 | 0.008 | 1.651 |
| Matsuyama | JPN | 569580 | 853 | 26.8 | 0.86 | 0.651 | 0.002 | 1.695 |
| Mclean | USA | 177917 | 3037 | 54.0 | 6.81 | 0.606 | 0.117 | 1.094 |
| Mclennan | USA | 226574 | 2578 | 80.6 | 3.17 | 0.66 | 0.01 | 1.245 |
| Melbourne | AU | 4095505 | 2840 | 16.5 | 0.88 | 0.921 | 0.003 | 1.988 |
| Memphis | USA | 882756 | 1879 | 60.1 | 2.33 | 0.549 | 0.033 | 1.042 |
| Merced | USA | 292925 | 4690 | 71.5 | 3.69 | 0.708 | 0.002 | 1.803 |
| Mesa | USA | 173061 | 6501 | 44.4 | 4.87 | 0.723 | 0.045 | 1.187 |
| Miami (Greater) | USA | 5934671 | 8476 | 34.3 | 1.15 | 0.711 | 0.006 | 1.452 |
| Middlesbrough | GBR | 442093 | 346 | 18.9 | 0.85 | 0.723 | 0.008 | 1.418 |
| Midland | USA | 154810 | 2042 | 42.0 | 3.51 | 0.872 | 0.001 | 2.139 |
| Milan | ITA | 4014526 | 1953 | 22.1 | 0.55 | 0.733 | 0.008 | 1.337 |
| Milton Keynes | GBR | 293420 | 306 | 14.5 | 0.86 | 0.829 | 0.003 | 1.9 |
| Milwaukee | USA | 918768 | 624 | 17.2 | 1.45 | 0.503 | 0.098 | 0.807 |
| Minneapolis | USA | 2046883 | 3506 | 26.8 | 2.52 | 0.817 | 0.07 | 1.042 |
| Minnehaha | USA | 179500 | 2095 | 45.6 | 4.13 | 0.71 | 0.006 | 1.588 |
| Miyazaki | JPN | 377500 | 637 | 37.0 | 1.07 | 0.834 | 0.009 | 1.231 |
| Mobile | USA | 375535 | 3040 | 99.8 | 2.93 | 0.598 | 0.006 | 1.274 |
| Monroe (In) | USA | 139892 | 1051 | 35.8 | 1.66 | 0.777 | 0.008 | 1.472 |
| Mons | BEL | 250777 | 538 | 18.3 | 2.04 | 0.117 | 1.198 | 0.164 |
| Monterey | USA | 405046 | 7507 | 52.7 | 4.17 | 0.639 | 0.012 | 1.488 |
| Montgomery (Al) | USA | 218200 | 1787 | 68.4 | 2.59 | 0.567 | 0.012 | 1.322 |

Supplementary Table 1: Cities' metrics: population, area, average proximity time, average annual $CO_2$ emissions for road transport per capita, correlation coefficient between proximity time and emissions inside the city, parameters of power-law fitting of the form of Eq. (2) in the main text.

| City | Country | Population | A (km²) | s (min) | $C_{pc}$ (t) | r | A | γ |
|---|---|---|---|---|---|---|---|---|
| Montgomery (Oh) | USA | 485928 | 1201 | 42.1 | 2.59 | 0.704 | 0.021 | 1.264 |
| Montreal | CAN | 3432357 | 1592 | 17.8 | 0.76 | 0.692 | 0.076 | 0.78 |
| Morioka | JPN | 265246 | 849 | 37.5 | 0.96 | 0.729 | 0.01 | 1.308 |
| Muenster | DEU | 342252 | 303 | 11.9 | 0.75 | 0.692 | 0.0 | 3.18 |
| Munich | DEU | 1609658 | 310 | 7.8 | 0.39 | 0.552 | 0.055 | 0.816 |
| Murcia | ESP | 460143 | 879 | 27.1 | 1.16 | 0.819 | 0.032 | 1.038 |
| Muscogee | USA | 184236 | 531 | 52.6 | 1.8 | 0.516 | 0.018 | 1.142 |
| Muskegon | USA | 157704 | 1349 | 62.1 | 2.9 | 0.279 | 0.394 | 0.435 |
| Nagano | JPN | 362020 | 789 | 23.3 | 1.5 | 0.836 | 0.003 | 1.876 |
| Nagasaki | JPN | 443237 | 444 | 51.4 | 0.49 | -0.153 | 1.124 | -0.24 |
| Nagoya | JPN | 7589542 | 4032 | 21.2 | 0.66 | 0.913 | 0.005 | 1.587 |
| Napa | USA | 141039 | 1587 | 72.5 | 2.66 | 0.668 | 0.034 | 1.178 |
| Napoli | ITA | 3397273 | 1282 | 18.9 | 0.31 | 0.544 | 0.003 | 1.457 |
| Nashville | USA | 356216 | 1599 | 63.5 | 2.14 | 0.74 | 0.005 | 1.453 |
| New Hanover | USA | 238045 | 468 | 55.1 | 1.2 | -0.277 | 6.39 | -0.514 |
| New Haven | USA | 857343 | 1607 | 34.0 | 2.0 | 0.648 | 0.146 | 0.733 |
| New Orleans | USA | 647984 | 1033 | 47.2 | 1.66 | 0.453 | 0.051 | 0.825 |
| New York | USA | 17410965 | 11604 | 19.9 | 1.01 | 0.666 | 0.026 | 1.142 |
| New York | USA | 17410965 | 11604 | 19.9 | 1.01 | 0.624 | 0.066 | 0.857 |
| Newcastle Upon Tyne | GBR | 864948 | 406 | 12.0 | 0.84 | 0.522 | 0.181 | 0.567 |
| Newport News | USA | 294549 | 303 | 33.9 | 1.4 | 0.884 | 0.0 | 4.28 |
| Nice | FRA | 660216 | 210 | 30.3 | 0.26 | 0.416 | 0.019 | 0.873 |
| Niort | FRA | 102560 | 499 | 38.6 | 2.23 | 0.26 | 0.754 | 0.37 |
| Nueces | USA | 339619 | 2027 | 61.2 | 1.93 | 0.496 | 0.063 | 0.979 |
| Numazu | JPN | 457340 | 449 | 21.0 | 0.79 | 0.741 | 0.015 | 1.296 |
| Nuremberg | DEU | 762520 | 327 | 11.0 | 0.56 | 0.929 | 0.037 | 1.15 |
| Nyiregyhaza | HUN | 134398 | 274 | 24.7 | 0.67 | 0.754 | 0.0 | 2.367 |
| Nîmes | FRA | 233749 | 538 | 35.3 | 0.98 | 0.449 | 0.062 | 0.8 |
| Obihiro | JPN | 157427 | 517 | 32.6 | 0.55 | 0.893 | 0.002 | 1.563 |
| Oita | JPN | 549530 | 624 | 25.7 | 0.71 | 0.651 | 0.022 | 1.054 |
| Okayama | JPN | 1137535 | 1144 | 27.7 | 0.81 | 0.433 | 0.092 | 0.696 |
| Oklahoma | USA | 1019132 | 3239 | 46.7 | 2.82 | 0.628 | 0.11 | 0.782 |
| Onondaga | USA | 430117 | 2061 | 40.0 | 4.0 | 0.679 | 0.061 | 1.106 |

Supplementary Table 1: Cities' metrics: population, area, average proximity time, average annual $CO_2$ emissions for road transport per capita, correlation coefficient between proximity time and emissions inside the city, parameters of power-law fitting of the form of Eq. (2) in the main text.

| City | Country | Population | A (km$^2$) | $s$ (min) | $C_{pc}$ (t) | $r$ | $A$ | $\gamma$ |
|---|---|---|---|---|---|---|---|---|
| Orange | USA | 1872129 | 2992 | 41.7 | 1.88 | 0.586 | 0.088 | 0.82 |
| Orebro | SWE | 150377 | 1494 | 31.7 | 1.13 | 0.467 | 0.03 | 1.071 |
| Osaka | JPN | 15490091 | 4785 | 12.8 | 0.39 | 0.809 | 0.008 | 1.489 |
| Oslo | NOR | 579375 | 452 | 13.8 | 0.16 | 0.953 | 0.013 | 1.058 |
| Ostrava | CZE | 390774 | 304 | 21.7 | 0.54 | 0.756 | 0.039 | 0.837 |
| Ottawa | CAN | 1341174 | 3200 | 17.1 | 1.57 | 0.88 | 0.059 | 1.116 |
| Oulu | FIN | 163758 | 2751 | 45.0 | 1.13 | 0.772 | 0.004 | 1.498 |
| Outagamie | USA | 184373 | 1630 | 51.8 | 4.59 | 0.856 | 0.141 | 0.899 |
| Palermo | ITA | 637654 | 189 | 28.2 | 0.24 | -0.205 | 3.022 | -1.203 |
| Paris | FRA | 9864887 | 2027 | 7.9 | 0.33 | 0.49 | 0.135 | 0.428 |
| Peoria | USA | 175124 | 1608 | 53.7 | 4.14 | 0.577 | 0.084 | 0.897 |
| Perugia | ITA | 163727 | 447 | 39.8 | 0.62 | 0.794 | 0.006 | 1.371 |
| Philadelphia | USA | 6000249 | 11146 | 39.4 | 2.0 | 0.662 | 0.194 | 0.66 |
| Philadelphia (Greater) | USA | 4238474 | 4427 | 25.9 | 1.43 | 0.674 | 0.093 | 0.848 |
| Phoenix | USA | 4548560 | 15320 | 30.0 | 1.48 | 0.815 | 0.008 | 1.468 |
| Pima | USA | 1085485 | 15966 | 46.6 | 1.96 | 0.728 | 0.004 | 1.563 |
| Pitt | USA | 192260 | 1614 | 77.3 | 2.36 | 0.504 | 0.012 | 1.224 |
| Pittsburgh | USA | 1102286 | 1927 | 31.9 | 2.56 | 0.771 | 0.028 | 1.24 |
| Polk | USA | 467947 | 1501 | 28.7 | 2.72 | 0.764 | 0.05 | 1.139 |
| Portland | USA | 1882994 | 4490 | 21.2 | 2.17 | 0.855 | 0.088 | 1.016 |
| Potter | USA | 252326 | 4195 | 53.0 | 4.48 | 0.747 | 0.001 | 2.256 |
| Prague | CZE | 1416104 | 532 | 11.4 | 0.41 | -0.466 | 0.846 | -0.33 |
| Providence | USA | 759269 | 1575 | 35.3 | 2.19 | 0.754 | 0.017 | 1.316 |
| Pueblo | USA | 164084 | 4463 | 54.6 | 3.9 | 0.747 | 0.003 | 1.656 |
| Pulaski | USA | 368090 | 2006 | 56.8 | 4.17 | 0.761 | 0.008 | 1.504 |
| Punta Gorda | USA | 170581 | 1375 | 131.2 | 2.79 | 0.596 | 0.0 | 2.147 |
| Quebec | CAN | 586593 | 463 | 19.2 | 1.22 | 0.877 | 0.012 | 1.504 |
| Quimper | FRA | 85256 | 290 | 34.6 | 1.18 | 0.919 | 0.008 | 1.4 |
| Racine | USA | 192951 | 875 | 39.1 | 2.8 | 0.716 | 0.001 | 1.925 |
| Ravenna | ITA | 138740 | 648 | 55.8 | 1.38 | 0.492 | 0.003 | 1.486 |
| Richland | USA | 429557 | 1938 | 52.9 | 2.54 | 0.741 | 0.01 | 1.399 |
| Riga | LVA | 639193 | 302 | 20.1 | 0.1 | -0.142 | 0.174 | -0.161 |
| Rochester (Mn) | USA | 154508 | 1666 | 39.2 | 6.1 | 0.824 | 0.064 | 1.205 |

Supplementary Table 1: Cities' metrics: population, area, average proximity time, average annual $CO_2$ emissions for road transport per capita, correlation coefficient between proximity time and emissions inside the city, parameters of power-law fitting of the form of Eq. (2) in the main text.

| City | Country | Population | A (km$^2$) | $s$ (min) | $C_{pc}$ (t) | $r$ | $A$ | $\gamma$ |
|---|---|---|---|---|---|---|---|---|
| Rochester (Ny) | USA | 705122 | 1715 | 35.8 | 2.37 | 0.578 | 0.051 | 1.03 |
| Rock | USA | 156828 | 1860 | 51.8 | 5.76 | 0.785 | 0.014 | 1.488 |
| Rome | ITA | 2681049 | 1239 | 13.6 | 0.63 | 0.894 | 0.016 | 1.35 |
| Rotterdam | NLD | 1287079 | 651 | 28.7 | 0.85 | -0.001 | 0.946 | -0.001 |
| Ruhr | DEU | 3636034 | 1887 | 11.6 | 0.84 | 0.766 | 0.061 | 1.08 |
| Sacramento | USA | 2252266 | 7994 | 40.3 | 1.83 | 0.682 | 0.008 | 1.45 |
| Saginaw | USA | 177878 | 2094 | 73.6 | 4.58 | 0.233 | 1.515 | 0.251 |
| Saintetienne | FRA | 371311 | 558 | 26.3 | 0.66 | 0.797 | 0.001 | 1.876 |
| Saintnazaire | FRA | 109789 | 217 | 44.2 | 0.5 | 0.629 | 0.094 | 0.48 |
| Salt Lake | USA | 1513628 | 2647 | 22.4 | 1.43 | 0.767 | 0.004 | 1.763 |
| San Antonio | USA | 2000475 | 3081 | 38.7 | 1.59 | 0.623 | 0.061 | 0.873 |
| San Diego | USA | 3227885 | 8484 | 31.1 | 1.66 | 0.823 | 0.003 | 1.708 |
| San Francisco (Greater) | USA | 5944926 | 7694 | 18.6 | 1.4 | 0.799 | 0.026 | 1.336 |
| San Joaquin | USA | 799145 | 3372 | 47.1 | 2.21 | 0.807 | 0.01 | 1.4 |
| Sangamon | USA | 189398 | 2213 | 43.7 | 5.24 | 0.725 | 0.019 | 1.424 |
| Santa Barbara | USA | 427005 | 4437 | 39.7 | 1.94 | 0.801 | 0.011 | 1.407 |
| Santa Cruz | USA | 240941 | 1130 | 26.7 | 2.24 | 0.662 | 0.033 | 1.194 |
| Sapporo | JPN | 1909233 | 1184 | 15.0 | 0.38 | 0.92 | 0.005 | 1.509 |
| Saragossa | ESP | 664638 | 898 | 23.0 | 0.75 | 0.727 | 0.009 | 1.345 |
| Sarasota | USA | 799735 | 3198 | 55.5 | 1.84 | 0.803 | 0.001 | 1.735 |
| Sassari | ITA | 121375 | 545 | 34.0 | 1.0 | 0.406 | 0.257 | 0.67 |
| Scott | USA | 297613 | 2341 | 51.8 | 4.39 | 0.674 | 0.134 | 0.924 |
| Seattle | USA | 3716435 | 13303 | 29.1 | 1.7 | 0.781 | 0.003 | 1.703 |
| Sebastian | USA | 121133 | 1216 | 64.7 | 3.16 | 0.828 | 0.024 | 1.151 |
| Sedgwick | USA | 513406 | 2592 | 56.0 | 2.63 | 0.687 | 0.005 | 1.491 |
| Sendai | JPN | 1186152 | 861 | 15.5 | 0.59 | 0.775 | 0.01 | 1.348 |
| Seoul | KOR | 21596984 | 3302 | 15.0 | 0.48 | 0.522 | 0.072 | 0.69 |
| Seville | ESP | 873772 | 560 | 14.6 | 0.54 | 0.764 | 0.011 | 1.381 |
| Shawnee | USA | 172756 | 1408 | 38.2 | 5.3 | 0.685 | 0.026 | 1.398 |
| Sheffield | GBR | 1103652 | 983 | 16.5 | 0.99 | 0.76 | 0.017 | 1.457 |
| Shimonoseki | JPN | 237927 | 706 | 43.6 | 1.28 | 0.399 | 0.005 | 1.208 |
| Shizuoka | JPN | 666357 | 1289 | 26.8 | 0.82 | 0.878 | 0.003 | 1.536 |
| Shunan | JPN | 180195 | 725 | 44.2 | 2.0 | 0.786 | 0.001 | 1.982 |

Supplementary Table 1: Cities' metrics: population, area, average proximity time, average annual $CO_2$ emissions for road transport per capita, correlation coefficient between proximity time and emissions inside the city, parameters of power-law fitting of the form of Eq. (2) in the main text.

| City | Country | Population | A (km$^2$) | s (min) | C$_{pc}$ (t) | r | A | γ |
|---|---|---|---|---|---|---|---|---|
| Sonoma | USA | 475316 | 3742 | 39.2 | 2.63 | 0.81 | 0.003 | 1.76 |
| Spokane | USA | 482919 | 4467 | 34.8 | 2.91 | 0.724 | 0.01 | 1.454 |
| St Johns | CAN | 116833 | 373 | 23.8 | 2.41 | 0.912 | 0.001 | 2.144 |
| Stanislaus | USA | 567139 | 3107 | 51.2 | 1.61 | 0.725 | 0.001 | 1.831 |
| Stark | USA | 343460 | 1494 | 48.7 | 2.95 | 0.503 | 0.122 | 0.759 |
| Stoke-On-Trent | GBR | 380793 | 303 | 19.9 | 0.78 | 0.839 | 0.003 | 1.754 |
| Stuttgart | DEU | 881210 | 349 | 9.1 | 0.5 | 0.729 | 0.052 | 0.87 |
| Summit | USA | 506281 | 1084 | 38.0 | 2.74 | 0.414 | 0.142 | 0.727 |
| Sumter | USA | 150098 | 1326 | 155.1 | 3.24 | 0.635 | 0.0 | 2.736 |
| Sutter | USA | 108824 | 1424 | 75.0 | 3.83 | 0.379 | 0.242 | 0.884 |
| Swansea | GBR | 253006 | 378 | 22.8 | 1.28 | 0.808 | 0.026 | 1.149 |
| Sydney | AU | 4687592 | 3654 | 19.4 | 0.65 | 0.589 | 0.04 | 0.884 |
| Szeged | HUN | 131179 | 282 | 20.7 | 1.09 | 0.732 | 0.001 | 2.09 |
| Takamatsu | JPN | 391252 | 369 | 20.8 | 0.68 | 0.996 | 0.009 | 1.403 |
| Takasaki | JPN | 791107 | 961 | 22.8 | 1.29 | 0.846 | 0.011 | 1.473 |
| Tallahassee | USA | 233118 | 1772 | 33.7 | 3.4 | 0.766 | 0.028 | 1.3 |
| Tampa-Hernando | USA | 215039 | 1199 | 110.4 | 2.2 | 0.705 | 0.0 | 1.991 |
| Tampa-Hillsborough | USA | 1443009 | 2595 | 40.5 | 1.63 | 0.734 | 0.008 | 1.315 |
| Tampa-Pinellas | USA | 867981 | 830 | 27.8 | 1.01 | -0.381 | 5.836 | -0.624 |
| Taranto | ITA | 150761 | 238 | 38.5 | 0.52 | 0.808 | 0.0 | 3.262 |
| Taylor | USA | 124926 | 2197 | 41.8 | 5.5 | 0.808 | 0.002 | 1.966 |
| Telford And Wrekin | GBR | 175094 | 289 | 21.2 | 0.93 | 0.539 | 0.07 | 0.877 |
| Terrebonne | USA | 114395 | 852 | 119.4 | 2.41 | 0.517 | 0.011 | 1.117 |
| Thurston | USA | 291264 | 1847 | 53.5 | 2.34 | 0.442 | 0.07 | 0.819 |
| Tokyo | JPN | 34597500 | 7192 | 12.5 | 0.29 | 0.851 | 0.007 | 1.424 |
| Tomakomai | JPN | 153132 | 553 | 36.8 | 1.25 | 0.816 | 0.005 | 1.436 |
| Toronto | CAN | 6838851 | 3244 | 14.5 | 0.68 | 0.889 | 0.017 | 1.315 |
| Toyama | JPN | 399280 | 1046 | 31.0 | 1.62 | 0.94 | 0.005 | 1.584 |
| Trois Rivieres | CAN | 141451 | 283 | 26.6 | 1.45 | 0.667 | 0.007 | 1.387 |
| Tulare | USA | 499535 | 9638 | 86.0 | 3.14 | 0.813 | 0.0 | 2.2 |
| Tulsa | USA | 611000 | 1470 | 32.0 | 2.63 | 0.088 | 1.892 | 0.082 |
| Tuscaloosa | USA | 198793 | 3181 | 82.5 | 4.34 | 0.673 | 0.046 | 1.011 |
| Utah | USA | 693699 | 4623 | 28.3 | 1.51 | 0.854 | 0.0 | 2.306 |

Supplementary Table 1: Cities' metrics: population, area, average proximity time, average annual $CO_2$ emissions for road transport per capita, correlation coefficient between proximity time and emissions inside the city, parameters of power-law fitting of the form of Eq. (2) in the main text.

| City | Country | Population | A (km$^2$) | $s$ (min) | $C_{pc}$ (t) | $r$ | $A$ | $\gamma$ |
|---|---|---|---|---|---|---|---|---|
| Utsunomiya | JPN | 495594 | 415 | 20.8 | 0.74 | 0.948 | 0.003 | 1.778 |
| Valencia | ESP | 962540 | 230 | 12.3 | 0.37 | 0.081 | 0.468 | 0.113 |
| Vancouver | CAN | 2410850 | 1429 | 12.9 | 0.73 | 0.853 | 0.097 | 0.759 |
| Vanderburgh | USA | 176034 | 608 | 50.8 | 2.08 | 0.94 | 0.0 | 2.814 |
| Vannes | FRA | 124016 | 450 | 37.4 | 0.78 | 0.534 | 0.079 | 0.641 |
| Vasteras | SWE | 145973 | 1031 | 33.1 | 0.88 | 0.351 | 0.078 | 0.744 |
| Ventura | USA | 863234 | 3462 | 34.5 | 1.78 | 0.834 | 0.007 | 1.493 |
| Verviers (Greater City) | BEL | 110311 | 286 | 18.9 | 2.3 | 0.952 | 0.002 | 2.295 |
| Vienna | AUT | 1706117 | 411 | 10.0 | 0.4 | 0.718 | 0.125 | 0.506 |
| Vilnius | LTU | 526328 | 399 | 21.5 | 0.34 | 0.693 | 0.063 | 0.562 |
| Virginia Beach | USA | 995459 | 1605 | 38.3 | 1.82 | 0.729 | 0.025 | 1.067 |
| Vitoria | ESP | 238569 | 275 | 14.5 | 0.79 | 0.817 | 0.004 | 1.945 |
| Volusia-Daytona Beach | USA | 525425 | 2740 | 60.0 | 2.21 | 0.69 | 0.006 | 1.358 |
| Wakayama | JPN | 379593 | 241 | 21.0 | 0.45 | 0.607 | 0.001 | 2.008 |
| Wake | USA | 1180179 | 2205 | 37.0 | 1.7 | 0.548 | 0.058 | 0.871 |
| Warsaw | POL | 1697605 | 515 | 13.1 | 0.49 | -0.187 | 0.691 | -0.19 |
| Washington (Greater) | USA | 6597578 | 7986 | 26.4 | 1.62 | 0.772 | 0.071 | 0.951 |
| Washoe | USA | 494598 | 8316 | 50.9 | 2.52 | 0.684 | 0.009 | 1.442 |
| Washtenaw | USA | 342858 | 1861 | 38.2 | 3.42 | 0.544 | 0.113 | 0.884 |
| Waveney | GBR | 117630 | 373 | 23.6 | 0.94 | 0.786 | 0.016 | 1.325 |
| Webb | USA | 305935 | 7390 | 63.8 | 2.43 | 0.825 | 0.0 | 2.163 |
| Weber | USA | 258410 | 1271 | 28.9 | 1.36 | 0.753 | 0.003 | 1.594 |
| Weld | USA | 330009 | 8760 | 74.9 | 3.35 | 0.838 | 0.001 | 1.86 |
| Wellington | NZL | 368709 | 298 | 16.5 | 0.39 | -0.408 | 6.602 | -1.22 |
| West Midlands Urban Area | GBR | 2760751 | 912 | 13.8 | 0.64 | 0.419 | 0.032 | 1.057 |
| Whatcom | USA | 219830 | 3443 | 50.0 | 3.48 | 0.753 | 0.0 | 2.102 |
| Wichita | USA | 120824 | 1461 | 66.4 | 4.78 | 0.746 | 0.0 | 2.259 |
| Winnebago (Il) | USA | 296623 | 1329 | 43.7 | 2.13 | 0.664 | 0.119 | 0.779 |
| Winnebago (Wi) | USA | 166131 | 1222 | 39.2 | 3.91 | 0.676 | 0.087 | 1.035 |
| Winnipeg | CAN | 725739 | 472 | 18.6 | 0.79 | 0.609 | 0.004 | 1.633 |
| Woodbury | USA | 110871 | 2903 | 52.3 | 7.21 | 0.726 | 0.01 | 1.58 |
| Wrexham | GBR | 141417 | 499 | 31.1 | 1.7 | 0.789 | 0.042 | 1.138 |
| Wycombe | GBR | 181258 | 324 | 19.1 | 1.37 | 0.941 | 0.004 | 1.922 |

Supplementary Table 1: Cities' metrics: population, area, average proximity time, average annual $CO_2$ emissions for road transport per capita, correlation coefficient between proximity time and emissions inside the city, parameters of power-law fitting of the form of Eq. (2) in the main text.

| City | Country | Population | A (km$^2$) | s (min) | C$_{pc}$ (t) | r | A | $\gamma$ |
|---|---|---|---|---|---|---|---|---|
| Yakima | USA | 237812 | 8809 | 77.1 | 5.2 | 0.815 | 0.0 | 2.213 |
| Yamagata | JPN | 234847 | 388 | 25.2 | 1.08 | 0.727 | 0.014 | 1.265 |
| Yellowstone | USA | 157740 | 4794 | 59.7 | 5.45 | 0.657 | 0.003 | 1.677 |
| Yokkaichi | JPN | 779940 | 1114 | 32.5 | 1.04 | 0.893 | 0.001 | 1.783 |
| York | GBR | 215660 | 270 | 14.4 | 0.73 | 0.624 | 0.161 | 0.57 |
| Yuma | USA | 222411 | 6282 | 89.8 | 3.08 | 0.702 | 0.002 | 1.61 |

Supplementary Table 2: Proximity optimization: population, average proximity time and road $CO_2$ emissions per capita before and after proximity optimization, percentage variation of the emissions after optimization defined as $\Delta = (C_{pc}^{opt} - C_{pc})/C_{pc}$, whole city road $CO_2$ emissions after optimization, for all cities considered. Data in this table may differ from the ones in Table 1 because urban boundaries considered may differ.

| City | Population | s (min) | C$_{pc}$ (t) | s$^{opt}$ (min) | C$_{pc}^{opt}$ (t) | $\Delta$(%) | C$^{opt}$ (t) |
|---|---|---|---|---|---|---|---|
| Lisbon | 1198074 | 9.74 | 0.27 | 6.44 | 0.16 | -41.7 | 190166 |
| Edinburgh | 503142 | 8.29 | 0.71 | 5.92 | 0.35 | -50.34 | 177769 |
| Helsinki | 1152724 | 12.97 | 0.33 | 9.43 | 0.21 | -35.92 | 247062 |
| Milan | 1202010 | 6.7 | 0.32 | 4.91 | 0.15 | -53.37 | 181057 |
| Prague | 1351524 | 11.18 | 0.41 | 8.12 | 0.45 | 10.12 | 605337 |
| Auckland | 1149490 | 14.02 | 0.52 | 9.74 | 0.61 | 18.16 | 701359 |
| Budapest | 1789480 | 11.2 | 0.33 | 7.33 | 0.33 | -2.04 | 582631 |
| Munich | 1609658 | 7.76 | 0.39 | 5.63 | 0.31 | -20.85 | 494926 |
| Vienna | 1915092 | 8.51 | 0.35 | 5.56 | 0.28 | -18.15 | 545673 |
| Sapporo | 1909233 | 14.99 | 0.38 | 11.11 | 0.24 | -36.14 | 466253 |
| Barcelona | 3144374 | 9.04 | 0.25 | 5.82 | 0.13 | -48.4 | 404069 |
| Warsaw | 1685680 | 10.07 | 0.48 | 6.42 | 0.52 | 8.11 | 883747 |
| Montreal | 1908991 | 13.16 | 0.48 | 8.17 | 0.4 | -15.89 | 772604 |
| Rotterdam | 1308069 | 37.34 | 0.84 | 21.58 | 0.84 | 0.35 | 1101119 |
| Athens | 3458092 | 11.1 | 0.34 | 9.04 | 0.29 | -13.6 | 1016839 |
| Fukuoka | 2387513 | 18.64 | 0.52 | 13.69 | 0.31 | -40.48 | 738270 |
| Milwaukee | 918767 | 17.17 | 1.41 | 10.59 | 0.97 | -30.68 | 894989 |
| Amsterdam | 1623135 | 18.61 | 0.89 | 12.66 | 0.73 | -17.9 | 1191015 |

13

Supplementary Table 2: Proximity optimization: population, average proximity time and road $CO_2$ emissions per capita before and after proximity optimization, percentage variation of the emissions after optimization defined as $\Delta = (C_{pc}^{opt} - C_{pc})/C_{pc}$, whole city road $CO_2$ emissions after optimization, for all cities considered. Data in this table may differ from the ones in Table 1 because urban boundaries considered may differ.

| City | Population | $s$ (min) | $C_{pc}$ (t) | $s^{opt}$ (min) | $C_{pc}^{opt}$ (t) | $\Delta(\%)$ | $C^{opt}$ (t) |
|---|---|---|---|---|---|---|---|
| Berlin | 3683057 | 8.42 | 0.41 | 5.73 | 0.24 | -40.08 | 898989 |
| Rome | 2681048 | 13.57 | 0.63 | 8.87 | 0.37 | -40.98 | 994484 |
| Madrid | 4934992 | 11.27 | 0.45 | 7.66 | 0.32 | -29.37 | 1563467 |
| Paris | 9865193 | 7.93 | 0.32 | 6.05 | 0.29 | -10.2 | 2864400 |
| Atlanta | 2995863 | 50.41 | 1.36 | 42.22 | 1.31 | -3.59 | 3925935 |
| Minneapolis | 2046883 | 26.75 | 2.43 | 19.85 | 1.91 | -21.41 | 3902298 |
| Osaka | 15284096 | 12.34 | 0.39 | 8.96 | 0.26 | -33.26 | 3951384 |
| Boston | 3546650 | 23.18 | 1.8 | 19.7 | 1.61 | -10.72 | 5707193 |
| Tokyo | 34387959 | 12.28 | 0.29 | 9.21 | 0.21 | -25.33 | 7323964 |
| Seoul | 21601204 | 14.98 | 0.47 | 11.22 | 0.39 | -16.22 | 8477075 |
| Dallas | 6760663 | 36.43 | 1.7 | 25.84 | 1.52 | -10.39 | 10303750 |

## Proximity/emissions relation within cities

Supplementary Figures from 1 to 27 display the relation between proximity time $s$, and $CO_2$ emissions at the intra-city level, i.e. at the emission grid scale. For every city we computed the correlation coefficient between the logarithm of emissions for road transport per capita, and the logarithm of proximity time $s$. We also fitted a power law of the form $C_{pc} \sim s^\gamma$, which is displayed in orange in the figures.

## Proximity and emissions before and after proximity optimisation

Supplementary Figures from 27 to 30 compare, on the map of the cities under study, proximity time and emissions in the real case and in the proximity optimised scenario.

## Proximity and emissions relative change after optimisation and modelling

Supplementary Figures from 31 to 40 display, on a map of the cities considered, the percentage variations of proximity and emissions after proximity optimisation. Absolute values of the same quantities before and after optimisation are also depicted in scatter plots, together with the power-law modelling of their relation.

While most cities decrease their $CO_2$ emissions for transport under proximity optimisation, some display the inverse behaviour: they increase their emissions. The cities following the latter behaviour are Prague, Warsaw and, more pronouncedly, Auckland. Rotterdam shows no relevant variations of its emissions. In these cities the two quantities of proximity time and emissions are either anti-correlated (it is the case of Prague, Warsaw and Auckland), or uncorrelated (it is the case of Rotterdam). We argue that a possible explanation for this behaviour is that external factors weigh more than the city's structure and functionality. For example in the case of Rotterdam the presence of the port could heavily influence transport-related emissions. The fitting of emissions vs proximity time among neighbourhoods of Prague is heavily influenced by a single neighbourhood displaying very high proximity time and very low emissions for road transport, which hosts an airport for a vast fraction of its area. Our filtering procedure did not highlight this neighbourhood as an outlier, and we made the choice not to remove it ad hoc, but seen its key role in driving alone the negative correlation between proximity time and emissions, results about Prague should be taken with caution. The

relationship between emissions and proximity time across Warsaw's and Auckland's neighbourhoods appears very scattered. Therefore, like the previous case, the resulting fits should be interpreted with caution.

## Emissions variation as a function of the scaling exponent between proximity time and emissions

Supplementary Figure 41 illustrates, for the cities under consideration, the relation between the exponent of the power law $C_{\text{pc}} \sim s^\gamma$ and the variation in emissions expected after proximity optimisation $C_{opt} - C$, where $C$ is the total amount of $CO_2$ emitted in 2021 for transport in the city and $C_{opt}$ is the one expected in the proximity optimised scenario.

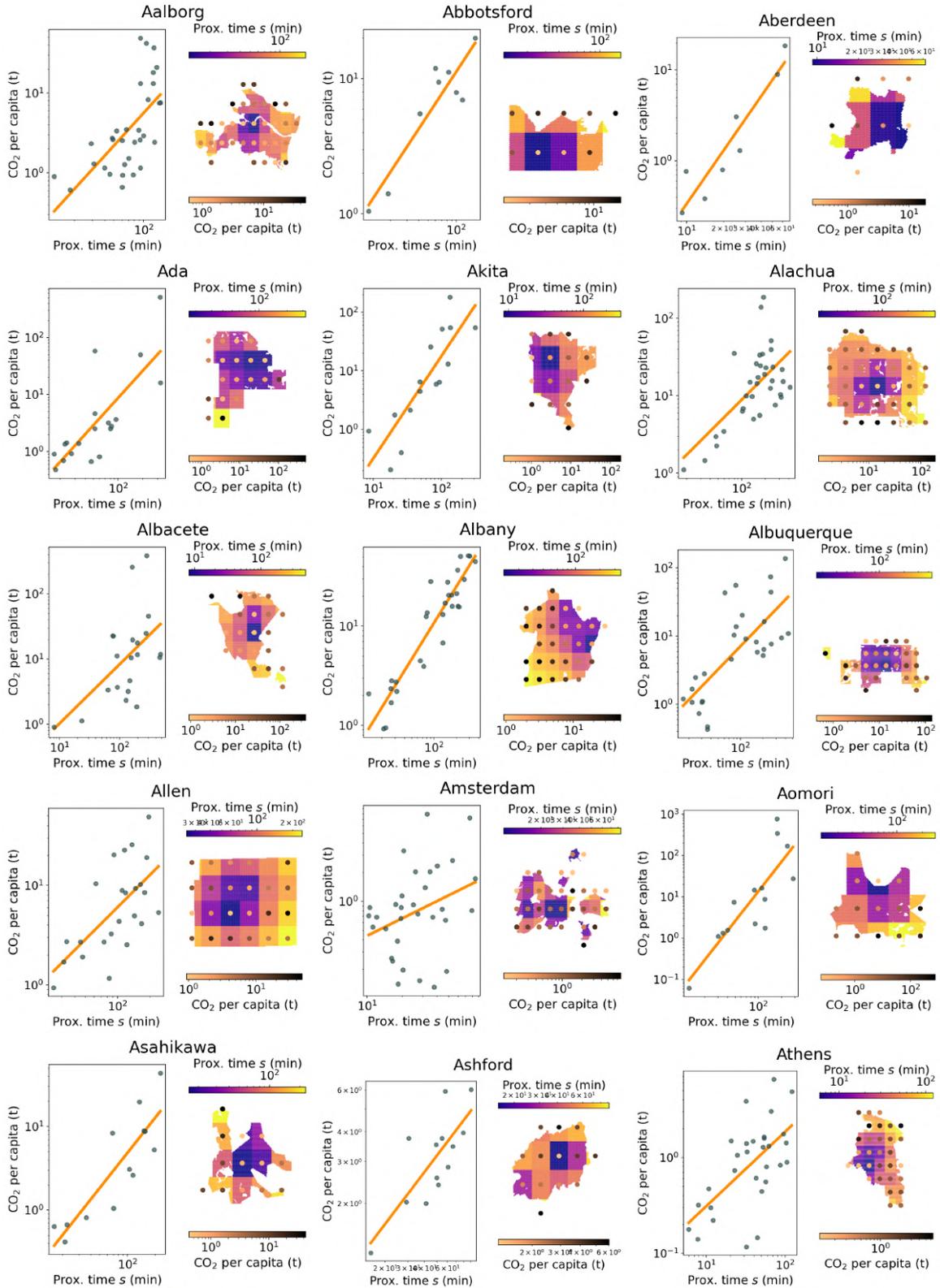

Supplementary Figure 1: **Accessibility/emissions relation at the intra-city level.** For each city, proximity time and per capita road transport emissions of each grid element are represented both as a color-coded map (right panel) and as individual data points in the scatterplot (left panel). The orange line in the scatterplot shows the best fit of a power-law relationship of the form $C_{pc} \sim s^{\gamma}$.

16

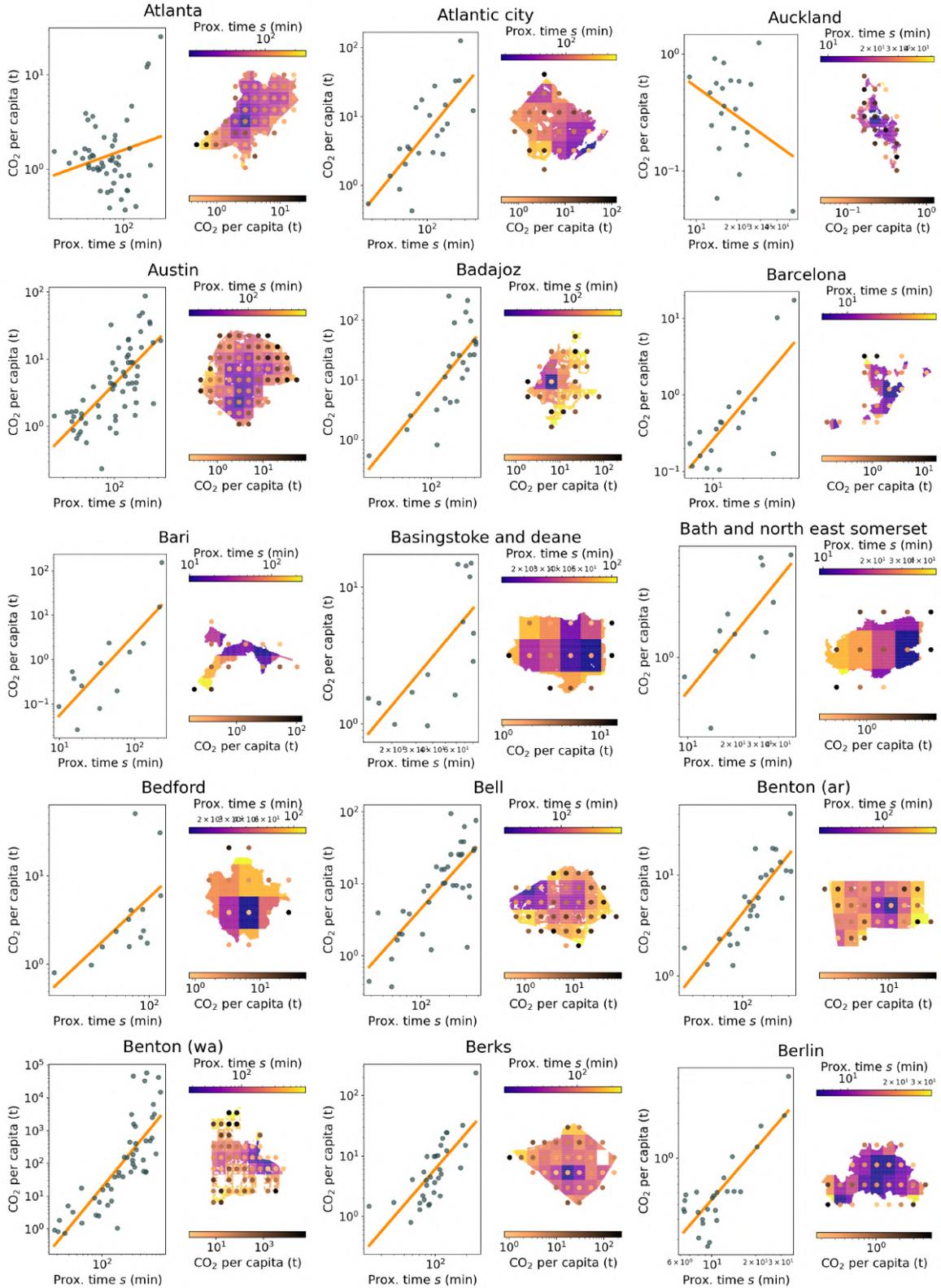

Supplementary Figure 2: **Accessibility/emissions relation at the intra-city level.** For each city, proximity time and per capita road transport emissions of each grid element are represented both as a color-coded map (right panel) and as individual data points in the scatterplot (left panel). The orange line in the scatterplot shows the best fit of a power-law relationship of the form $C_{pc} \sim s^{\gamma}$.

17

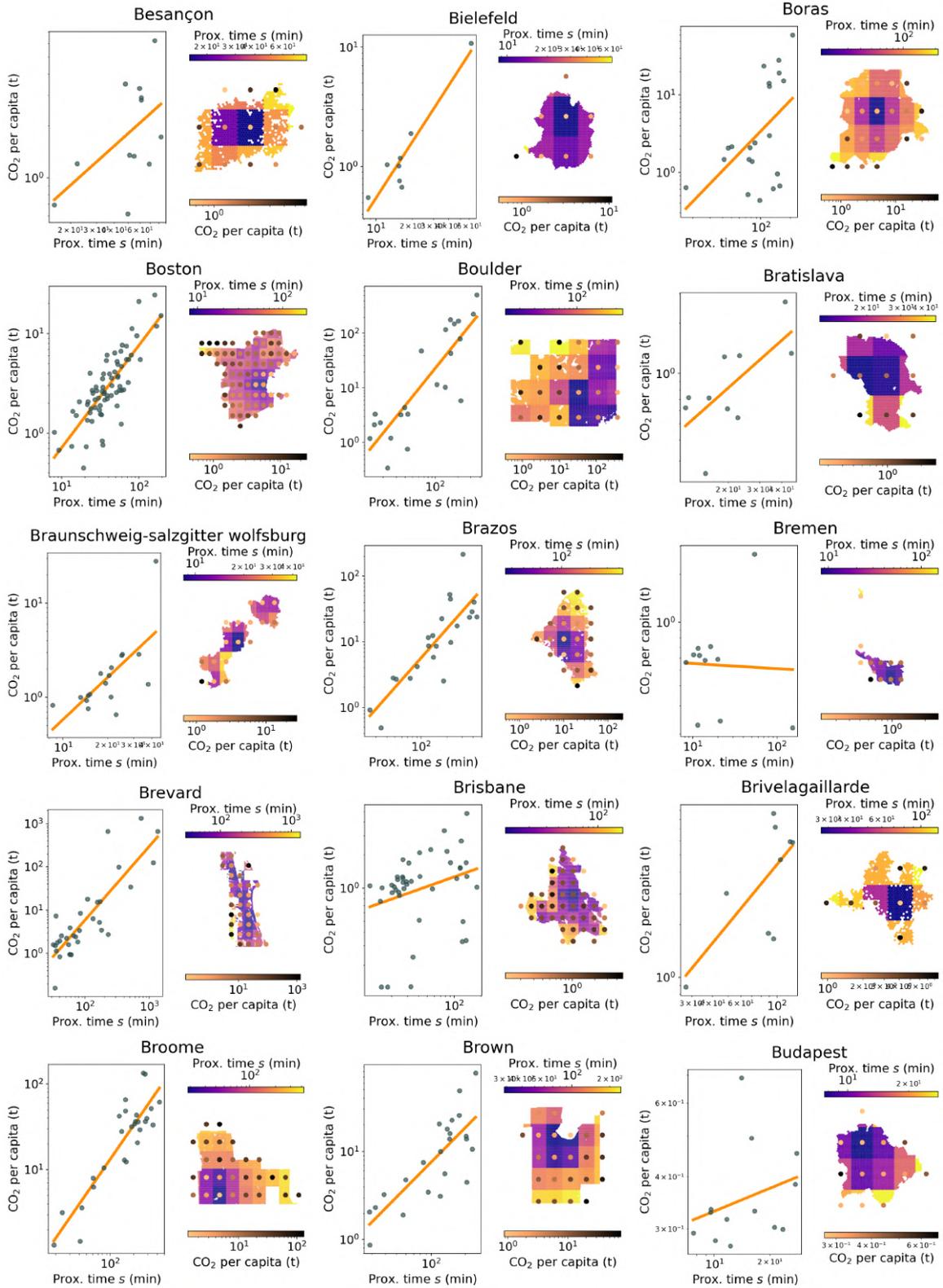

Supplementary Figure 3: **Accessibility/emissions relation at the intra-city level.** For each city, proximity time and per capita road transport emissions of each grid element are represented both as a color-coded map (right panel) and as individual data points in the scatterplot (left panel). The orange line in the scatterplot shows the best fit of a power-law relationship of the form $C_{pc} \sim s^{\gamma}$.

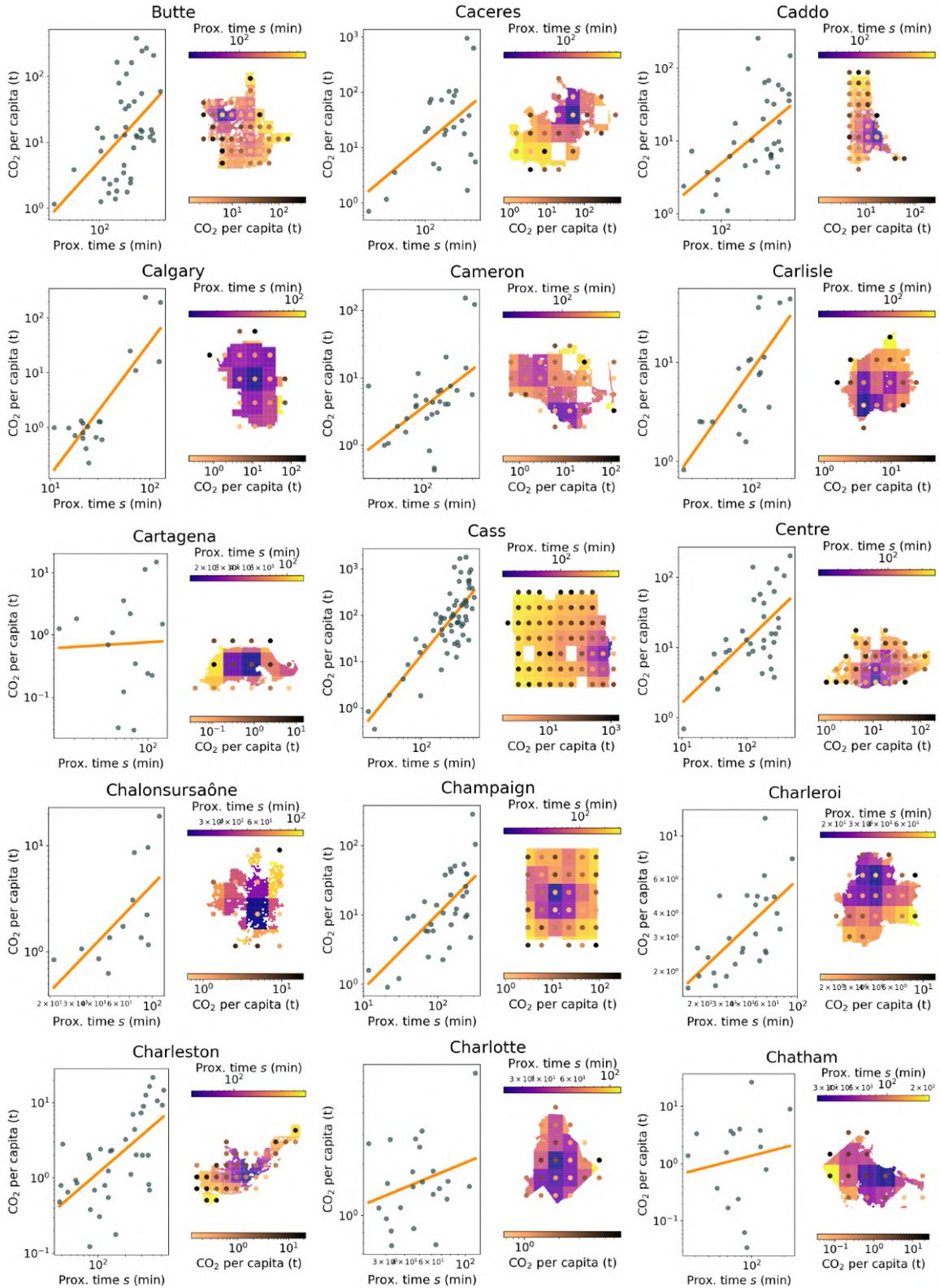

Supplementary Figure 4: **Accessibility/emissions relation at the intra-city level.** For each city, proximity time and per capita road transport emissions of each grid element are represented both as a color-coded map (right panel) and as individual data points in the scatterplot (left panel). The orange line in the scatterplot shows the best fit of a power-law relationship of the form $C_{pc} \sim s^\gamma$.

19

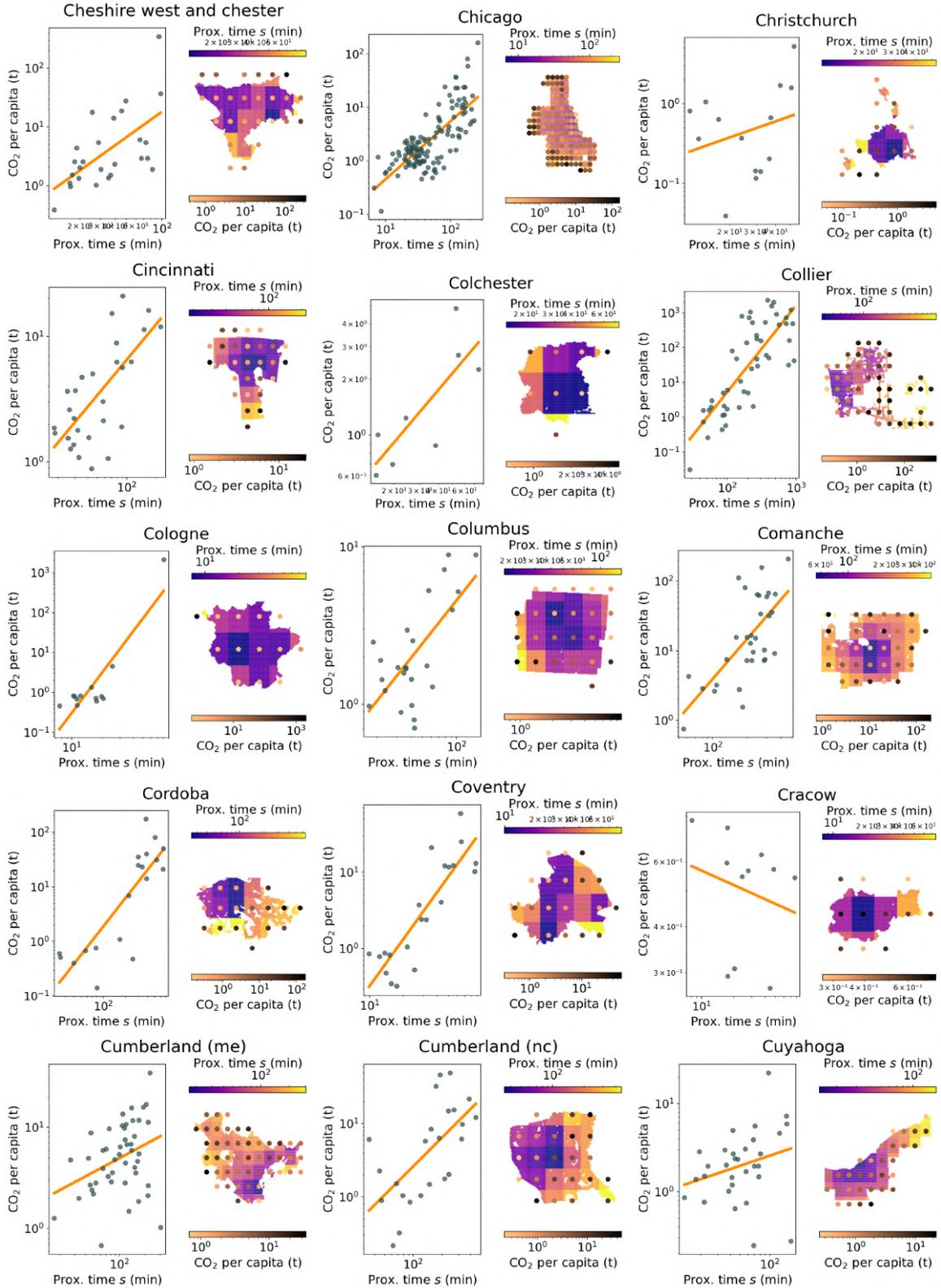

Supplementary Figure 5: **Accessibility/emissions relation at the intra-city level.** For each city, proximity time and per capita road transport emissions of each grid element are represented both as a color-coded map (right panel) and as individual data points in the scatterplot (left panel). The orange line in the scatterplot shows the best fit of a power-law relationship of the form $C_{pc} \sim s^\gamma$.

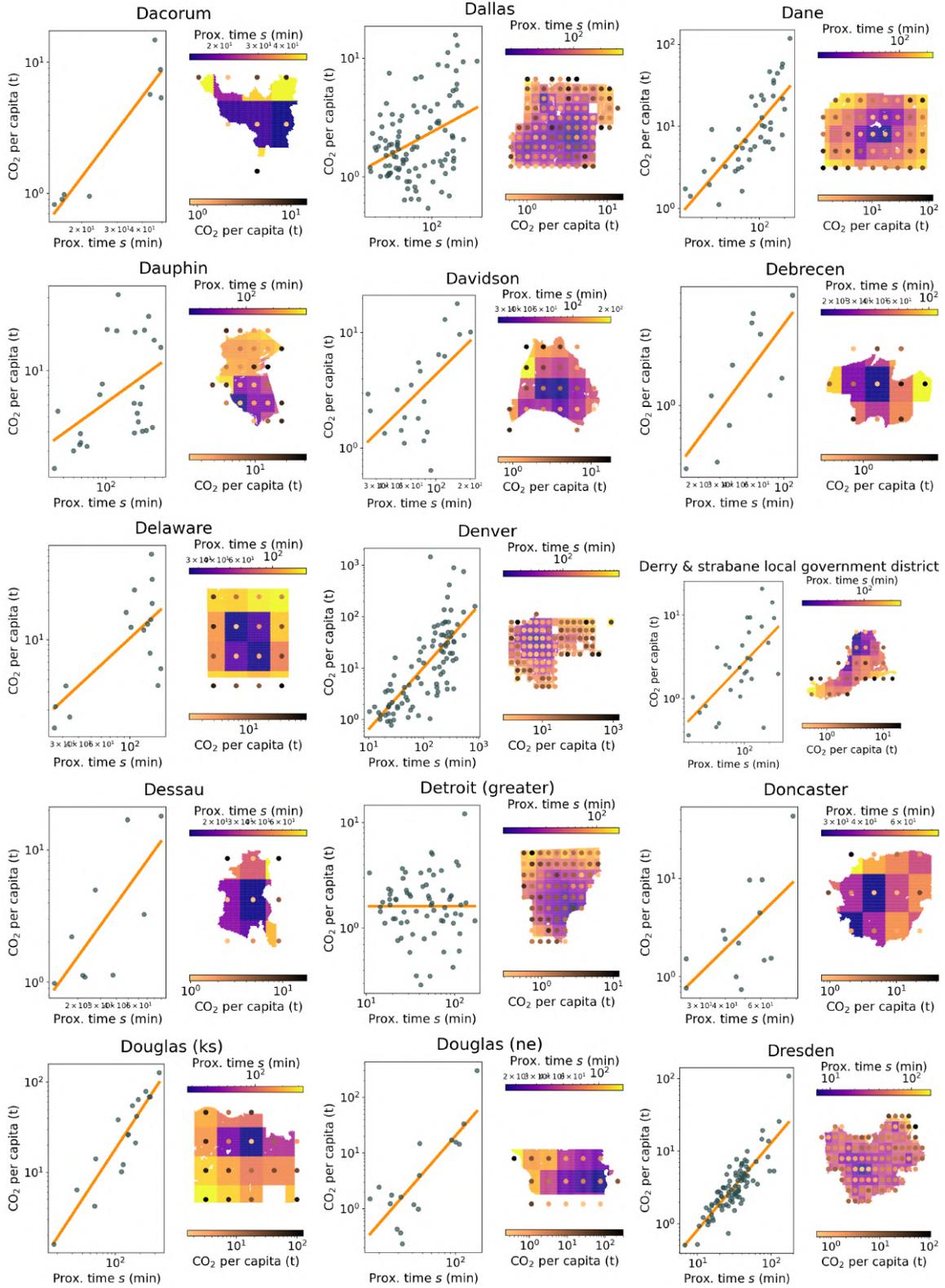

Supplementary Figure 6: **Accessibility/emissions relation at the intra-city level.** For each city, proximity time and per capita road transport emissions of each grid element are represented both as a color-coded map (right panel) and as individual data points in the scatterplot (left panel). The orange line in the scatterplot shows the best fit of a power-law relationship of the form $C_{pc} \sim s^\gamma$.

21

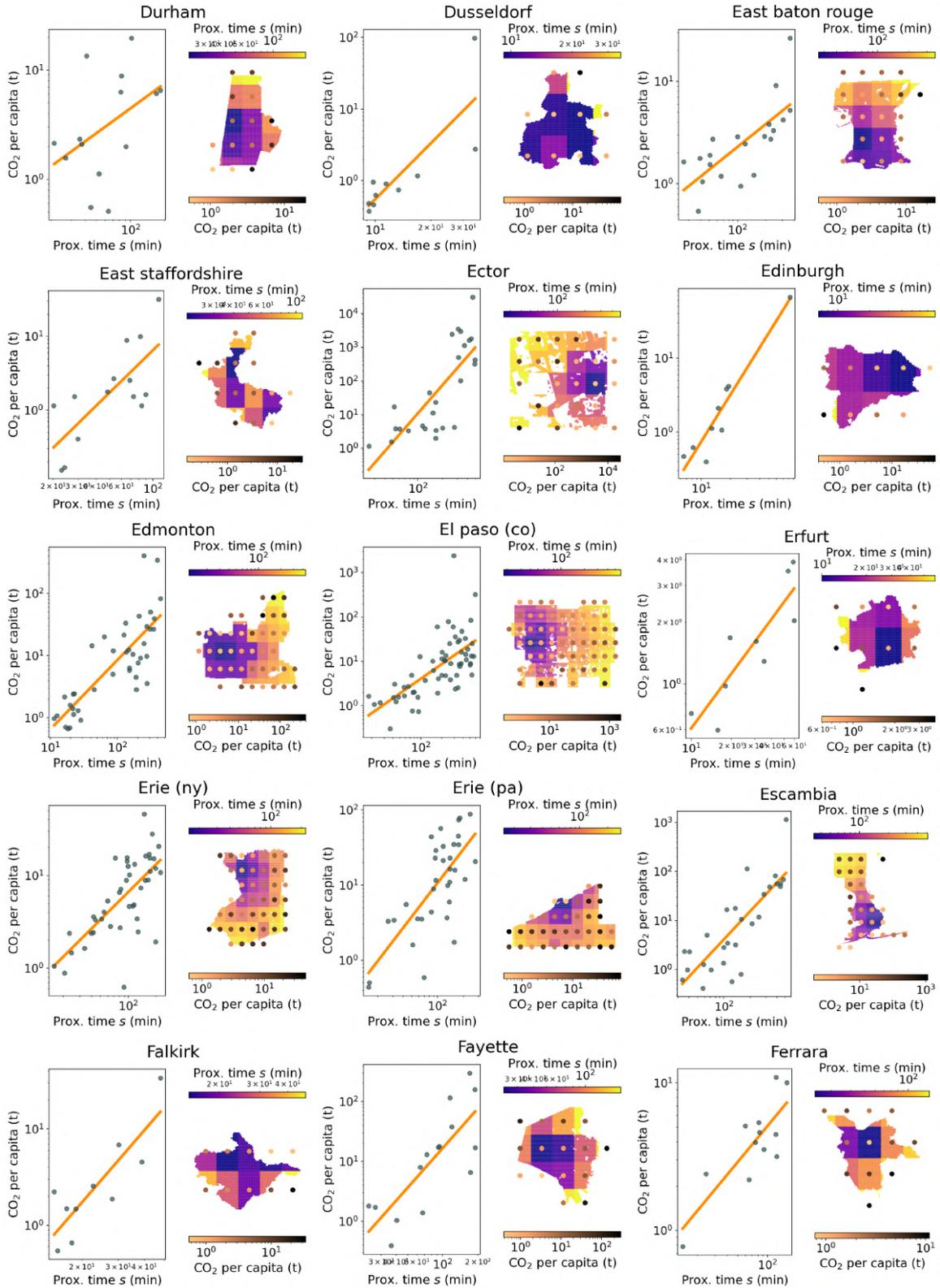

Supplementary Figure 7: **Accessibility/emissions relation at the intra-city level.** For each city, proximity time and per capita road transport emissions of each grid element are represented both as a color-coded map (right panel) and as individual data points in the scatterplot (left panel). The orange line in the scatterplot shows the best fit of a power-law relationship of the form $C_{pc} \sim s^\gamma$.

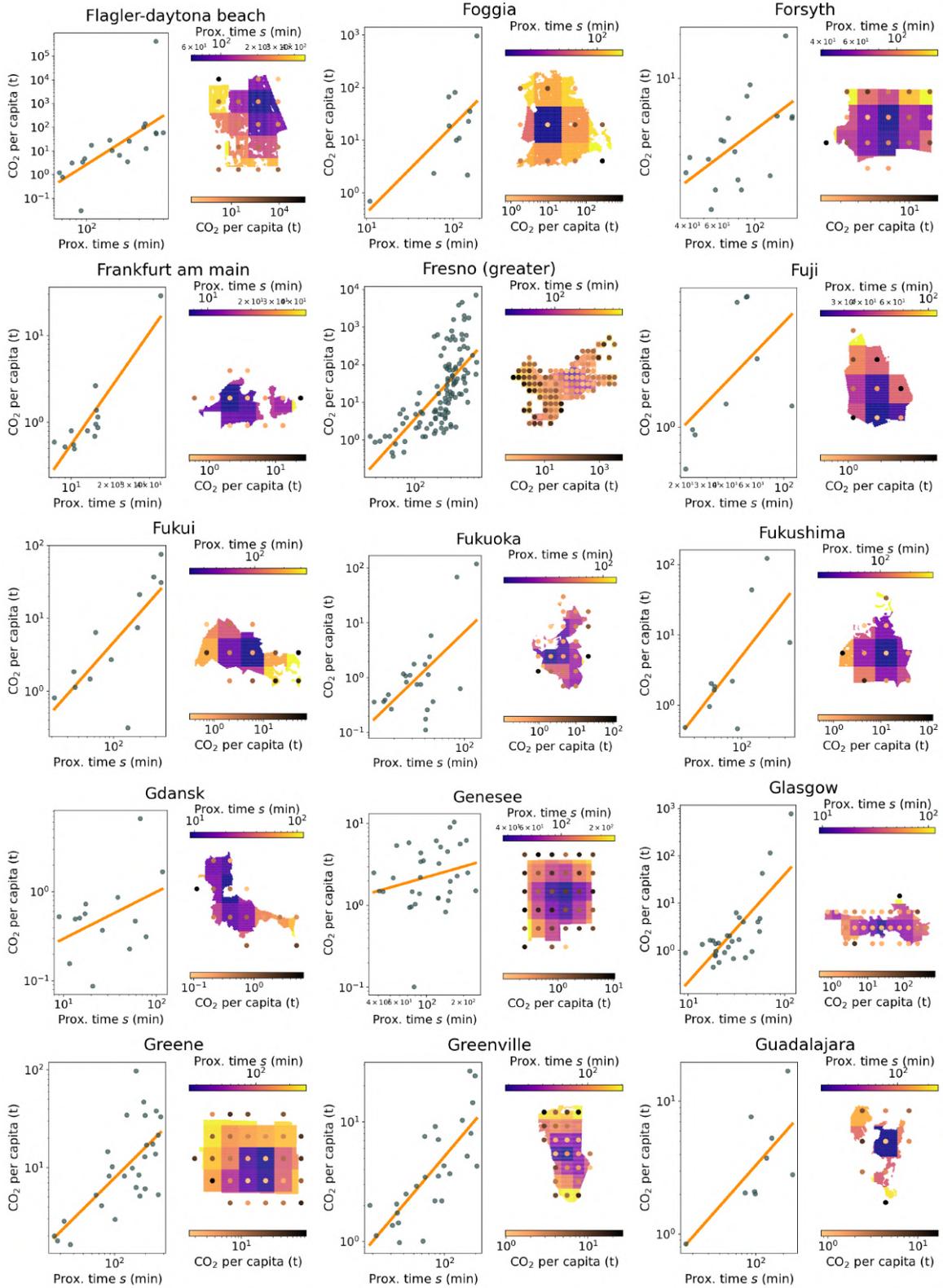

Supplementary Figure 8: **Accessibility/emissions relation at the intra-city level.** For each city, proximity time and per capita road transport emissions of each grid element are represented both as a color-coded map (right panel) and as individual data points in the scatterplot (left panel). The orange line in the scatterplot shows the best fit of a power-law relationship of the form $C_{pc} \sim s^\gamma$.

23

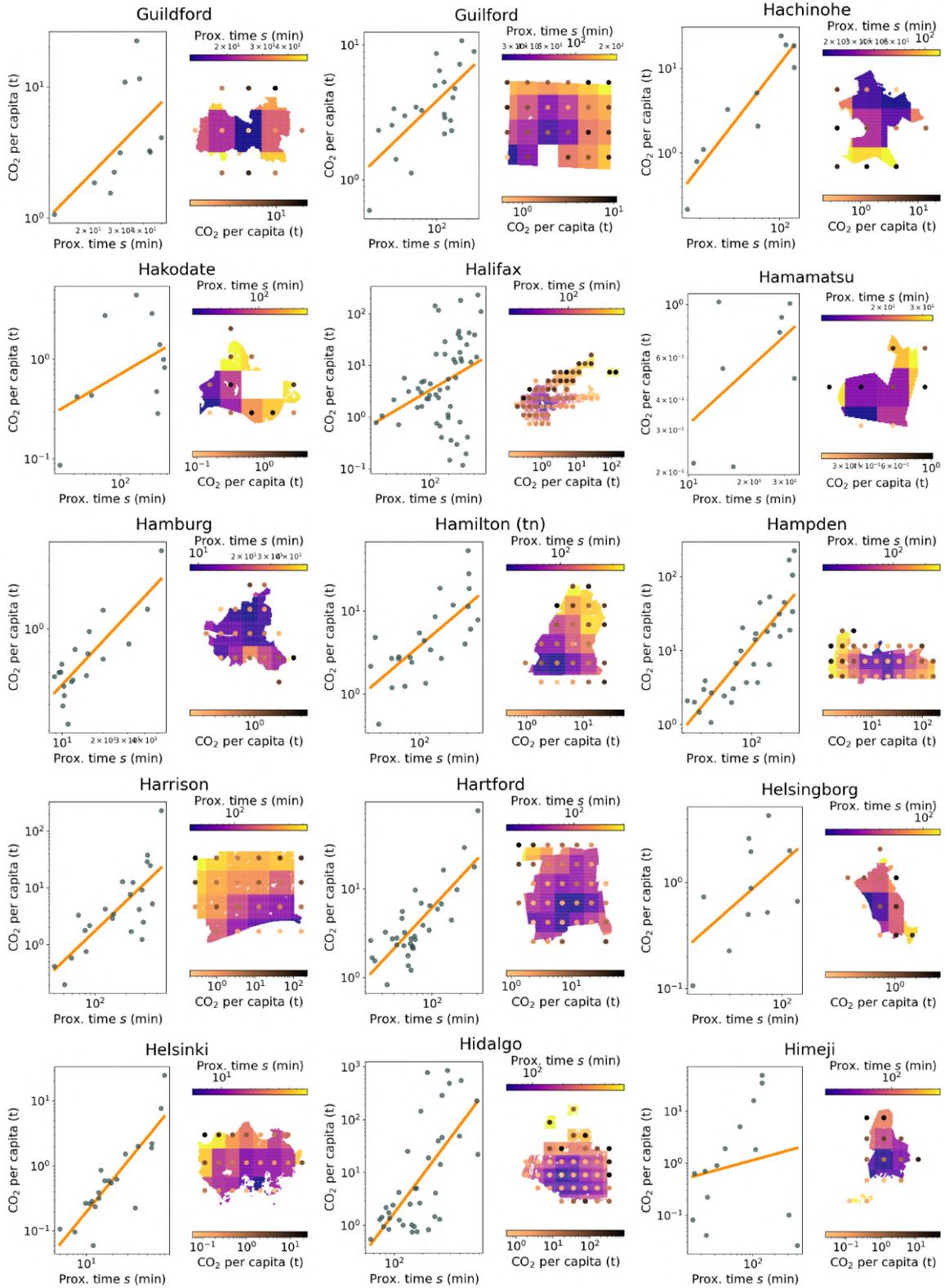

Supplementary Figure 9: **Accessibility/emissions relation at the intra-city level.** For each city, proximity time and per capita road transport emissions of each grid element are represented both as a color-coded map (right panel) and as individual data points in the scatterplot (left panel). The orange line in the scatterplot shows the best fit of a power-law relationship of the form $C_{pc} \sim s^{\gamma}$.

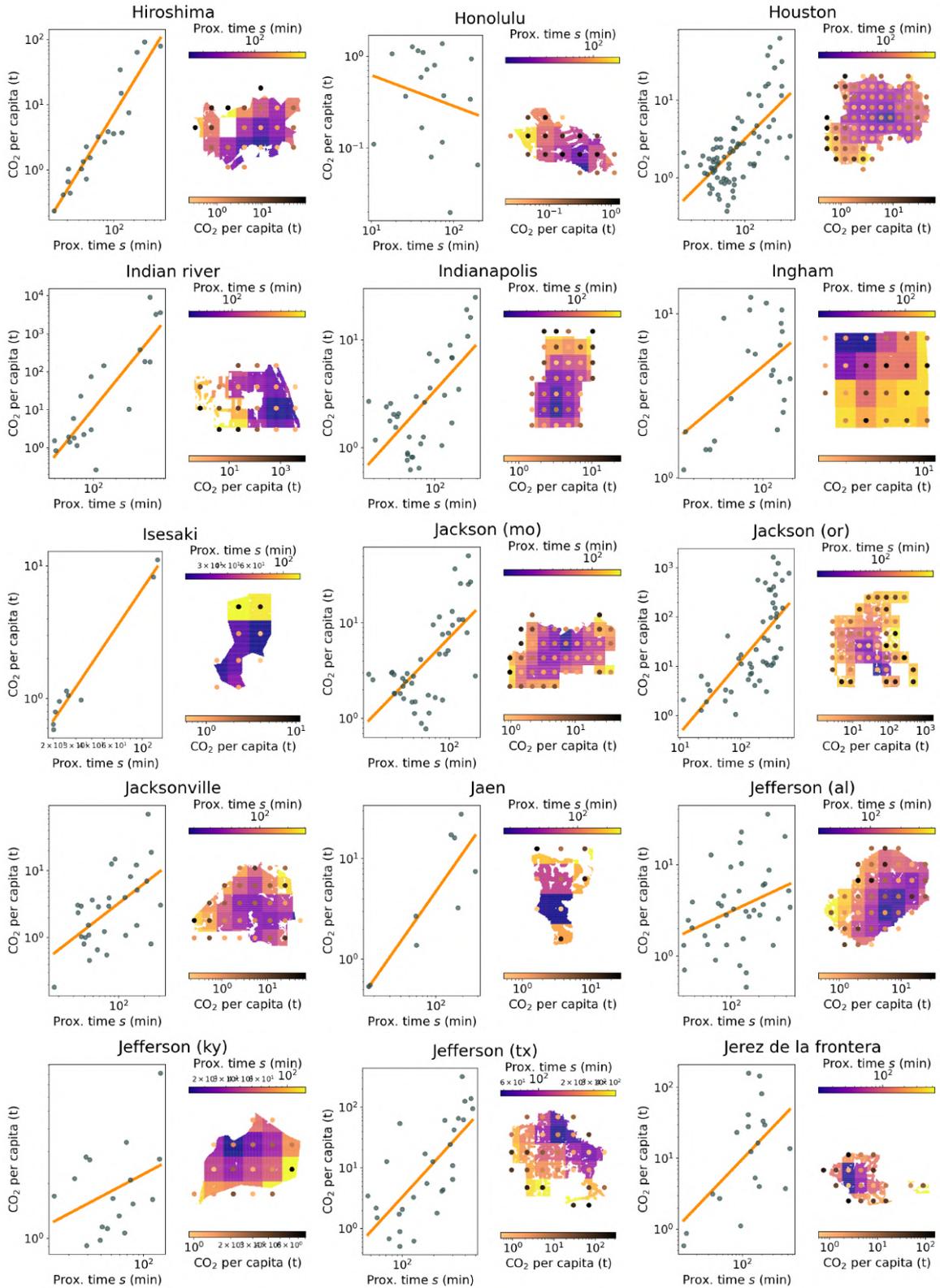

Supplementary Figure 10: **Accessibility/emissions relation at the intra-city level.** For each city, proximity time and per capita road transport emissions of each grid element are represented both as a color-coded map (right panel) and as individual data points in the scatterplot (left panel). The orange line in the scatterplot shows the best fit of a power-law relationship of the form $C_{pc} \sim s^{\gamma}$.

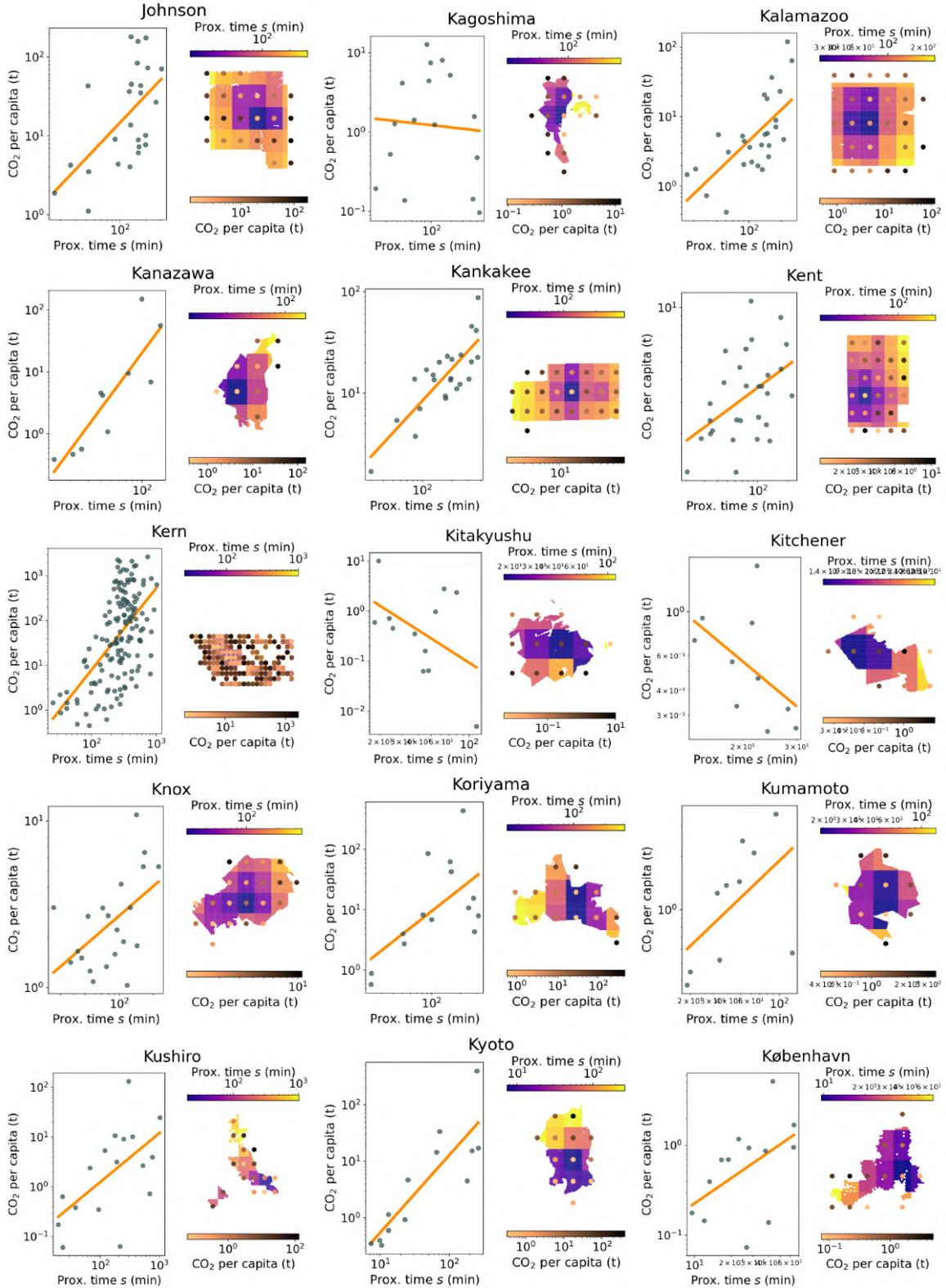

Supplementary Figure 11: **Accessibility/emissions relation at the intra-city level.** For each city, proximity time and per capita road transport emissions of each grid element are represented both as a color-coded map (right panel) and as individual data points in the scatterplot (left panel). The orange line in the scatterplot shows the best fit of a power-law relationship of the form $C_{pc} \sim s^{\gamma}$.

26

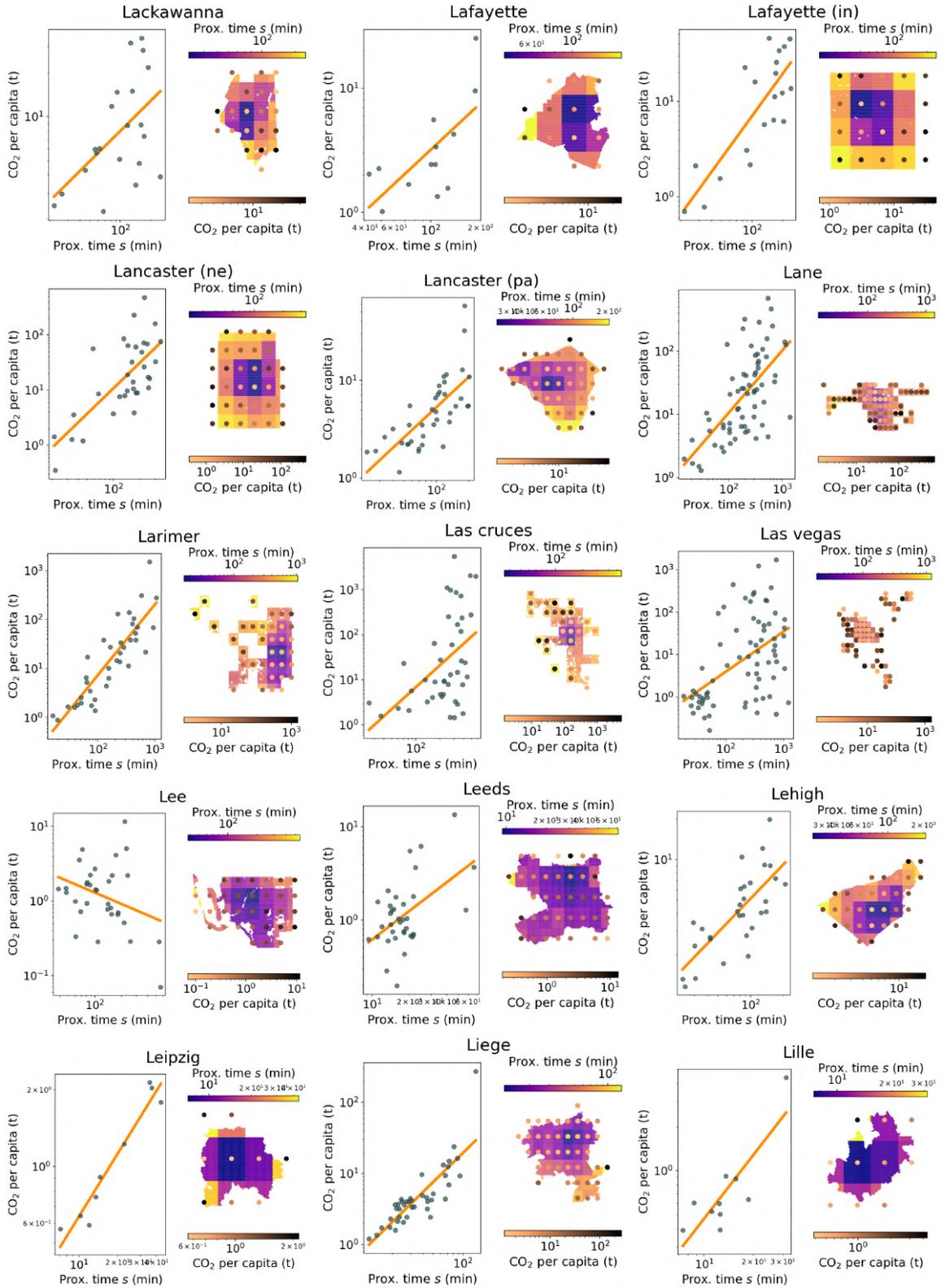

Supplementary Figure 12: **Accessibility/emissions relation at the intra-city level.** For each city, proximity time and per capita road transport emissions of each grid element are represented both as a color-coded map (right panel) and as individual data points in the scatterplot (left panel). The orange line in the scatterplot shows the best fit of a power-law relationship of the form $C_{pc} \sim s^\gamma$.

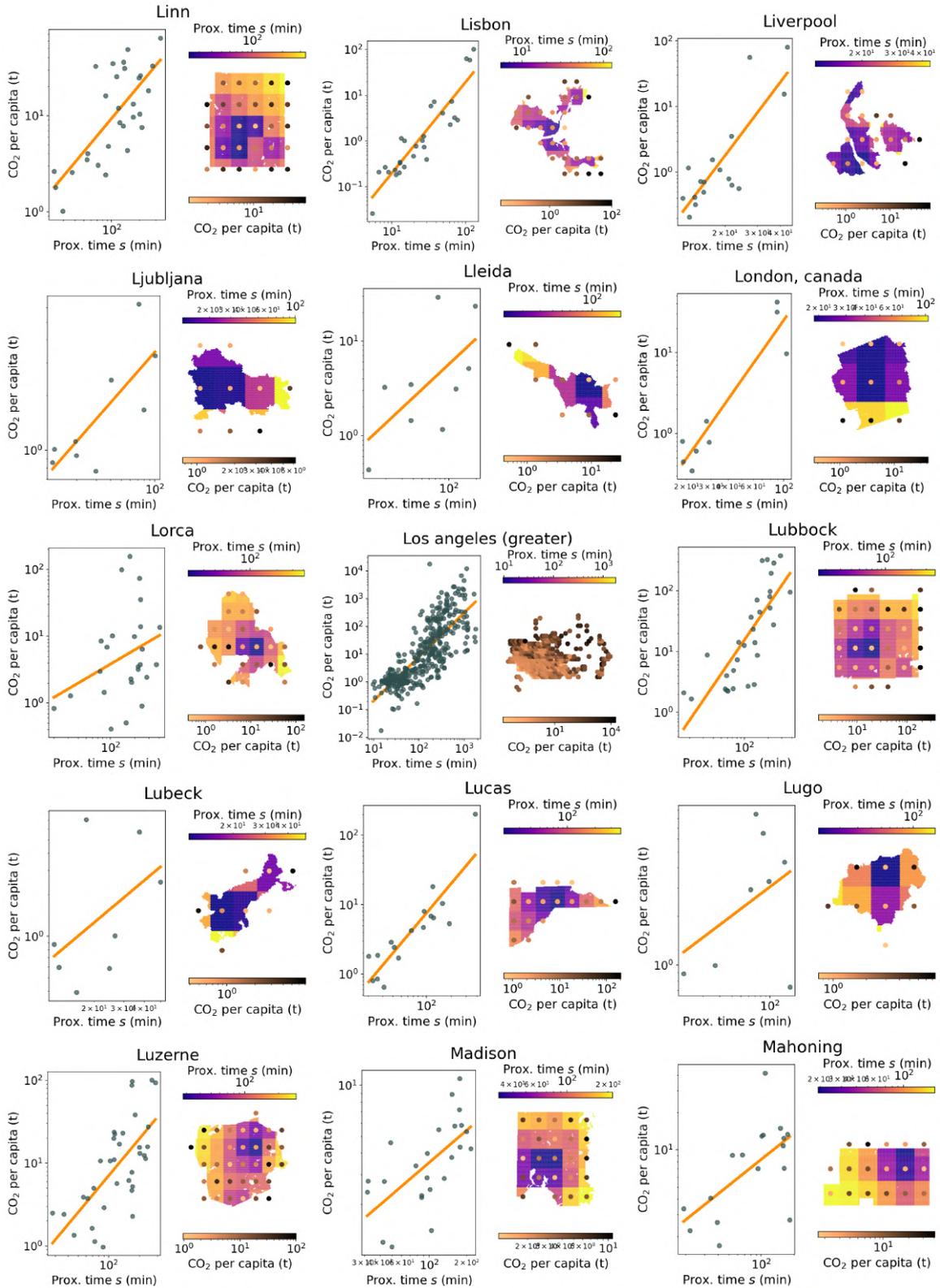

Supplementary Figure 13: **Accessibility/emissions relation at the intra-city level.** For each city, proximity time and per capita road transport emissions of each grid element are represented both as a color-coded map (right panel) and as individual data points in the scatterplot (left panel). The orange line in the scatterplot shows the best fit of a power-law relationship of the form $C_{pc} \sim s^\gamma$.

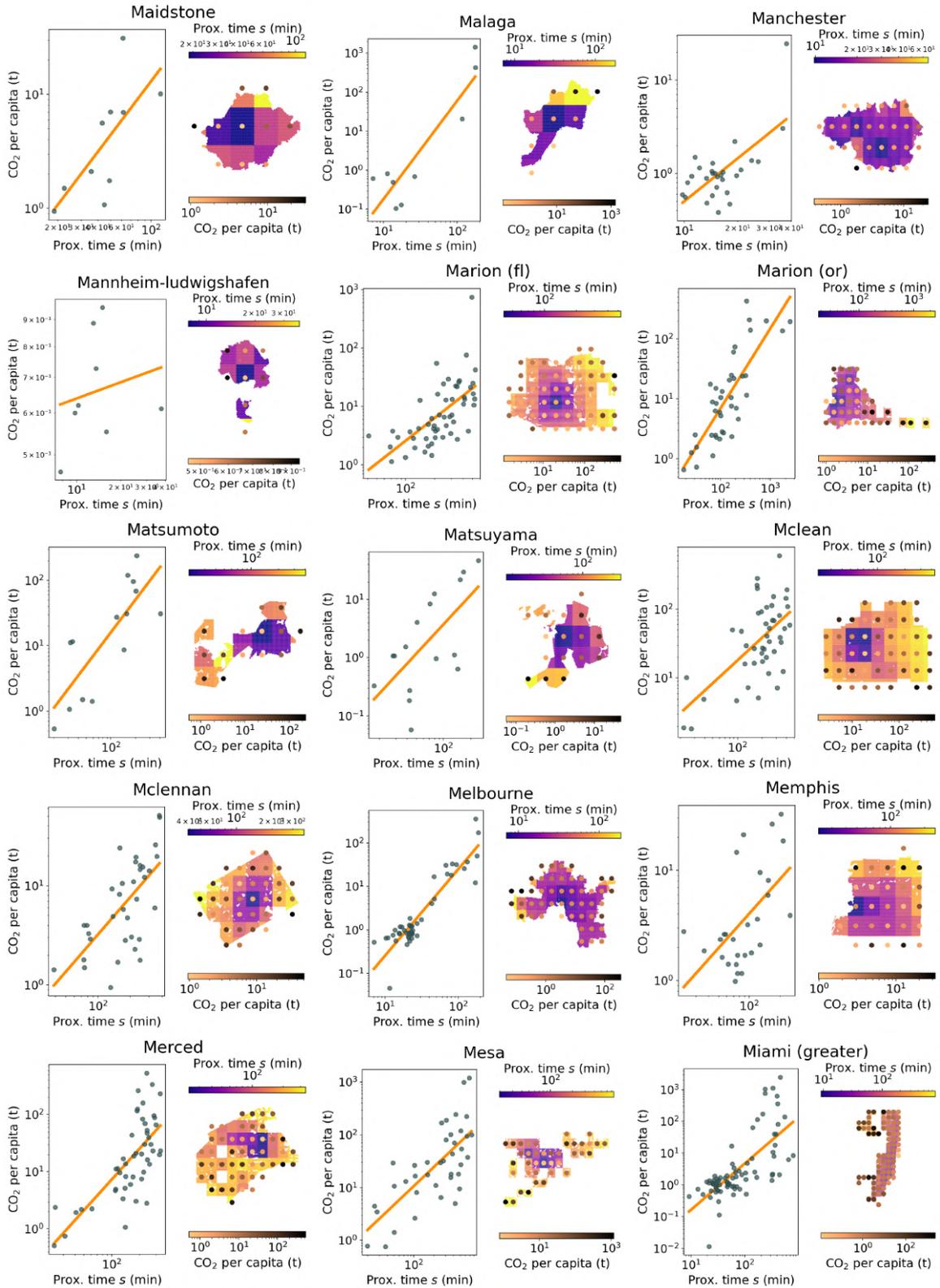

Supplementary Figure 14: **Accessibility/emissions relation at the intra-city level.** For each city, proximity time and per capita road transport emissions of each grid element are represented both as a color-coded map (right panel) and as individual data points in the scatterplot (left panel). The orange line in the scatterplot shows the best fit of a power-law relationship of the form $C_{pc} \sim s^{\gamma}$.

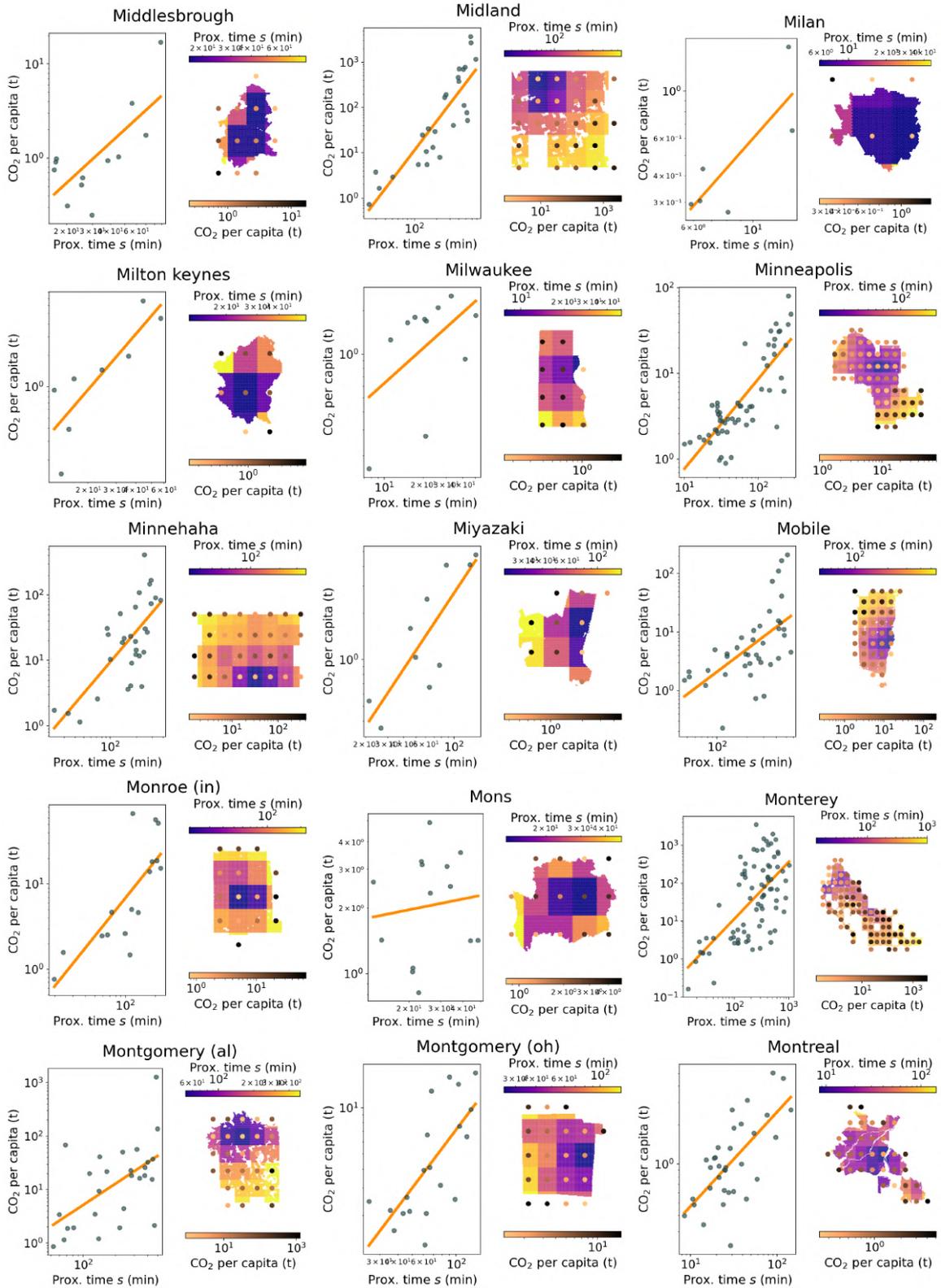

Supplementary Figure 15: **Accessibility/emissions relation at the intra-city level.** For each city, proximity time and per capita road transport emissions of each grid element are represented both as a color-coded map (right panel) and as individual data points in the scatterplot (left panel). The orange line in the scatterplot shows the best fit of a power-law relationship of the form $C_{pc} \sim s^\gamma$.

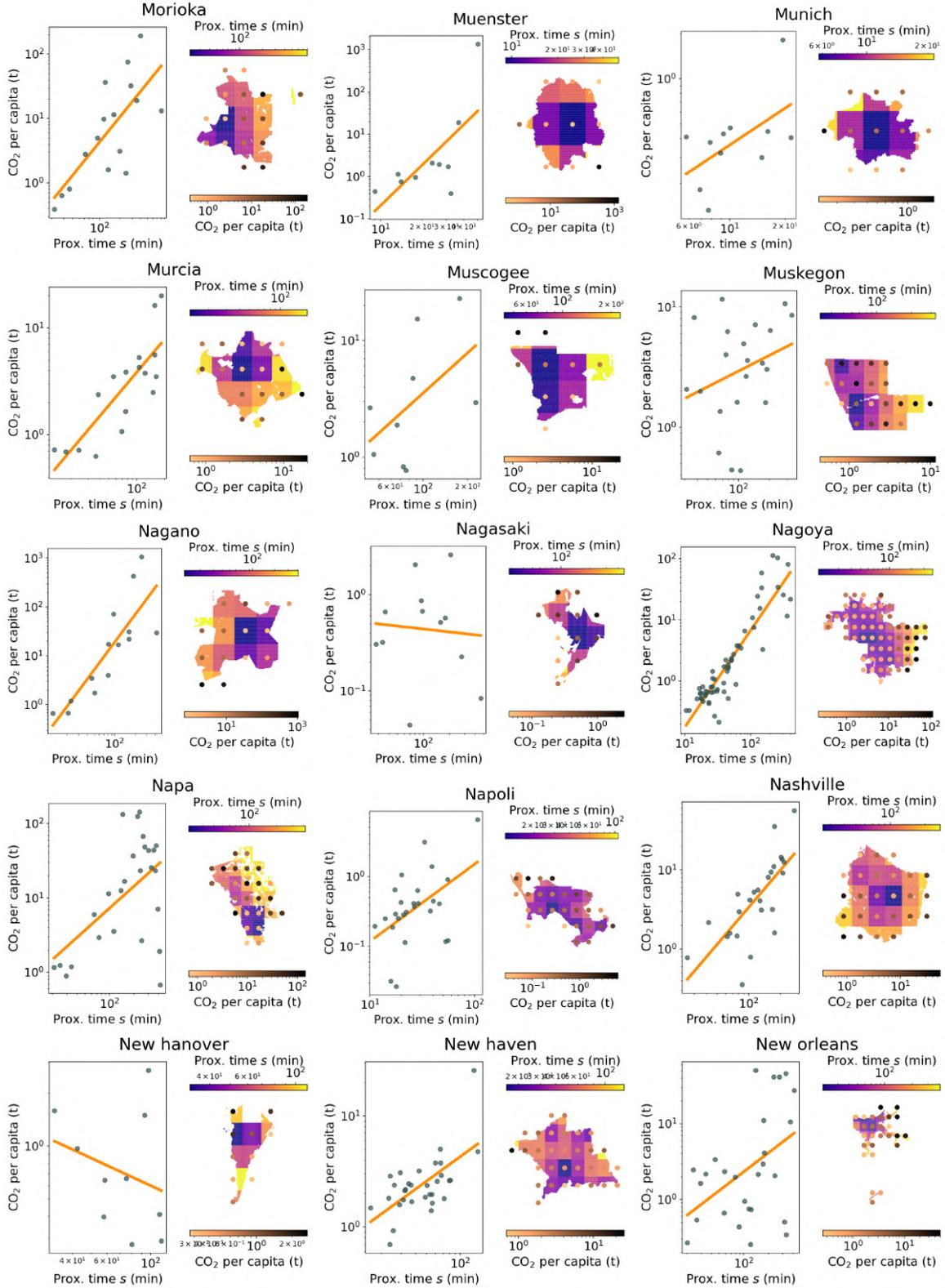

Supplementary Figure 16: **Accessibility/emissions relation at the intra-city level.** For each city, proximity time and per capita road transport emissions of each grid element are represented both as a color-coded map (right panel) and as individual data points in the scatterplot (left panel). The orange line in the scatterplot shows the best fit of a power-law relationship of the form $C_{pc} \sim s^\gamma$.

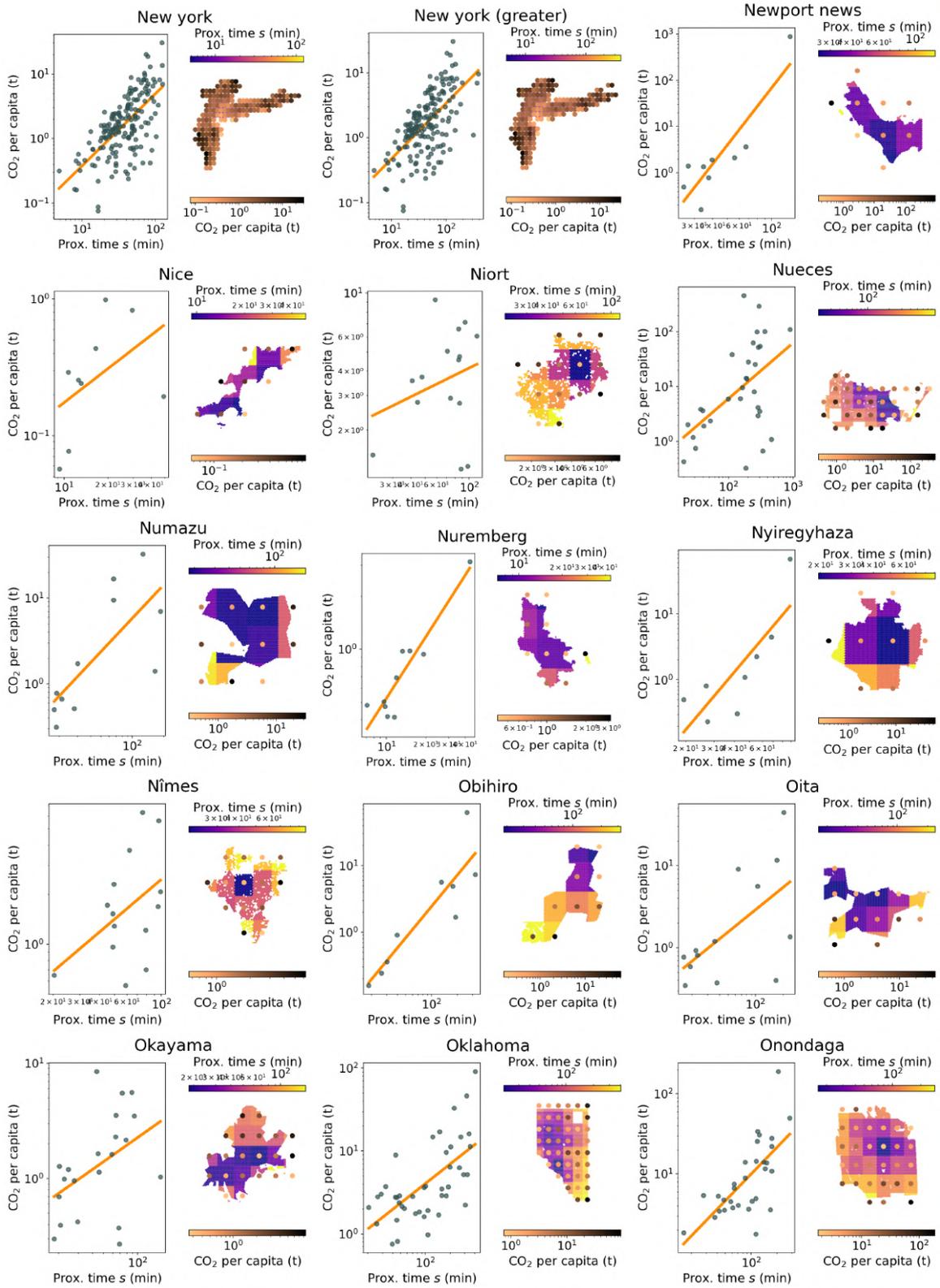

Supplementary Figure 17: **Accessibility/emissions relation at the intra-city level.** For each city, proximity time and per capita road transport emissions of each grid element are represented both as a color-coded map (right panel) and as individual data points in the scatterplot (left panel). The orange line in the scatterplot shows the best fit of a power-law relationship of the form $C_{pc} \sim s^{\gamma}$.

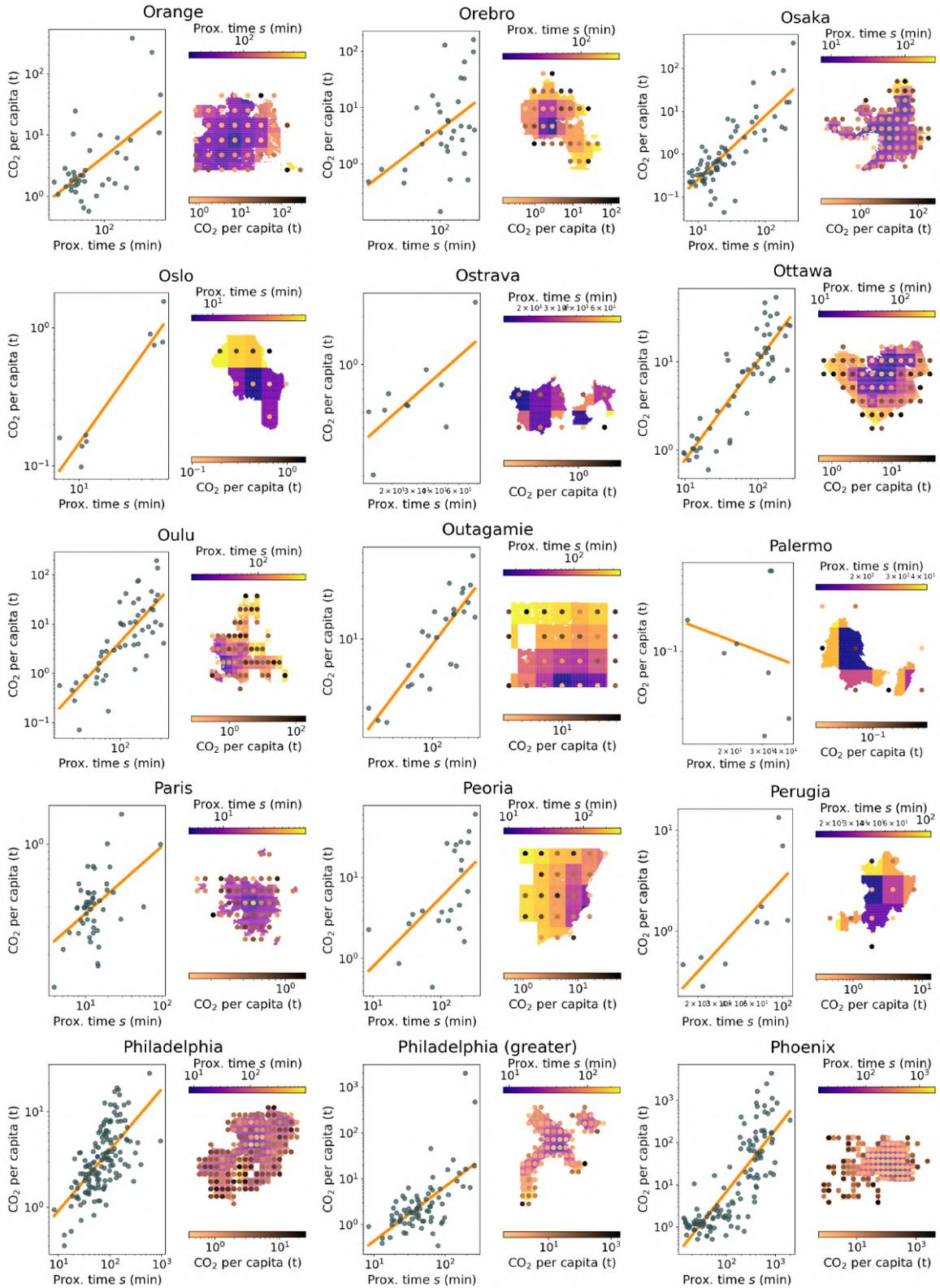

Supplementary Figure 18: **Accessibility/emissions relation at the intra-city level.** For each city, proximity time and per capita road transport emissions of each grid element are represented both as a color-coded map (right panel) and as individual data points in the scatterplot (left panel). The orange line in the scatterplot shows the best fit of a power-law relationship of the form $C_{pc} \sim s^\gamma$.

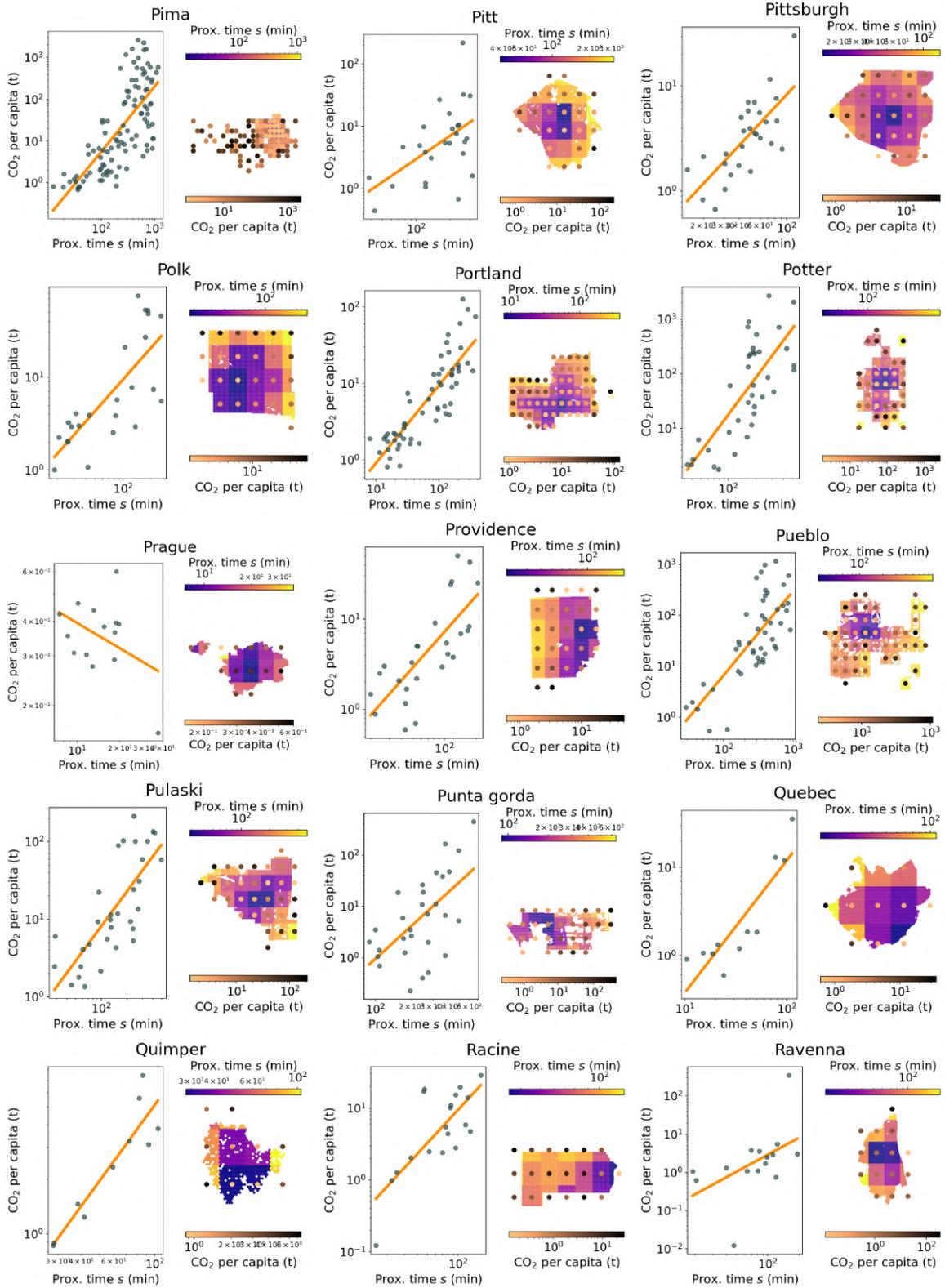

Supplementary Figure 19: **Accessibility/emissions relation at the intra-city level.** For each city, proximity time and per capita road transport emissions of each grid element are represented both as a color-coded map (right panel) and as individual data points in the scatterplot (left panel). The orange line in the scatterplot shows the best fit of a power-law relationship of the form $C_{pc} \sim s^\gamma$.

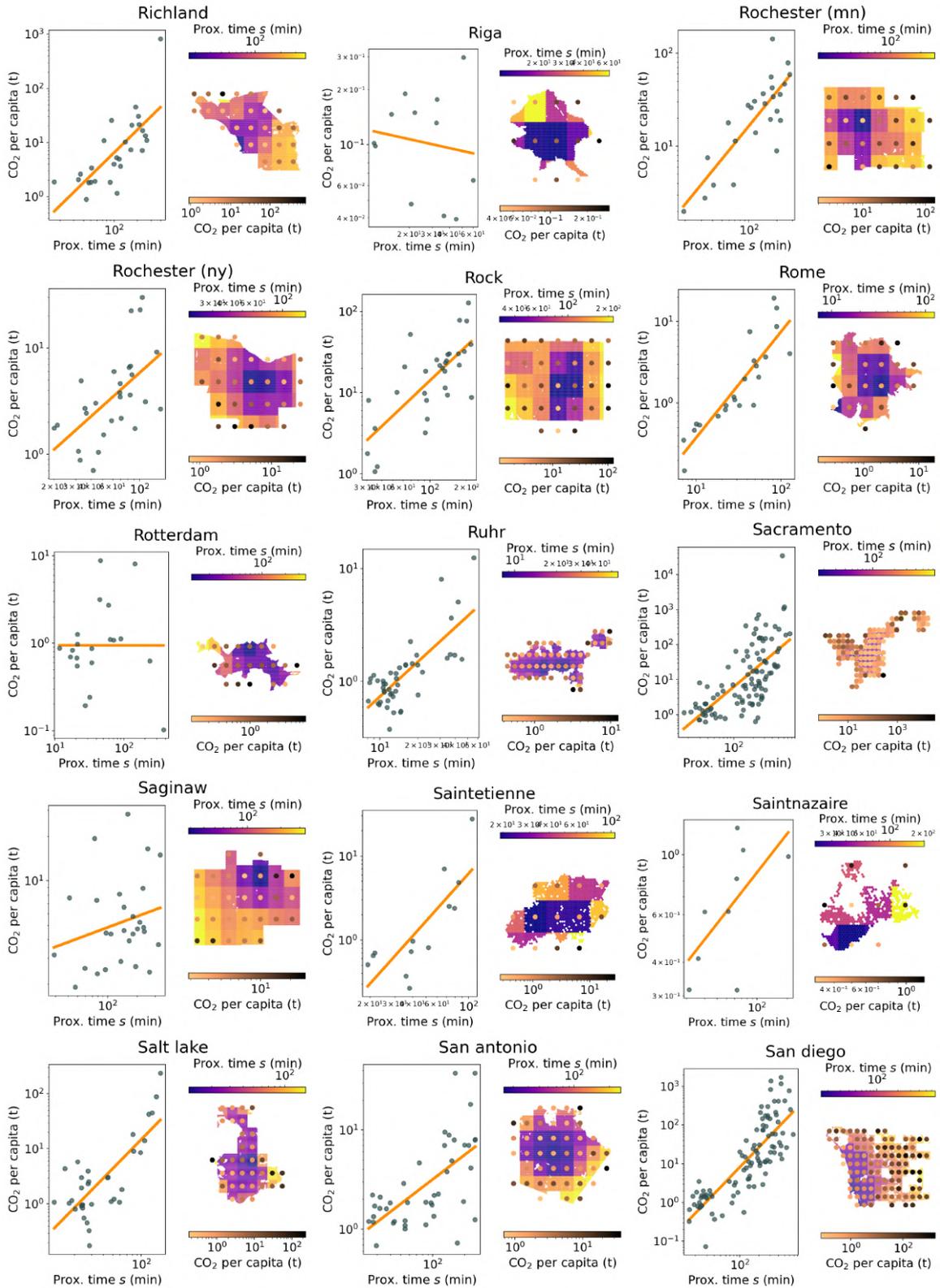

Supplementary Figure 20: **Accessibility/emissions relation at the intra-city level.** For each city, proximity time and per capita road transport emissions of each grid element are represented both as a color-coded map (right panel) and as individual data points in the scatterplot (left panel). The orange line in the scatterplot shows the best fit of a power-law relationship of the form $C_{pc} \sim s^{\gamma}$.

35

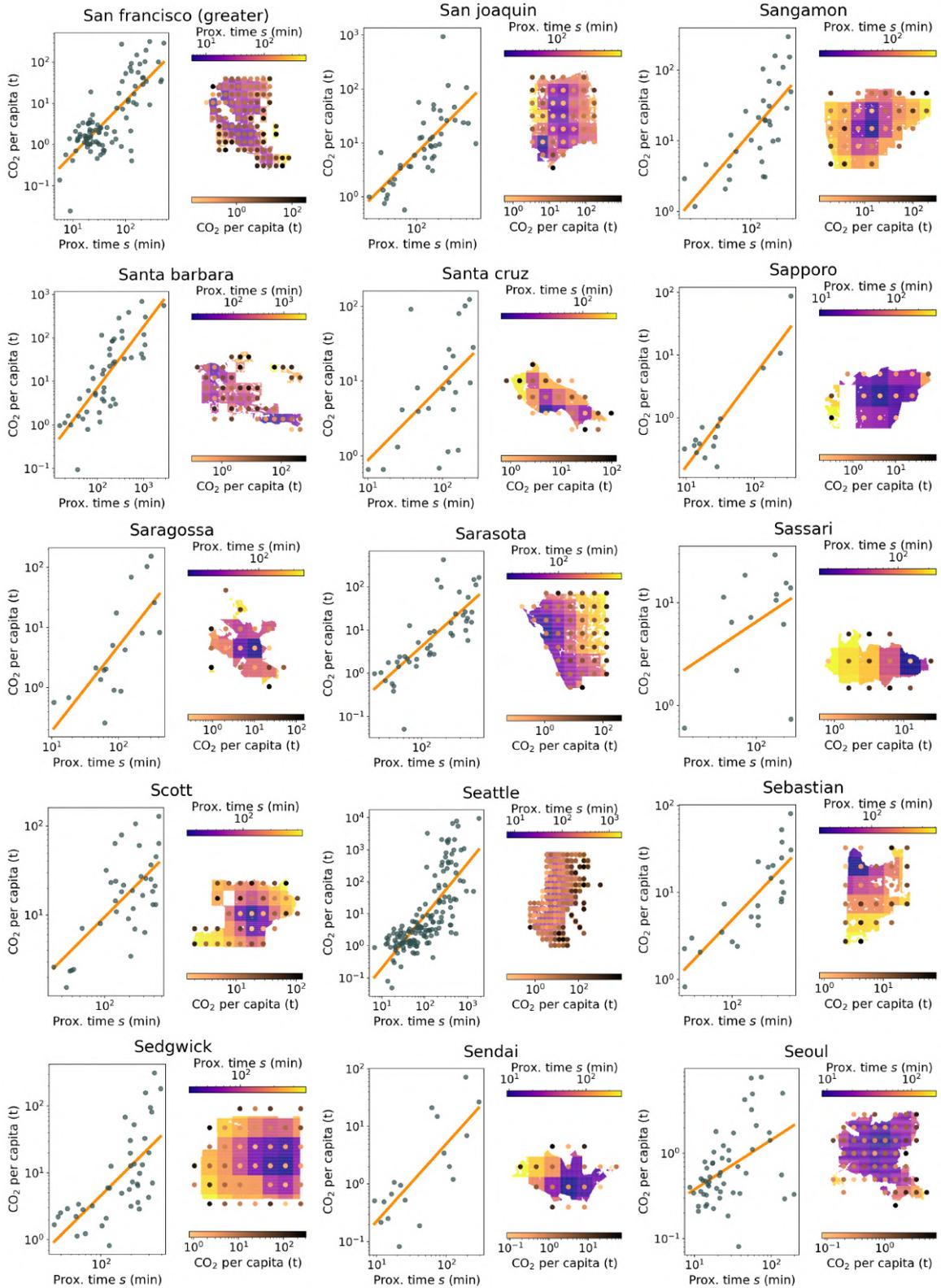

Supplementary Figure 21: **Accessibility/emissions relation at the intra-city level.** For each city, proximity time and per capita road transport emissions of each grid element are represented both as a color-coded map (right panel) and as individual data points in the scatterplot (left panel). The orange line in the scatterplot shows the best fit of a power-law relationship of the form $C_{pc} \sim s^\gamma$.

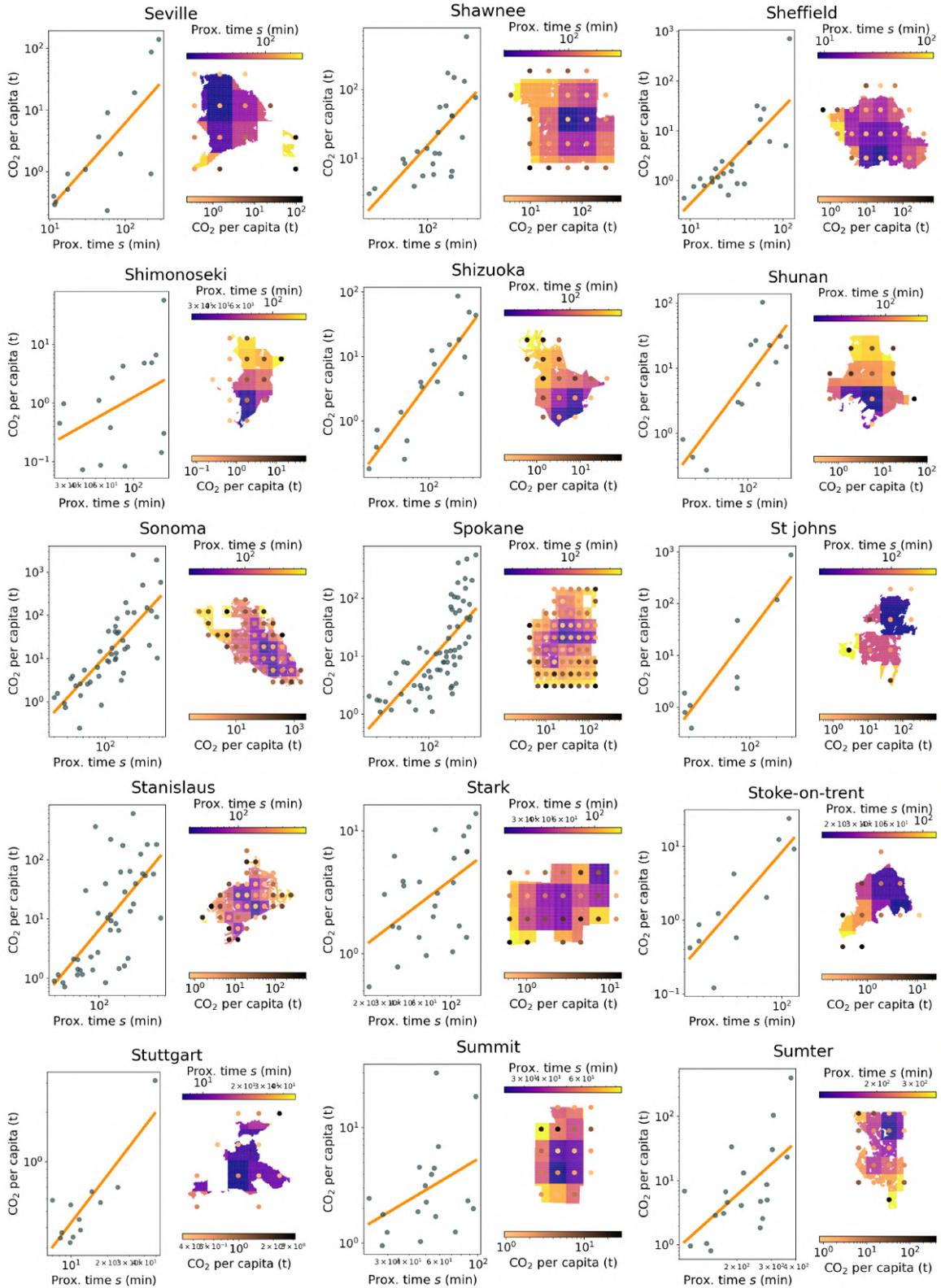

Supplementary Figure 22: **Accessibility/emissions relation at the intra-city level.** For each city, proximity time and per capita road transport emissions of each grid element are represented both as a color-coded map (right panel) and as individual data points in the scatterplot (left panel). The orange line in the scatterplot shows the best fit of a power-law relationship of the form $C_{pc} \sim s^\gamma$.

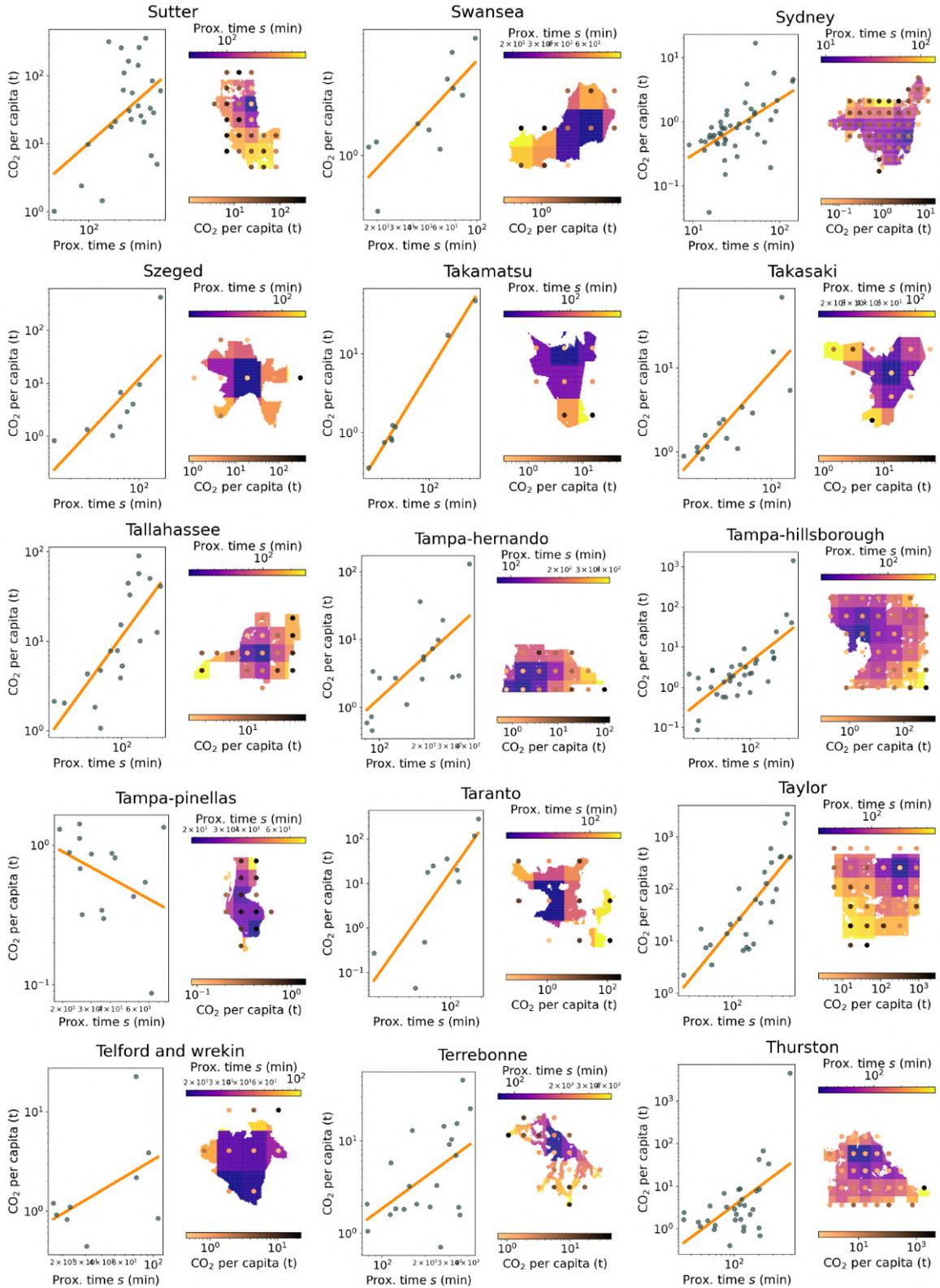

Supplementary Figure 23: **Accessibility/emissions relation at the intra-city level.** For each city, proximity time and per capita road transport emissions of each grid element are represented both as a color-coded map (right panel) and as individual data points in the scatterplot (left panel). The orange line in the scatterplot shows the best fit of a power-law relationship of the form $C_{pc} \sim s^\gamma$.

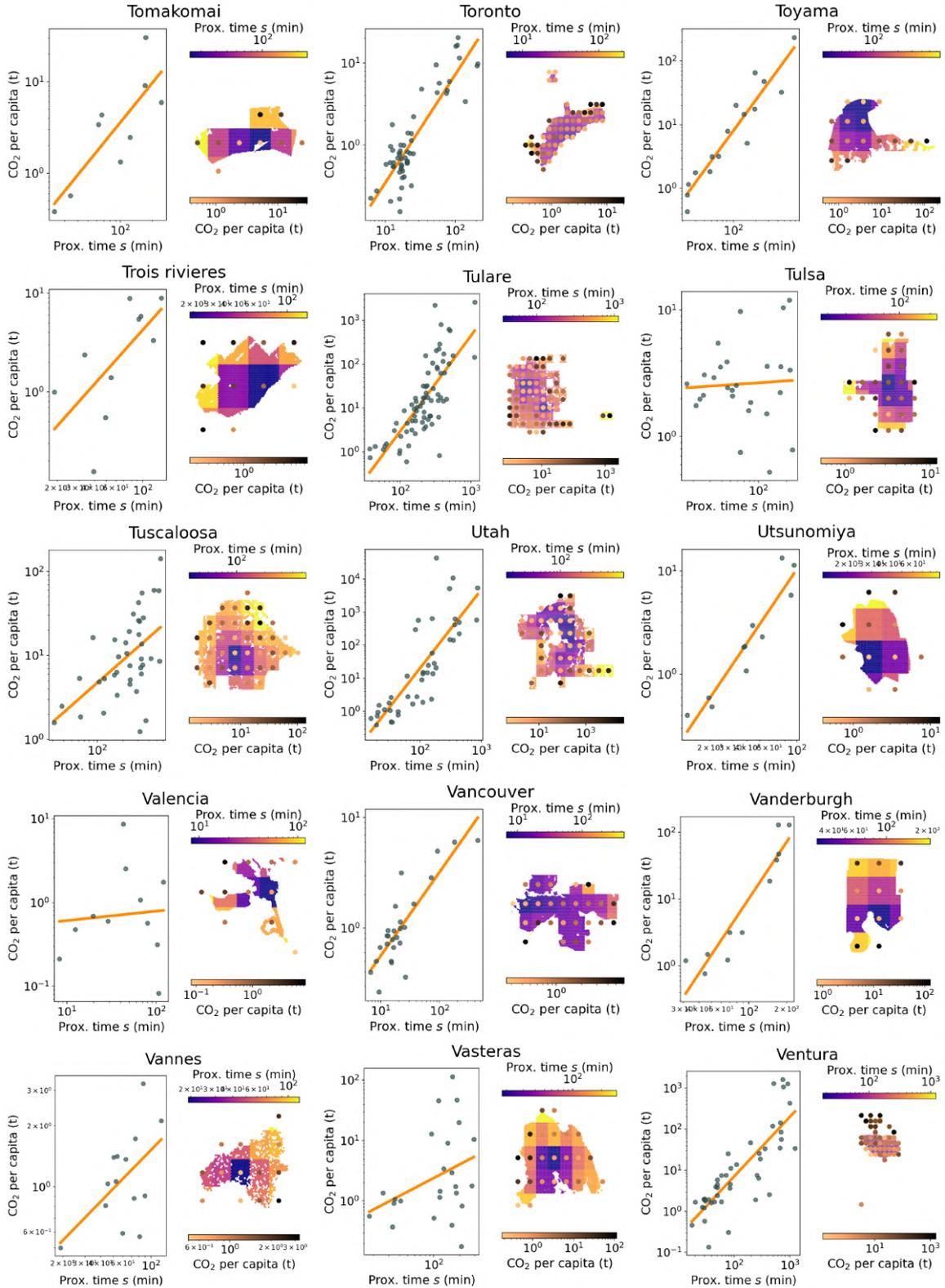

Supplementary Figure 24: **Accessibility/emissions relation at the intra-city level.** For each city, proximity time and per capita road transport emissions of each grid element are represented both as a color-coded map (right panel) and as individual data points in the scatterplot (left panel). The orange line in the scatterplot shows the best fit of a power-law relationship of the form $C_{pc} \sim s^\gamma$.

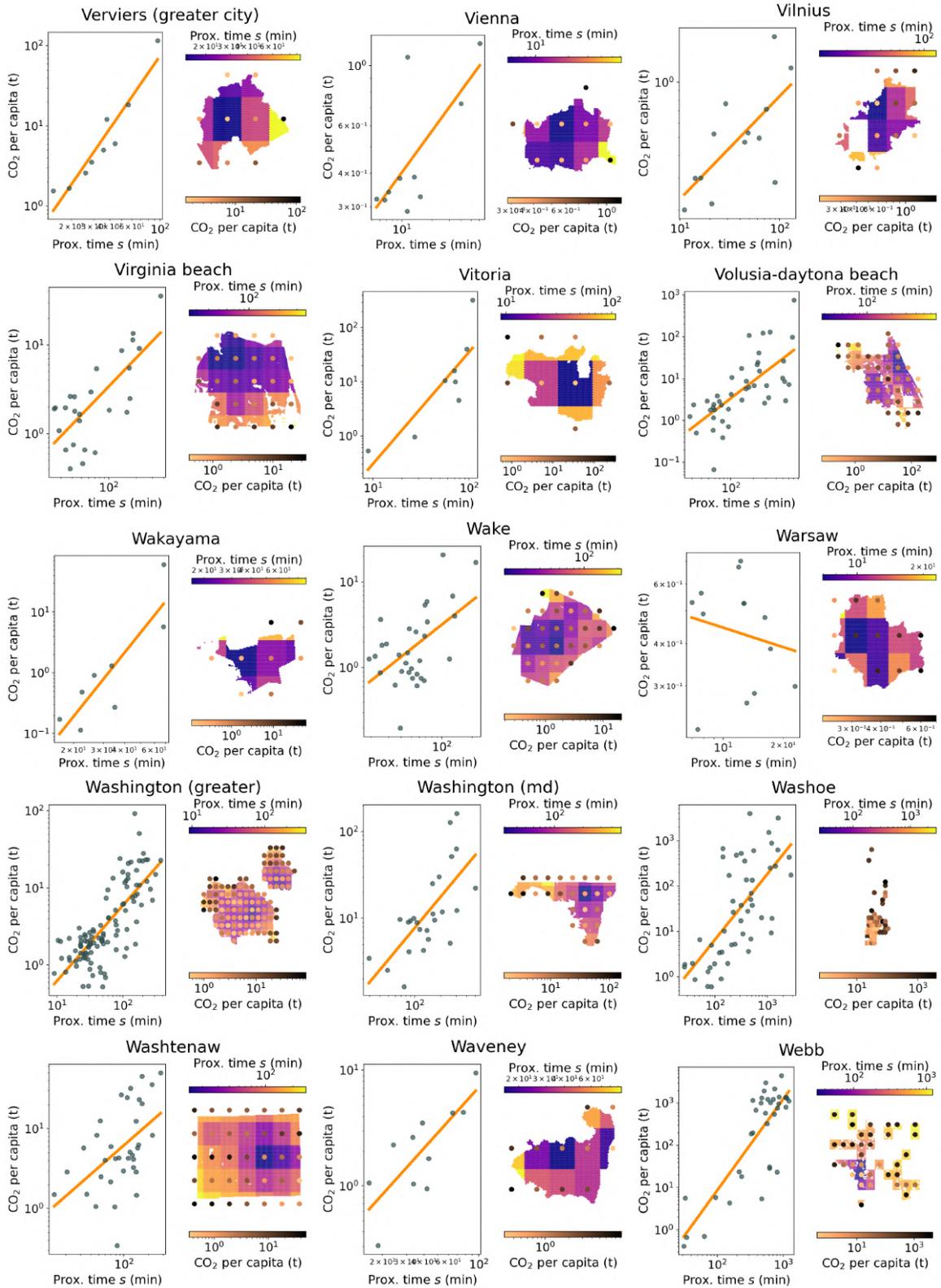

Supplementary Figure 25: **Accessibility/emissions relation at the intra-city level.** For each city, proximity time and per capita road transport emissions of each grid element are represented both as a color-coded map (right panel) and as individual data points in the scatterplot (left panel). The orange line in the scatterplot shows the best fit of a power-law relationship of the form $C_{pc} \sim s^\gamma$.

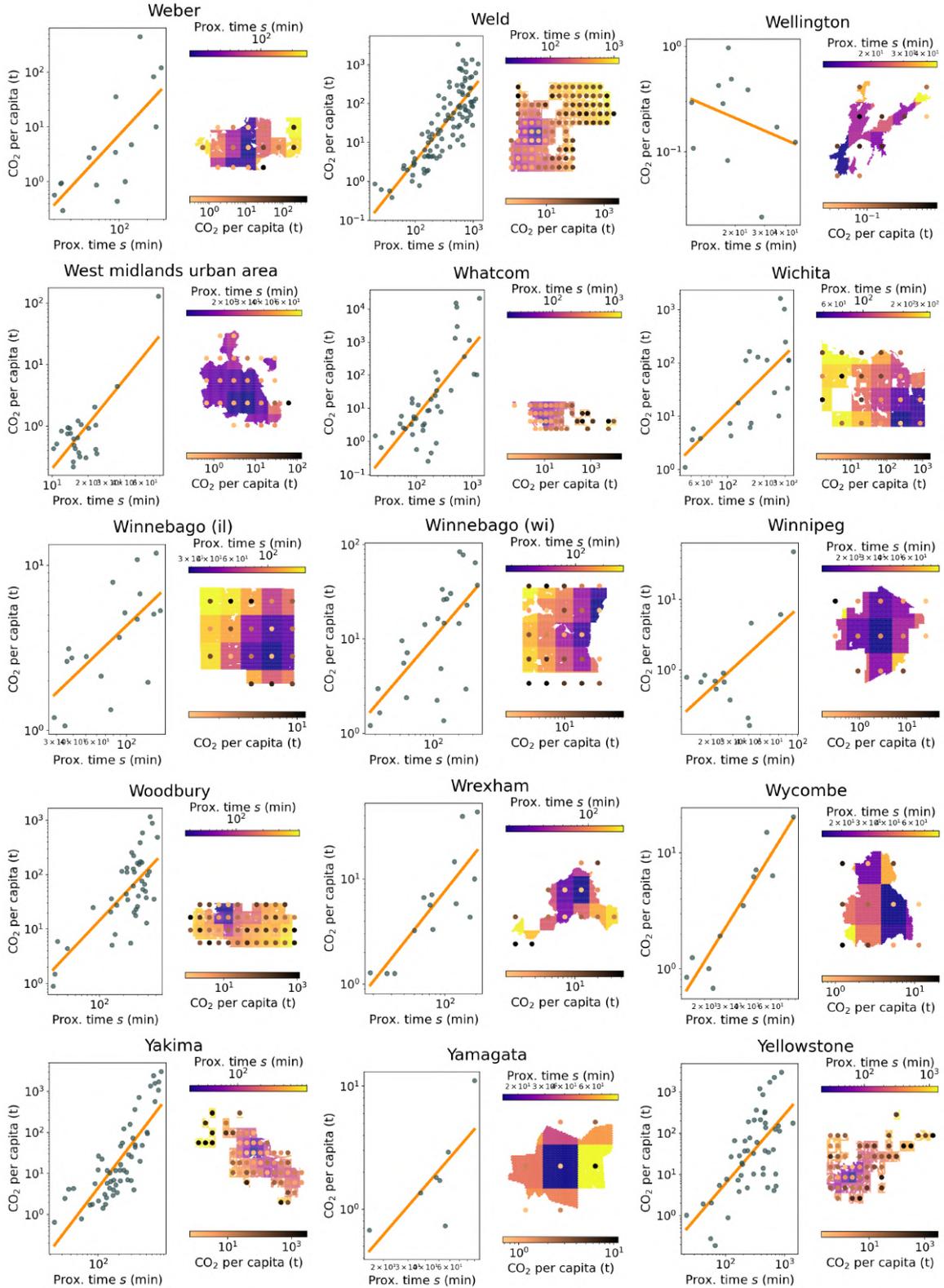

Supplementary Figure 26: **Accessibility/emissions relation at the intra-city level.** For each city, proximity time and per capita road transport emissions of each grid element are represented both as a color-coded map (right panel) and as individual data points in the scatterplot (left panel). The orange line in the scatterplot shows the best fit of a power-law relationship of the form $C_{pc} \sim s^\gamma$.

41

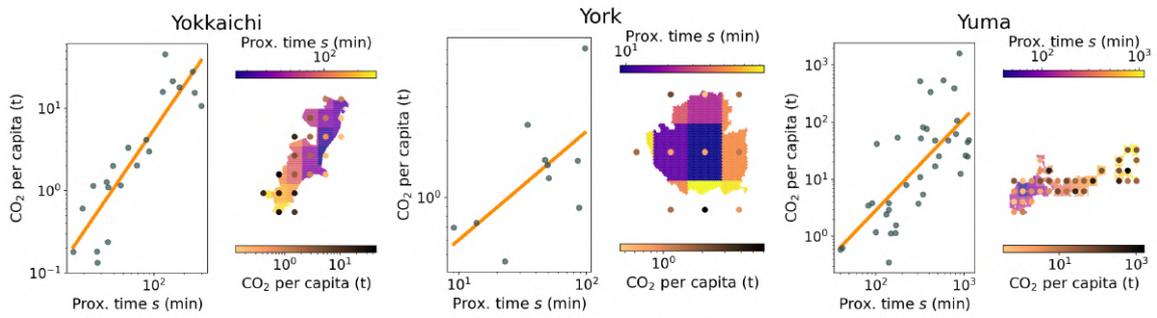

Supplementary Figure 27: **Accessibility/emissions relation at the intra-city level.** For each city, proximity time and per capita road transport emissions of each grid element are represented both as a color-coded map (right panel) and as individual data points in the scatterplot (left panel). The orange line in the scatterplot shows the best fit of a power-law relationship of the form $C_{pc} \sim s^\gamma$.

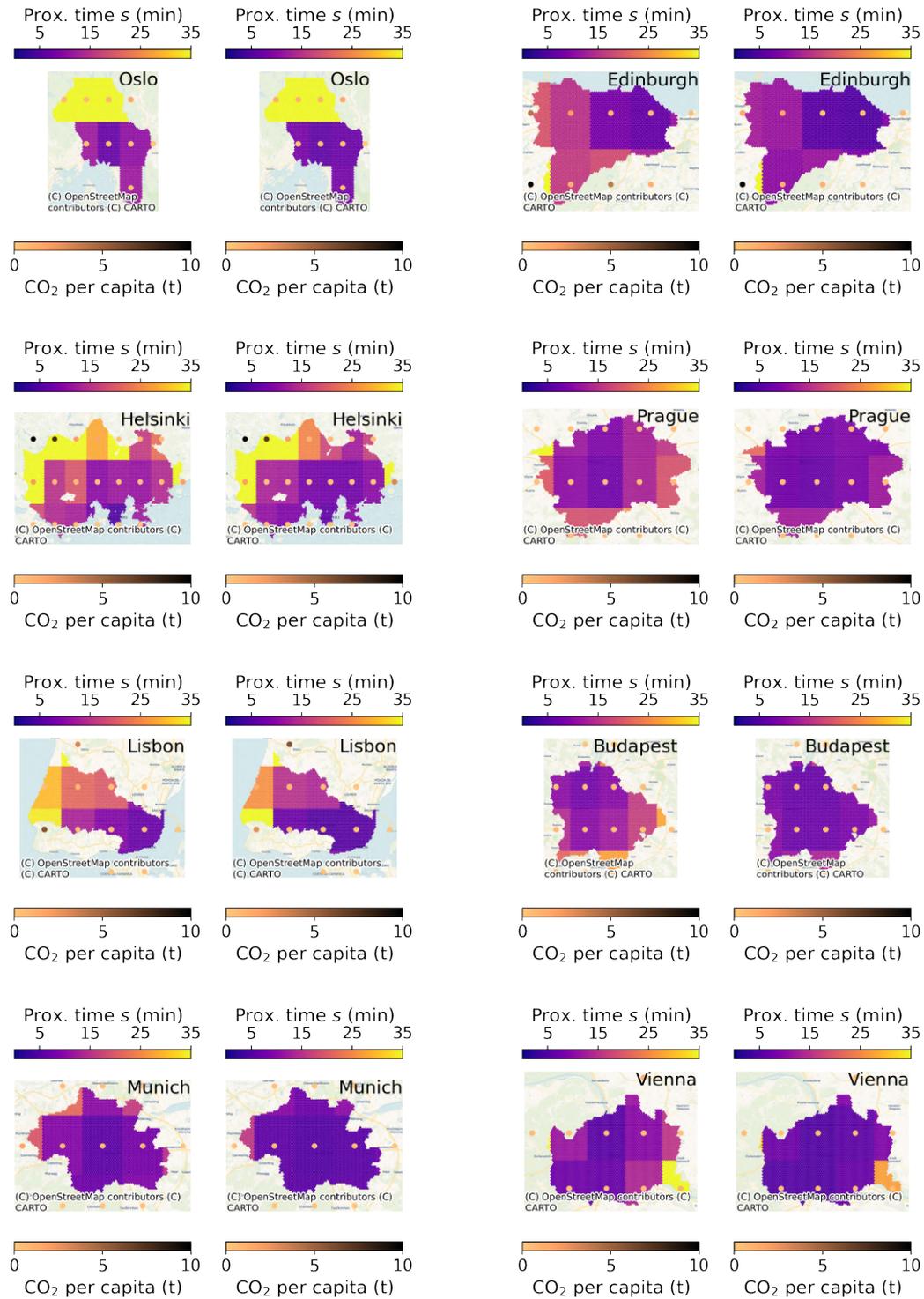

Supplementary Figure 28: **Comparison between actual and optimized scenarios for proximity time and road emissions.** Maps showing population-weighted averages of accessibility and emissions inside cities, as they are now and in the scenario of services moved to locations where they would provide the greatest accessibility on foot to the widest public, following the 15-minute planning ideal.

43

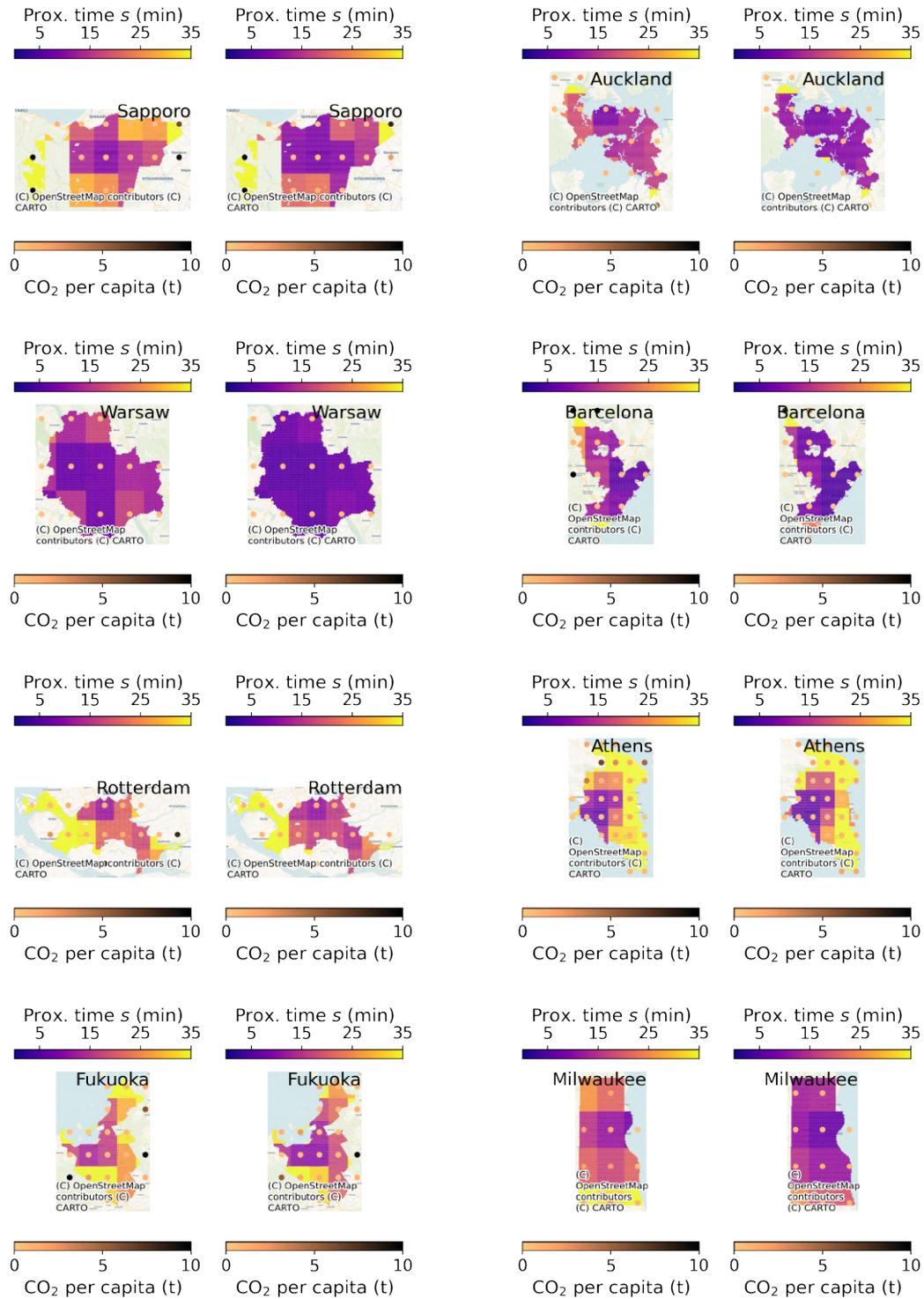

Supplementary Figure 29: **Comparison between actual and optimized scenarios for proximity time and road emissions.** Maps showing population-weighted averages of accessibility and emissions inside cities, as they are now and in the scenario of services moved to locations where they would provide the greatest accessibility on foot to the widest public, following the 15-minute planning ideal.

44

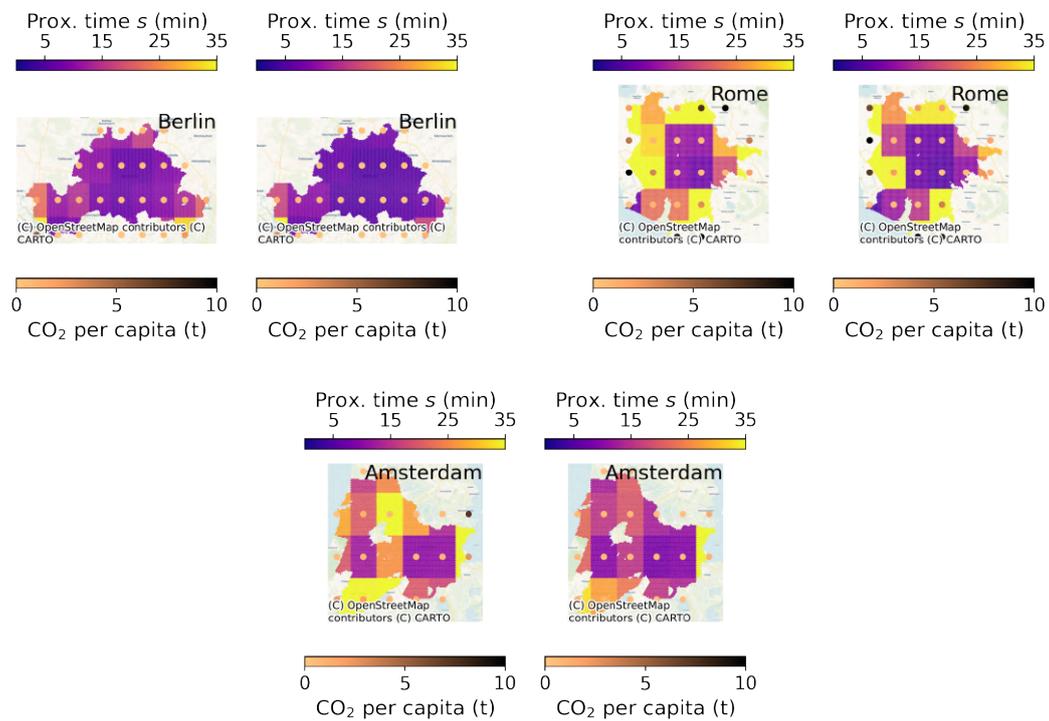

Supplementary Figure 30: **Comparison between actual and optimized scenarios for proximity time and road emissions.** Maps showing population-weighted averages of accessibility and emissions inside cities, as they are now and in the scenario of services moved to locations where they would provide the greatest accessibility on foot to the widest public, following the 15-minute planning ideal.

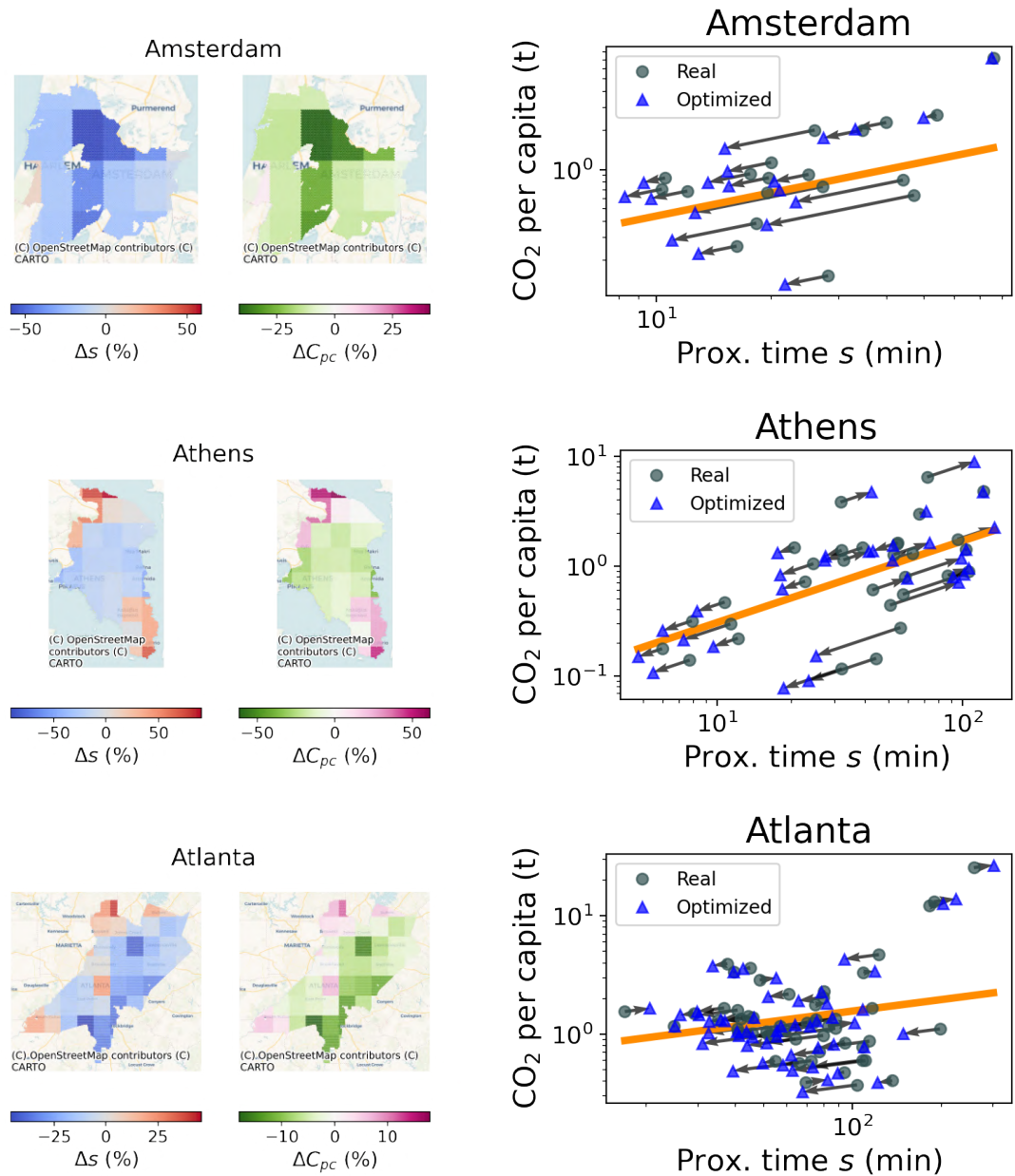

Supplementary Figure 31: **Changes in proximity time and expected road emissions under relocation of services for optimizing foot accessibility.** On the left of each row, maps showing percentage variations of respectively proximity time $s$ and $CO_2$ emissions $C_{pc}$ after optimizing for accessibility by relocating POIs inside cities, in order to provide the greatest accessibility to the widest public. On the right of each row, circles represent the elements of the rectangular grid displayed on the left at present, while triangles represent the accessibility-optimized scenario. Arrows connect markers representing the same element of the grind, in the two scenarios. The orange line represents a power-law fit which models the present relation between emissions and proximity time, for each city.

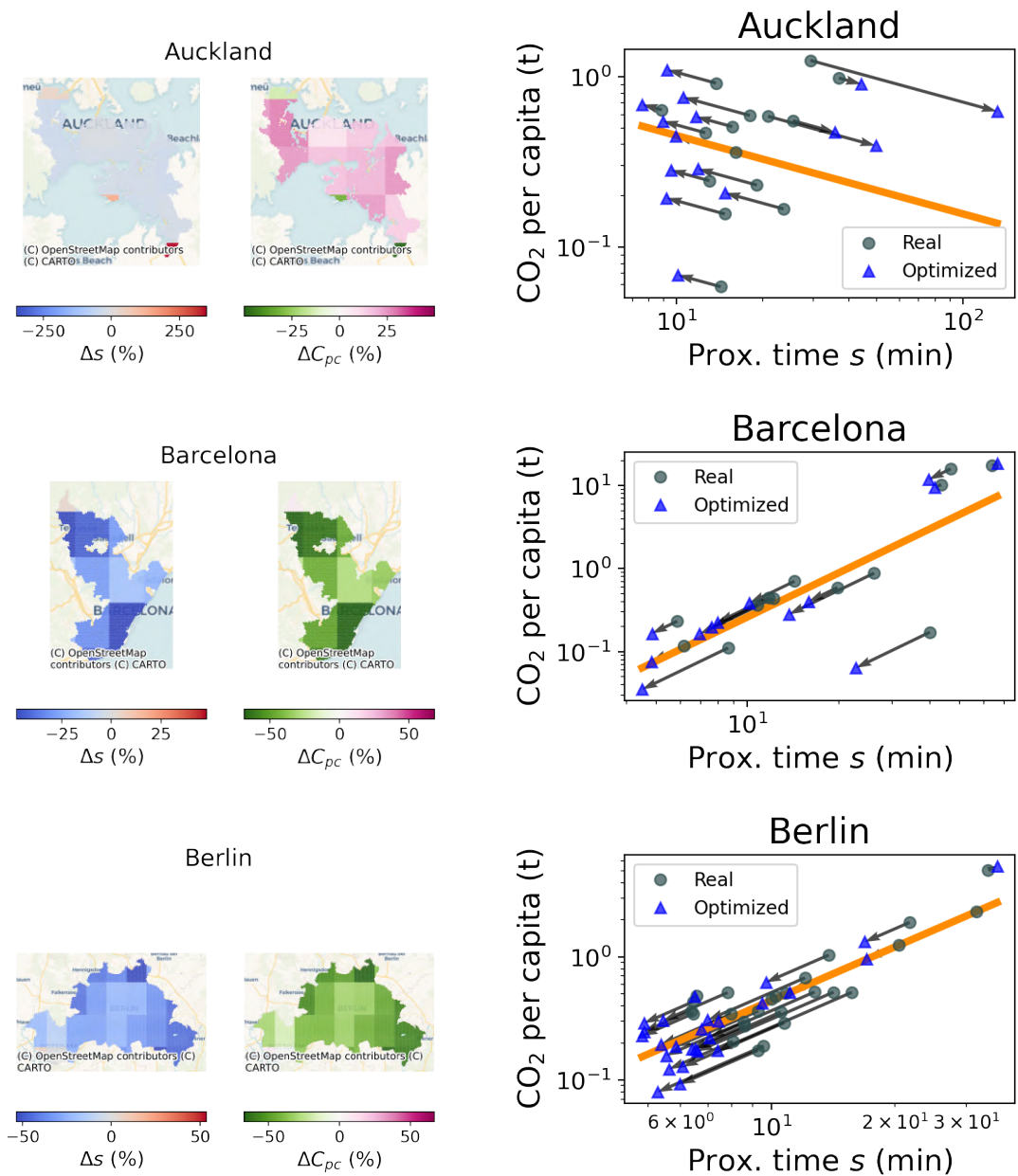

Supplementary Figure 32: **Changes in proximity time and expected road emissions under relocation of services for optimizing foot accessibility.** On the left of each row, maps showing percentage variations of respectively proximity time $s$ and $CO_2$ emissions $C_{pc}$ after optimizing for accessibility by relocating POIs inside cities, in order to provide the greatest accessibility to the widest public. On the right of each row, circles represent the elements of the rectangular grid displayed on the left at present, while triangles represent the accessibility-optimized scenario. Arrows connect markers representing the same element of the grind, in the two scenarios. The orange line represents a power-law fit which models the present relation between emissions and proximity time, for each city.

47

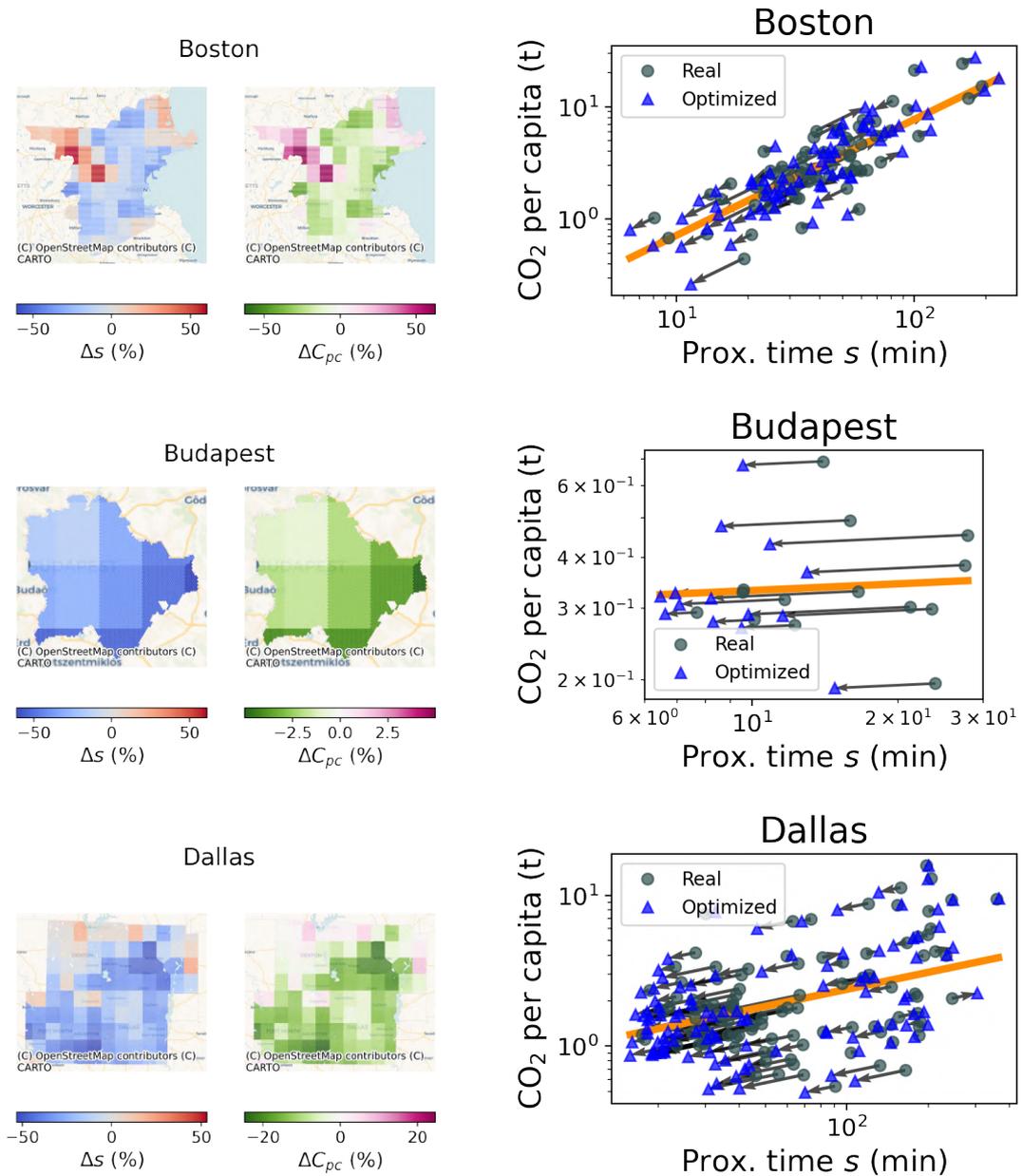

Supplementary Figure 33: **Changes in proximity time and expected road emissions under relocation of services for optimizing foot accessibility.** On the left of each row, maps showing percentage variations of respectively proximity time $s$ and $CO_2$ emissions $C_{pc}$ after optimizing for accessibility by relocating POIs inside cities, in order to provide the greatest accessibility to the widest public. On the right of each row, circles represent the elements of the rectangular grid displayed on the left at present, while triangles represent the accessibility-optimized scenario. Arrows connect markers representing the same element of the grind, in the two scenarios. The orange line represents a power-law fit which models the present relation between emissions and proximity time, for each city.

48

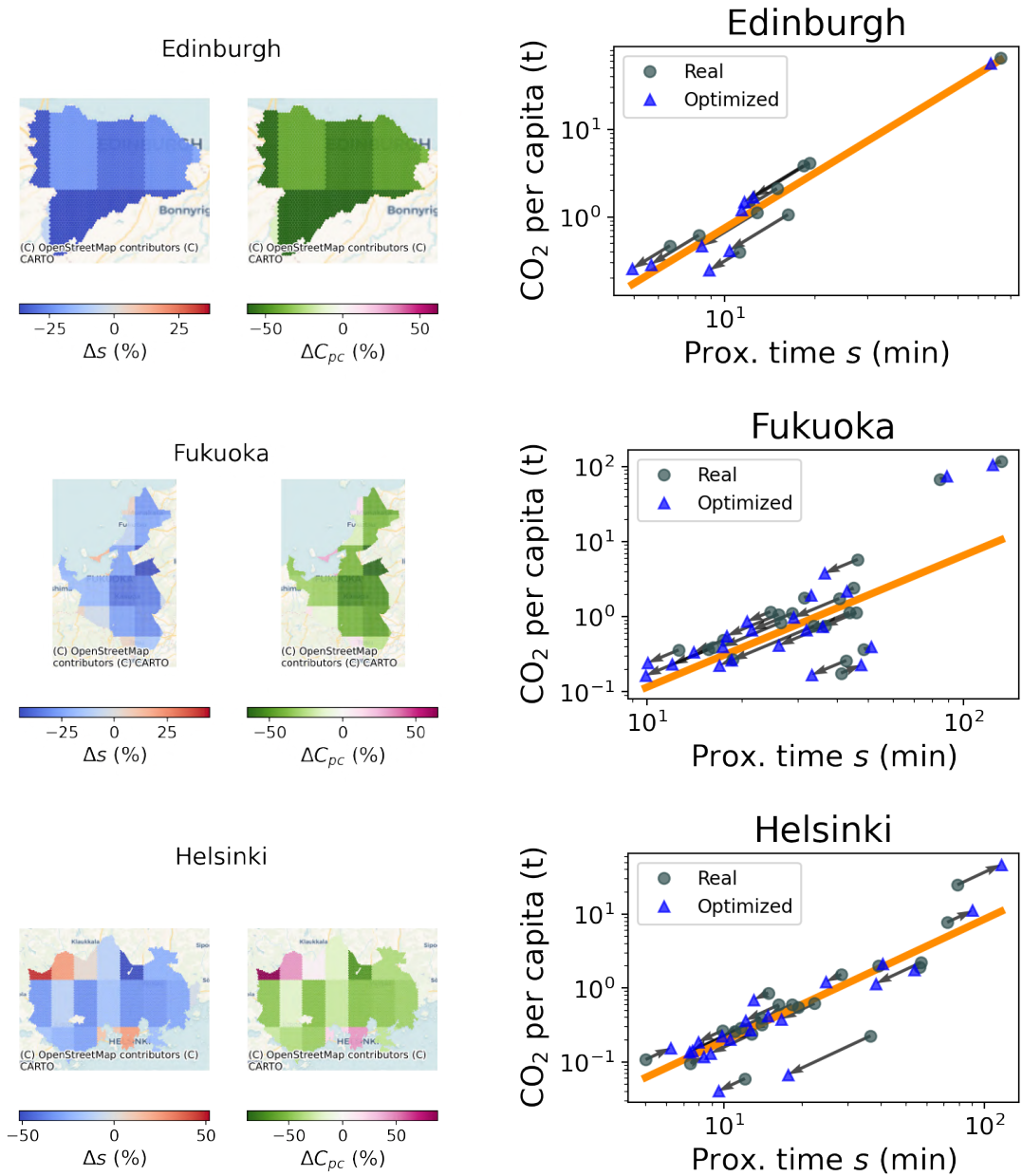

Supplementary Figure 34: **Changes in proximity time and expected road emissions under relocation of services for optimizing foot accessibility.** On the left of each row, maps showing percentage variations of respectively proximity time $s$ and $CO_2$ emissions $C_{pc}$ after optimizing for accessibility by relocating POIs inside cities, in order to provide the greatest accessibility to the widest public. On the right of each row, circles represent the elements of the rectangular grid displayed on the left at present, while triangles represent the accessibility-optimized scenario. Arrows connect markers representing the same element of the grind, in the two scenarios. The orange line represents a power-law fit which models the present relation between emissions and proximity time, for each city.

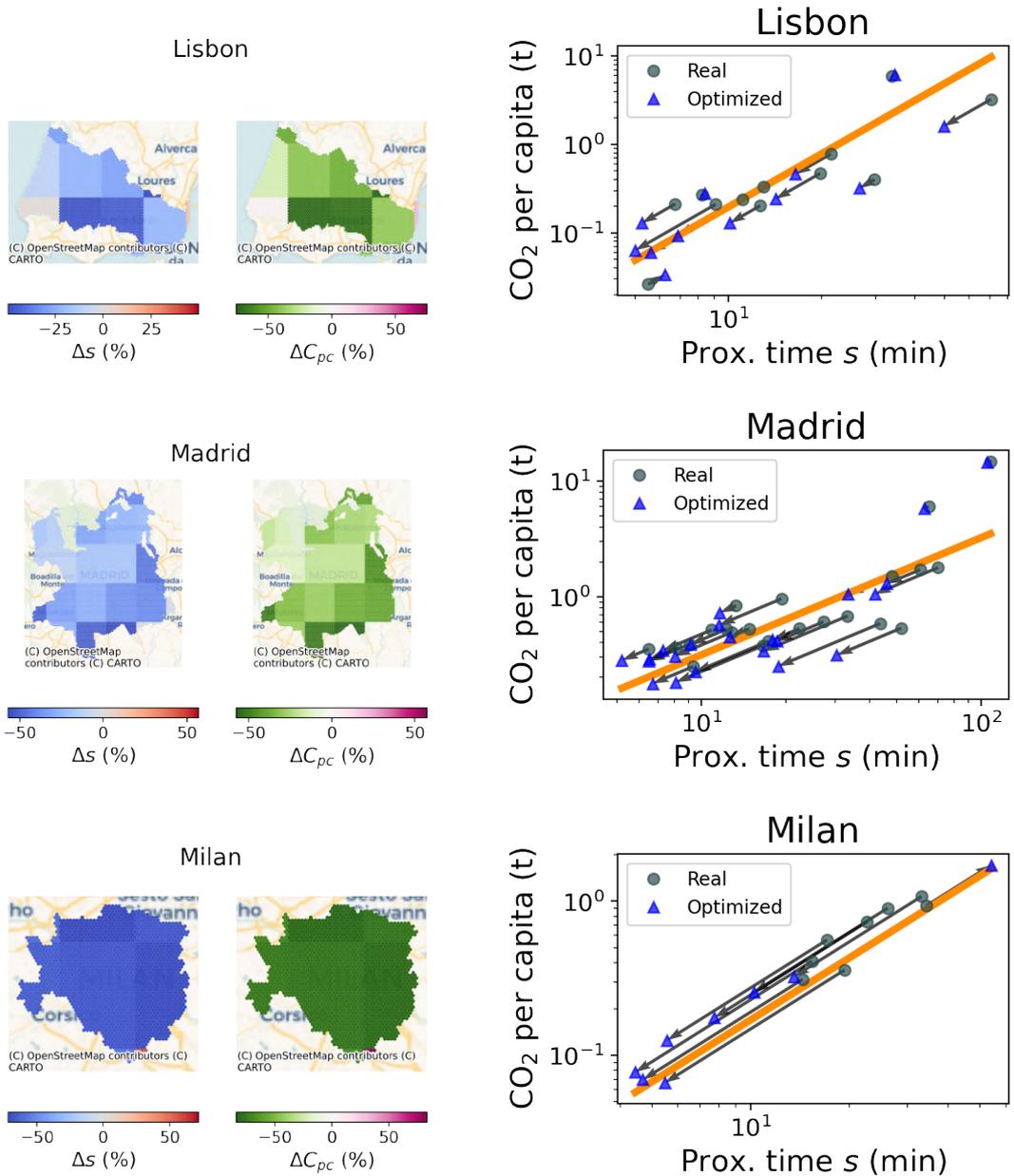

Supplementary Figure 35: **Changes in proximity time and expected road emissions under relocation of services for optimizing foot accessibility.** On the left of each row, maps showing percentage variations of respectively proximity time $s$ and $CO_2$ emissions $C_{pc}$ after optimizing for accessibility by relocating POIs inside cities, in order to provide the greatest accessibility to the widest public. On the right of each row, circles represent the elements of the rectangular grid displayed on the left at present, while triangles represent the accessibility-optimized scenario. Arrows connect markers representing the same element of the grind, in the two scenarios. The orange line represents a power-law fit which models the present relation between emissions and proximity time, for each city.

50

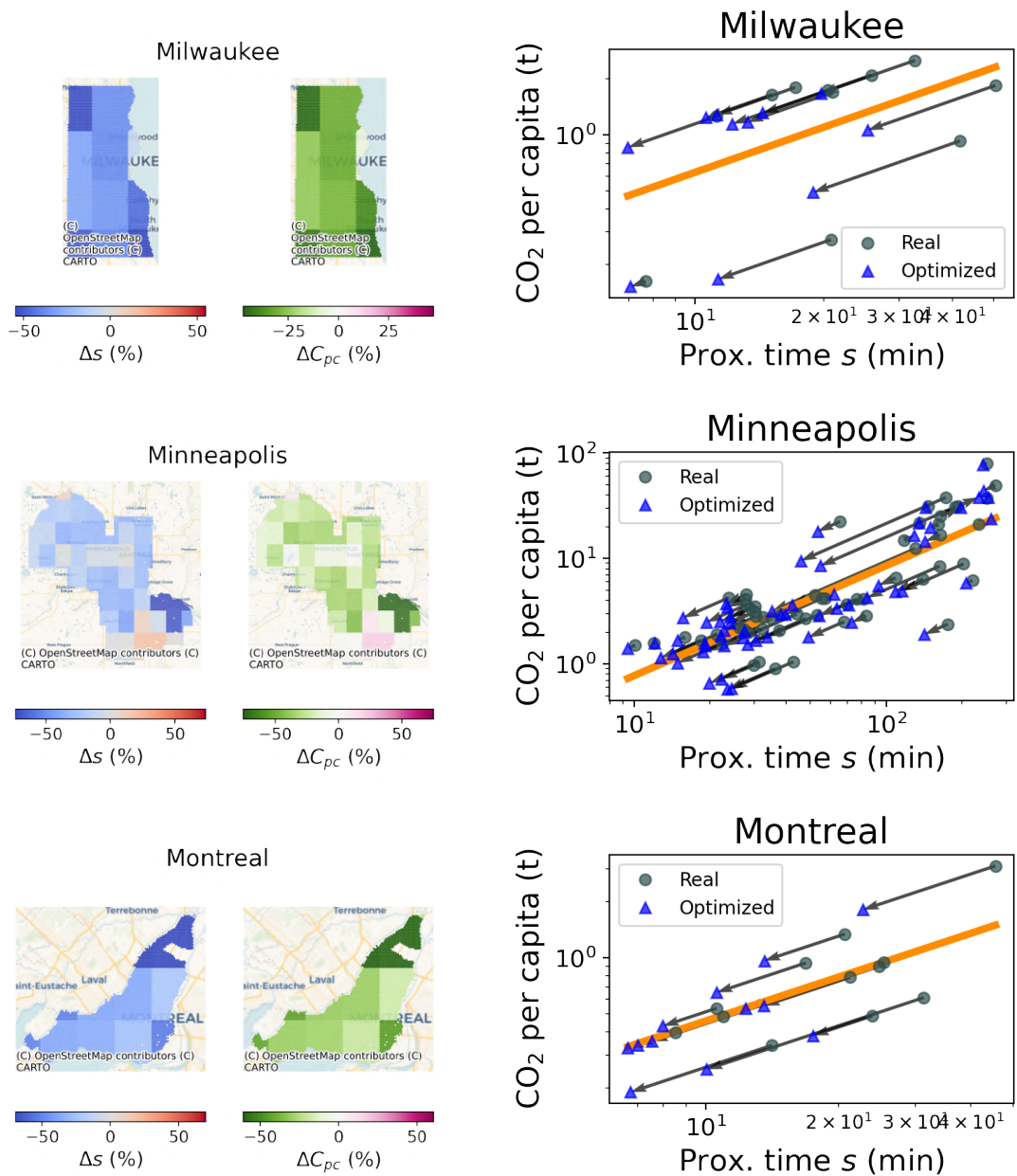

Supplementary Figure 36: **Changes in proximity time and expected road emissions under relocation of services for optimizing foot accessibility.** On the left of each row, maps showing percentage variations of respectively proximity time $s$ and $CO_2$ emissions $C_{pc}$ after optimizing for accessibility by relocating POIs inside cities, in order to provide the greatest accessibility to the widest public. On the right of each row, circles represent the elements of the rectangular grid displayed on the left at present, while triangles represent the accessibility-optimized scenario. Arrows connect markers representing the same element of the grind, in the two scenarios. The orange line represents a power-law fit which models the present relation between emissions and proximity time, for each city.

51

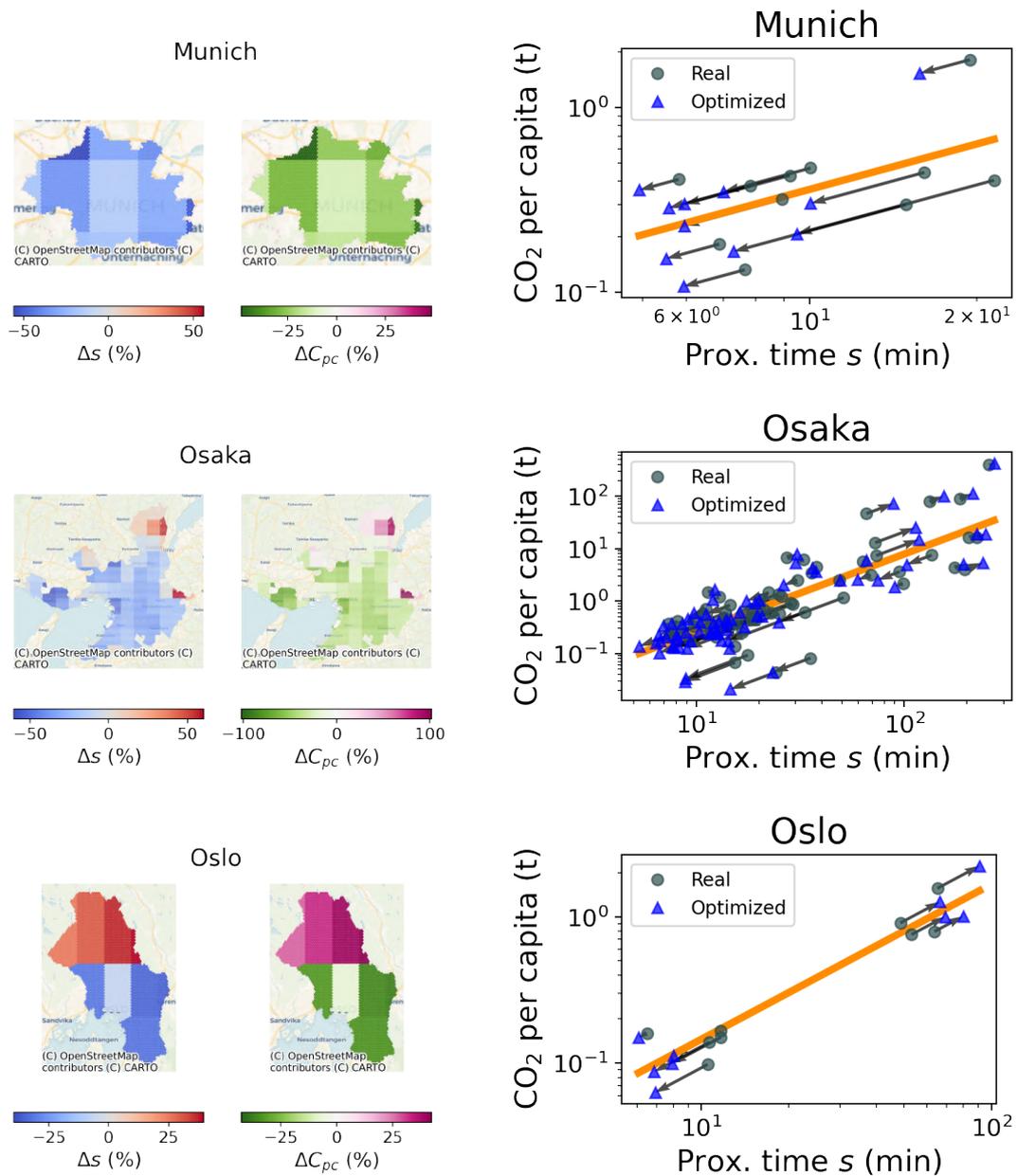

Supplementary Figure 37: **Changes in proximity time and expected road emissions under relocation of services for optimizing foot accessibility.** On the left of each row, maps showing percentage variations of respectively proximity time $s$ and $CO_2$ emissions $C_{pc}$ after optimizing for accessibility by relocating POIs inside cities, in order to provide the greatest accessibility to the widest public. On the right of each row, circles represent the elements of the rectangular grid displayed on the left at present, while triangles represent the accessibility-optimized scenario. Arrows connect markers representing the same element of the grind, in the two scenarios. The orange line represents a power-law fit which models the present relation between emissions and proximity time, for each city.

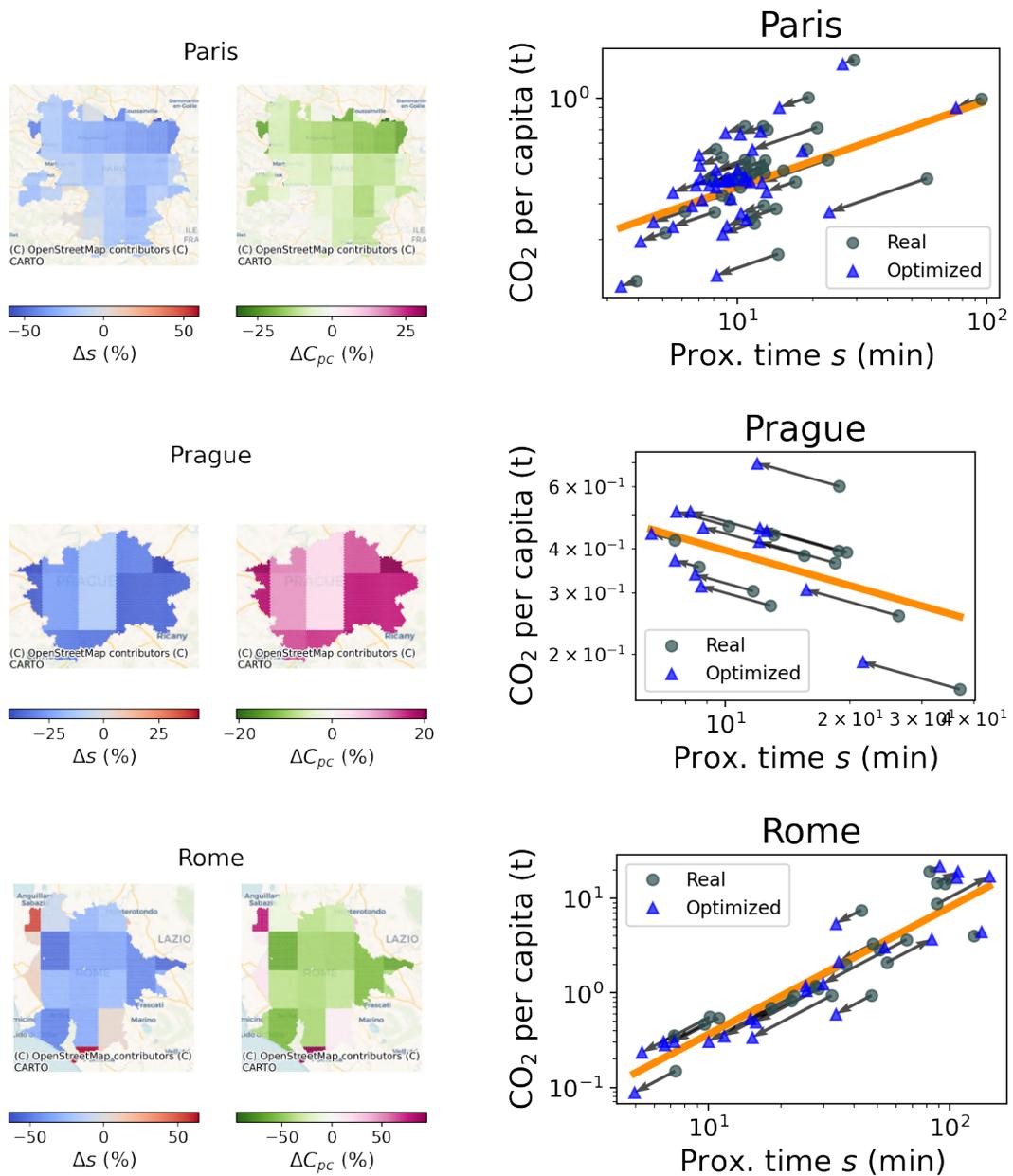

Supplementary Figure 38: **Changes in proximity time and expected road emissions under relocation of services for optimizing foot accessibility.** On the left of each row, maps showing percentage variations of respectively proximity time $s$ and $CO_2$ emissions $C_{pc}$ after optimizing for accessibility by relocating POIs inside cities, in order to provide the greatest accessibility to the widest public. On the right of each row, circles represent the elements of the rectangular grid displayed on the left at present, while triangles represent the accessibility-optimized scenario. Arrows connect markers representing the same element of the grind, in the two scenarios. The orange line represents a power-law fit which models the present relation between emissions and proximity time, for each city.

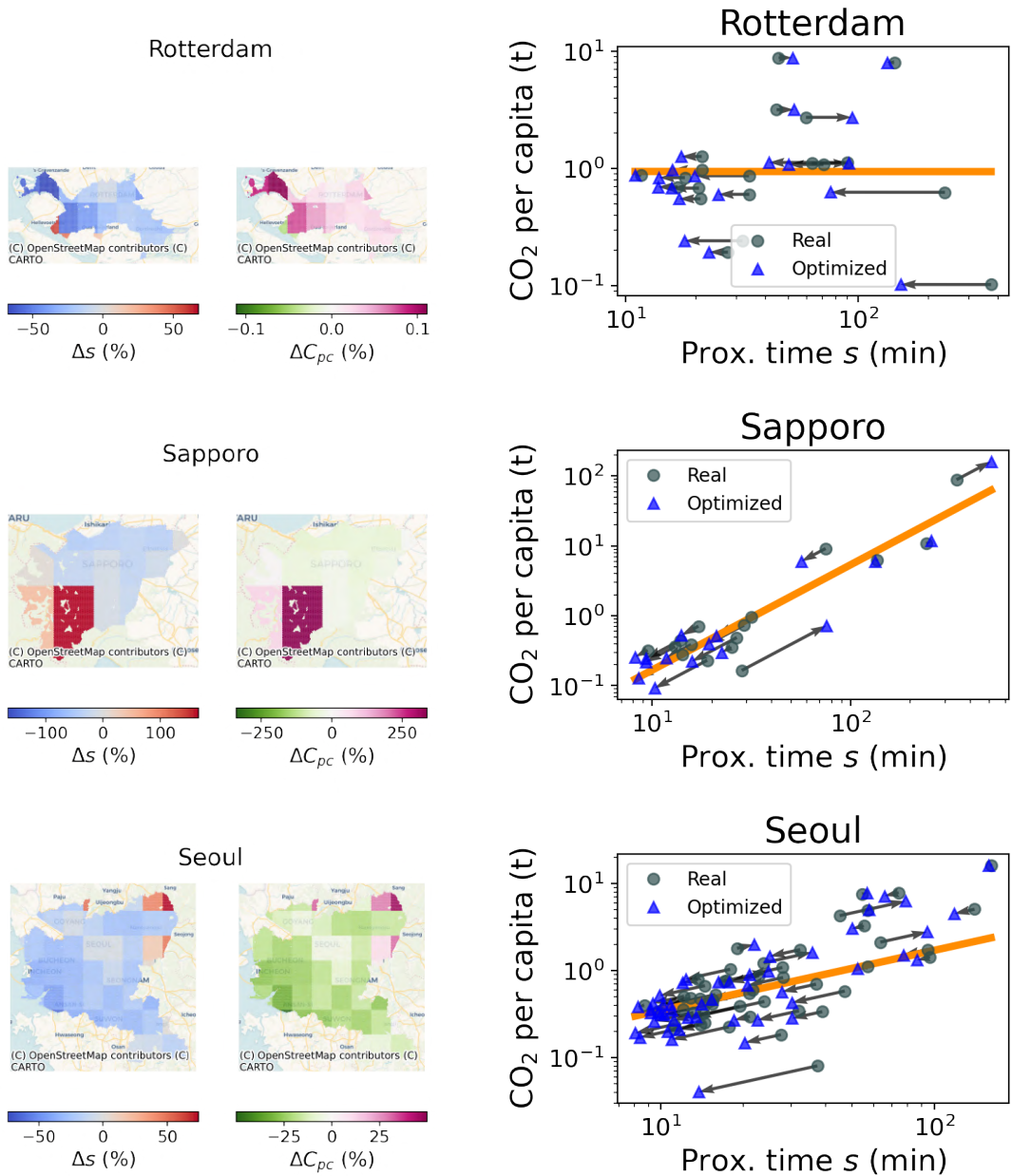

Supplementary Figure 39: **Changes in proximity time and expected road emissions under relocation of services for optimizing foot accessibility.** On the left of each row, maps showing percentage variations of respectively proximity time $s$ and $CO_2$ emissions $C_{pc}$ after optimizing for accessibility by relocating POIs inside cities, in order to provide the greatest accessibility to the widest public. On the right of each row, circles represent the elements of the rectangular grid displayed on the left at present, while triangles represent the accessibility-optimized scenario. Arrows connect markers representing the same element of the grind, in the two scenarios. The orange line represents a power-law fit which models the present relation between emissions and proximity time, for each city.

54

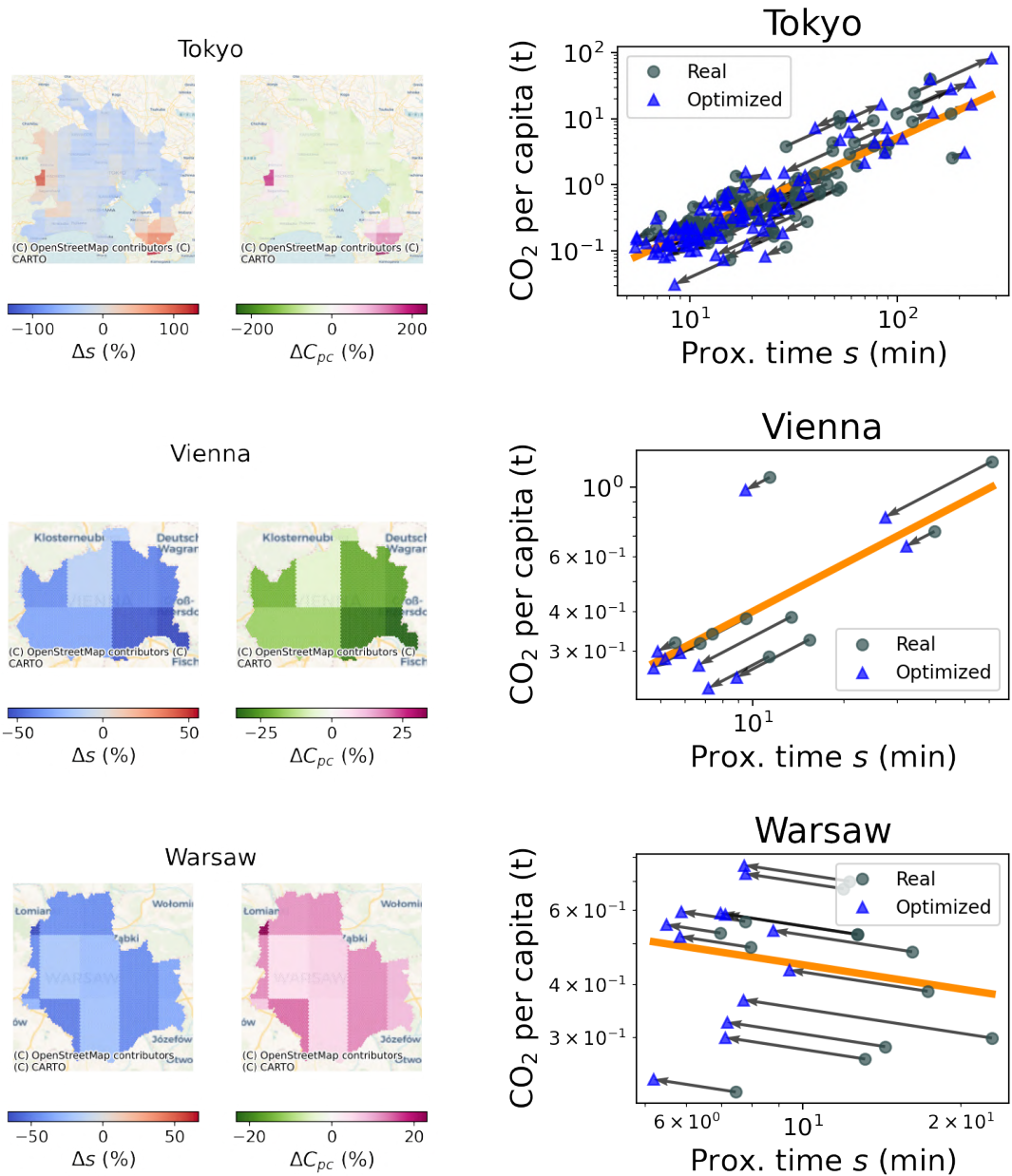

Supplementary Figure 40: **Changes in proximity time and expected road emissions under relocation of services for optimizing foot accessibility.** On the left of each row, maps showing percentage variations of respectively proximity time $s$ and $CO_2$ emissions $C_{pc}$ after optimizing for accessibility by relocating POIs inside cities, in order to provide the greatest accessibility to the widest public. On the right of each row, circles represent the elements of the rectangular grid displayed on the left at present, while triangles represent the accessibility-optimized scenario. Arrows connect markers representing the same element of the grind, in the two scenarios. The orange line represents a power-law fit which models the present relation between emissions and proximity time, for each city.

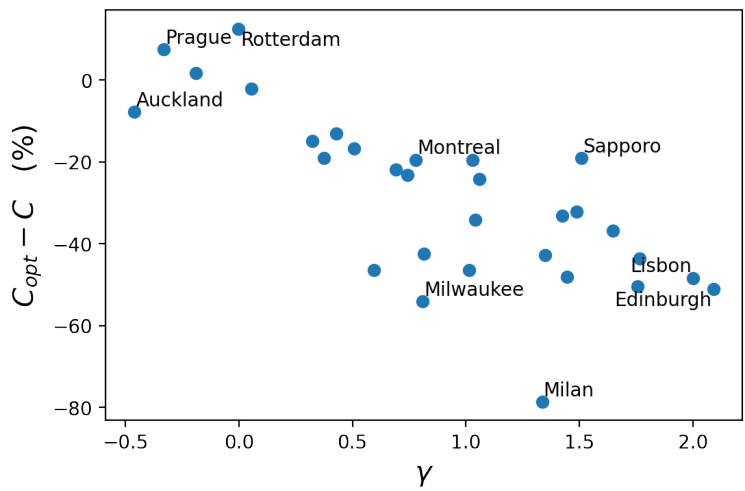

Supplementary Figure 41: **Relation between scaling exponent and emissions variation after foot accessibility optimization.** Global $CO_2$ emissions variation after optimization as a function of the scaling exponent between accessibility and $CO_2$ emissions, inside the city.


\end{document}